\documentclass{tlp}

\RequirePackage{graphics} \RequirePackage{graphicx}
\RequirePackage{alltt}
\RequirePackage{url}

\newcommand{\com}{\rule[0.15ex]{0.4pt}{7pt}}

\newcommand{\false}{{\it false}}
\newcommand{\simp}{{\; \Leftrightarrow \;}}

\def\bq{\begin{quote}}
\def\eq{\end{quote}}

\newcommand{\prop}{{\; \Rightarrow\; }}

\newtheorem{notation}[subsubsection]{Notation}
\newtheorem{example}[subsubsection]{Example}

\newtheorem{corollaire}[subsubsection]{Corollary}
\newtheorem{definition1}[subsubsection]{Definition}
\newtheorem{propriete}[subsubsection]{Property}
\newtheorem{remarque}[subsubsection]{Remark}
\newtheorem{theoreme}[subsubsection]{Theorem}

\newcommand{\vrai}{\mathit{true}}
\newcommand{\fini}{\mathit{finite}}
\newcommand{\faux}{\mathit{false}}
\newcommand{\Ff}{ F}
\newcommand{\Vv}{ V}

  \title[Theory and Practice of Logic Programming]
        {Theory of Finite or Infinite Trees Revisited }

   \author[Khalil Djelloul,  Thi-Bich-Hanh Dao and  Thom Fr\"uhwirth]
         {KHALIL DJELLOUL \\ Faculty of computer science\\ University of Ulm       \\Germany
           \and  THI-BICH-HANH DAO\\
            Laboratoire d'informatique fondamentale d'Orleans\\ Universite d'Orleans \\ France
            \and
           THOM FR\"UHWIRTH\\
            Faculty of computer science\\ University of Ulm\\       Germany  }

\begin{document}

\submitted{15 October 2006}
\revised{6 Mars  2007 }
\accepted{27 June  2007 }

\label{firstpage}

\maketitle
To appear in Theory and Practice of Logic Programming (TPLP)  
\begin{abstract}

We present in this paper a first-order axiomatization of an extended theory $T$
of finite or infinite trees, built on a signature containing an infinite set of
function symbols and a relation $\fini(t)$ which enables to distinguish between
finite or infinite trees. We show that $T$ has at least one model and prove its
completeness by giving not only a decision procedure, but a full first-order
constraint solver which gives clear and explicit solutions for any first-order
constraint satisfaction problem in $T$.  The solver is given in the form of 16
rewriting rules which transform any first-order constraint $\varphi$ into an
equivalent disjunction $\phi$ of simple formulas such that $\phi$ is either the
formula $\vrai$ or the formula $\faux$ or a formula having at least one free
variable, being equivalent neither to $\vrai$ nor to $\faux$ and where the
solutions of the free variables are expressed in a clear and explicit way. The
correctness of our rules implies the completeness of $T$.  We also describe an
implementation of our algorithm in CHR (Constraint Handling Rules) and compare
the performance with an implementation in C++ and that of a recent decision
procedure for decomposable theories.

\end{abstract}


\section{Introduction}

The algebra of finite or infinite trees plays a fundamental role in
computer science: it is a model for data structures, program schemes and program executions.  As early as 1930, J. Herbrand~\cite{herb} gave an 
informal description of an algorithm for unifying finite terms, that is solving equations in finite trees.  A. Robinson~\cite{rob43} rediscovered a similar algorithm when he introduced the resolution procedure for first-order 
logic in 1965.  Some algorithms
with better complexities have been proposed after by M.S. Paterson
and M.N.Wegman~\cite{pat41} and A. Martelli and U.
Montanari~\cite{mat39}.   A good synthesis on this field can be found in the paper of J.P. Jouannaud and C. Kirchner \cite{29}. Solving conjunctions of
equations on infinite trees has been studied by G.
Huet~\cite{hue}, by A. Colmerauer~\cite{col7} and by J.
Jaffar~\cite{jaf28}.  Solving conjunctions of equations
and disequations on finite or infinite trees has been
studied by H.J.
Burckert~\cite{bur6} and A. Colmerauer~\cite{col84}. An incremental algorithm for solving
conjunctions of equations and disequations on rational trees has then 
been proposed  by V.Ramachandran and P. Van
Hentenryck~\cite{ram42} and  a quasi-linear incremental algorithm for testing entailment and disentailment over rational trees has been given by A. Podelski and P. Van Roy~\cite{podelski}. 

On the other hand, K.L. Clark has  proposed a complete axiomatization  of the equality theory, also called  Clark equational theory CET, and gave intuitions about a complete axiomatization of the theory of finite trees \cite{clark}.  B. Courcelle has studied the properties of
infinite trees in the scope of recursive program
schemes~\cite{cou1,cou2} and A. Colmerauer  has described the
execution of Prolog II, III and IV programs in terms of solving
equations and disequations in the algebra of finite or infinite trees 
~\cite{col84,Colmerauer90,ben}.

Concerning quantified constraints,  solving  universally quantified
disequations on finite trees has been  studied by D.A. Smith \cite{45} and there exist some decision procedures which transform any first-order formula into a Boolean combination of quantified conjunctions of atomic formulas using elimination of quantifiers. In the case of finite
trees we can refer to A. Malcev~\cite{mal38}, K.
Kunen~\cite{kun31} and H.
Comon~\cite{com,com14,com15}. For infinite trees,  we can refer to the work of
 H. Comon~\cite{com,com13} and M. Maher \cite{Maher}. 
 
M. Maher has axiomatized all the cases by complete first-order
theories \cite{Maher}. In particular, he has introduced the theory $\cal T$ 
 of finite or infinite trees built on an infinite set $\Ff$ of
function symbols and showed its completeness using a decision procedure which transforms any first-order  formula $\varphi$ into a Boolean combination $\phi$ of quantified conjunctions of atomic formulas.   If $\varphi$ does not contain free variables then $\phi$ is either the formula $\vrai$ or $\false$. 

K. Djelloul has then presented in \cite{moitplp} the class of  decomposable theories and proved that the theory of finite or infinite trees is decomposable. He has also given a decision procedure in the form of five rewriting rules which, for any decomposable theory, transforms any first-order formula $\varphi$ into an equivalent conjunction $\phi$ of solved formulas easily
transformable into a Boolean combination of existentially
quantified conjunctions of atomic formulas. In particular, if
$\varphi$ has no free variables then $\phi$ is either the formula
$\vrai$ or $\neg\vrai$. 

Unfortunately, all the preceding decision procedures are not able to solve complex first-order constraint satisfaction problems in $\cal T$. In fact, these algorithms are only  basic decision procedures and not  full first-order  constraint solvers: they do not warrant that the solutions of the free variables of a solved formula are expressed in a clear and explicit way and can even produce, starting from a formula $\varphi$ which contains free  variables, an equivalent solved formula $\phi$ having  free variables  but being always false or always true in $\cal T$. The appropriate solved formula of $\varphi$ in this case  should be the formula $\faux$ or the formula $\vrai$ instead of $\phi$.  If we use for example the decision procedure of \cite{moitplp} to solve the following formula $\varphi$
\[\neg(\exists y\, x=f(y)\wedge \neg(\exists zw\, x=f(z)\wedge w=f(w))),\]
then we get the following solved\footnote{$\phi$ is solved according to Definition 4.2.4 of \cite{moitplp}} formula $\phi$
 \[\neg(\exists y\, x=f(y)\wedge\neg(\exists z\, x=f(z))).\] The problem is that this formula contains free variables but is always true in the theory of finite or infinite trees. In fact, it is equivalent to \[\neg(\exists y\,  x=f(y)\wedge\neg(\exists z\, x=f(y)\wedge x=f(z))),\] i.e. to \[\neg(\exists y\, x=f(y)\wedge\neg(x=f(y)\wedge(\exists z\,  z=y))),\] thus to \[\neg(\exists y\, x=f(y)\wedge\neg( x=f(y))),\] which is finally equivalent to $\vrai$. As a consequence, the solved formula of $\varphi$ should be $\vrai$ instead of $\phi$. This is a good example which shows the limits of the decision procedures in solving first-order constraints having at least one free variable. 
 
 Much more elaborated algorithms are then needed,  specially when we want to induce solved formulas  expressing   
solutions of complex first-order constraint satisfaction problems in the theory of finite or infinite. Of course, our goal in these kinds of problems is not only to know
  if there exist  solutions or not, but to express these solutions in the
 form of  a solved first-order formula $\phi$ which is either the formula $\vrai$ (i.e. the problem is always satisfiable) or the formula $\faux$ (i.e. the problem is always unsatisfiable) or a simple formula which is neither equivalent to $\vrai$ nor to $\faux$ and where the solutions of  the free variables are expressed in a clear and explicit way.  Algorithms which are able to produce such a formula $\phi$ are  called \emph{first-order constraint solvers}.

We have then presented in \cite{moi3}, not only a decision procedure, but a full first-order  constraint solver in the theory $\cal T$ of finite or infinite trees, in the form of 11 rewriting rules, which gives clear and explicit solutions for any first-order constraint  satisfaction problem in $\cal T$. The intuitions behind this algorithm come from the works of T. Dao in \cite{dao1} where many elegant properties of the theory of finite or infinite trees were given. As far as we know, this is the first algorithm which is able  to do a such work in $\cal T$. 

This  is an extended and detailed  version with full proofs of our previous work on the theory $\cal T$ of finite or infinite trees \cite{moi3}. Moreover, in this paper we  extend  the signature of $\cal T$  by the relation $\fini(t)$ which forces the term $t$ to be a finite tree. Then we extend Maher's axiomatization by two new axioms and show its completeness by giving an extended version of our previous first-order  constraint solver  \cite{moi3}. We also describe a CHR (Constraint Handling Rules) implementation of our rules and compare the performances with those obtained using  a C++ implementation of our solver and the decision procedure for decomposable theories \cite{moitplp}.

\subsection*{Overview of the paper}
This paper is organized in five sections followed by a conclusion.
This introduction is the first section. In  section 2, we introduce the structure of finite or infinite trees and give formal definitions of  trees, finites trees, infinite trees and rational trees. We end this section by presenting particular algebras which handle finite or infinite trees. 

In section 3,  after a brief recall on first-order  logic, we present the five axioms of our extended theory\footnote{We have chosen to denote by $\cal T$ the Maher's theory of finite or infinite trees and by $T$ our extended theory of finite or infinite trees.} $T$ of finite or infinite trees built on a signature containing not only an infinite set of function symbols, but also a relation $\fini(t)$ which enables to distinguish between finite or infinite trees. We then extend  the algebras given at the end of section 2 by the relation $\fini(t)$ and show that these extended  algebras are models of $T$. In particular, we show that the models of sets of nodes, of finite or infinite trees and of rational trees are models of $T$. 

In section 4, we present structured formulas that we call \emph{working formulas} and give some of their properties. These working formulas are extensions of those  given in \cite{moitplp}. We  also introduce  the notion of reachable variables and show that there exist particular formulas which have only quantified reachable variables, do not accept  elimination of quantifiers and cannot  be simplified any further. Such formulas are called \emph{general solved formulas}.  We then present  16 rewriting rules which handle  working formulas and transform an initial working formula into  an equivalent  conjunction of final working formulas from which we can extract easily an equivalent conjunction of general solved formulas. We end this section by a full first-order constraint solver in $T$. This algorithm uses, among other things, our 16 rules and transforms any first-order  formula $\varphi$ into a disjunction $\phi$ of simple formulas  such that $\phi$ is either the formula $\vrai$ or the formula $\faux$ or a  formula having at least one free variable, being equivalent neither to $\vrai$ nor to $\faux$ and where the solutions of the free variables are expressed in a clear and explicit way. The correctness of our algorithm implies the completeness of $T$.  

Finally, in section 5, we give a series of benchmarks. Our algorithm was
implemented in C++ and CHR~\cite{Fru98,book,chrsite}. The C++ implementation is
able to solve formulas of a two player game involving 80 nested alternated
quantifiers.  Even if the C++ implementation is fastest, we found interesting to
see how we can translate our algorithm into CHR rules.  Using this high-level
approach, we will be able to quickly prototype optimizations and variations of
our algorithm and hope to parallelize it.
We also compare the performances with those of C++ implementation of the
decision procedure for decomposable theories\footnote{In \cite{moitplp}, we have
shown that the Maher's theory ${\cal T}$ of finite or infinite trees is
decomposable. We can show easily using a similar proof that our extended theory
$\emph{T}$ is also decomposable.}  \cite{moitplp}.  

The axiomatization of $T$, the proof that $T$ has at least one model, the 16
rewriting rules, the proof of the correctness of our rules, the first-order
constraint solver in $T$, the completeness of $T$, the CHR implementation, the
two player game and the benchmarks are new contributions in this paper.

\section{The structure of finite or infinite trees}
\subsection{What is a tree?}\label{ex}
Trees are well known objects in the computer science world. Here are some of them: 

\includegraphics*[width=13cm]{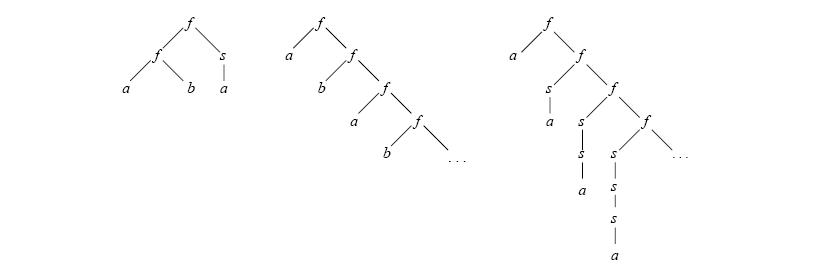}$\\$
their nodes are labeled by the symbols a,b,f,s of respective arities 0,0,2,1. While  the first tree is a \emph{finite tree}, i.e. it has a finite set of nodes, the two others are \emph{infinite trees}, i.e. they have an infinite set of nodes. 

Let us now number from $1$ to $n$ and from left to right the branches that connect each node $l$ to his $n$ sons.  We get:

\includegraphics*[width=13cm]{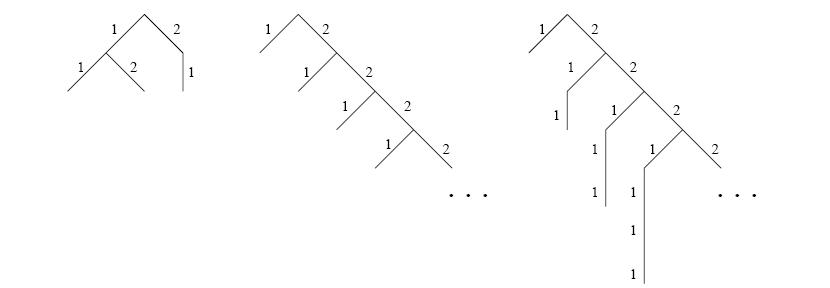}$\\$
 Each node $c$ labeled by $l$ can now be seen as a pair $(p,l)$ where $p$ is the position of the node, i.e. the smallest series of positive integers that we meet if we move from the root of the tree to the node $c$. Thus, the preceding trees can be represented by the following sets of nodes: 
\[\begin{array}{c}
\{(\varepsilon,f),(1,f),(2,s),(11,a),(12,b),(21,a)\}\\
\{(\varepsilon,f),(1,a),(2,f),(21,b),(22,f),(221,a),(222,f),(2221,b),...\}\\
\left\{\begin{array}{c} (\varepsilon,f),(1,a),(2,f),(21,s),(22,f),(211,a),(221,s),(222,f),\\(2211,s),(2221,s),(2222,f),(22111,a),(22211,s),(222111,s),(2221111,a),...\end{array}\right\}
\end{array}\]

Let us now formalize all the preceding statements. Let $L$ be a (possibly infinite) set. Its elements are called \emph{labels}. To each label $l\in L$ is linked a non-negative integer called \emph{arity of $l$}. An $n$-ary label is a label of arity $n$. A \emph{position} is a word built on strictly positive integers (the empty word is denoted by $\varepsilon$). Let $p$ be a position and $l$ a label. The pair $(p,l)$ is called \emph {node} and its \emph{depth} is the length\footnote{ As usual, the length of the empty word $\varepsilon$ is $0$.} of $p$. An $n$-ary node is a node whose label is of arity $n$. A \emph{root} is a  node of depth $0$. The \emph{row} of an $n$-ary node, with $n\neq 0$, is the last integer of its position. We say that $c$ is the father of $c'$ or $c'$ is the son of $c$ if $c$ and $c'$ are nodes whose positions are respectively of the form $i_1...i_k$ and $i_1...i_ki_{k+1}$, where the $i_j$'s are strictly positive integers and $k$ a (possibly null\footnote{Of course, for $k=0,\,i_{1}...i_{k}$ is reduced to $\varepsilon$.}) positive integer. Let us denote by $N$ the set of the nodes labeled by elements of $L$.

\begin{definition1}
A node $c$ of $N$ is called \emph{arborescent} in a sub-set $N_1$ of $N$ if $N_1\neq\emptyset$ and either $c\not\in N_1$, or $c\in N_1$ and the two following conditions hold: \begin{itemize}
\item  $N_1-\{c\}$ does not contain any node whose position is the same than those of  $c$,
\item
$c$ is either a root or the son of an $n$-ary node of $N_1$ which has exactly $n$ sons in $N_1$ of respective rows $1,...,n$. 
\end{itemize}
\end{definition1}

We can now define formally a \emph{tree}:
\begin{definition1}\label{arbre}
A \emph{tree} $tr$ is a sub-set of $N$ such that each element of $N$ is arborescent in $tr$. A \emph{finite} tree is a tree whose set of nodes is finite. An \emph{infinite} tree is a tree whose set of nodes is infinite.
\end{definition1}

Let us now define the notion of subtree: 
\begin{definition1}
Let $tr$ be a tree. The subtree linked to a node $(i_1...i_k,l)$ of $tr$ is the set of the nodes of the form $(i_{k+1}...i_{k+n},l')$ with $(i_{1}...i_{k+n},l')\in tr$ and\footnote{Of course, for $n=0,\,(i_{k+1}...i_{k+n},l')$ is reduced to $(\varepsilon,l')$.}  $n\geq 0$.  We call  \emph{subtree of $tr$} a subtree linked to one of the nodes of $tr$.  A subtree of $tr$ of depth $k$ is a subtree  linked to a node of $tr$ of depth $k$.

\end{definition1}
From Definition \ref{arbre}, we deduce that each subtree of a tree $tr$ is also a tree. 
\begin{definition1}
A rational tree is a tree whose set of subtrees is a finite set.
\end{definition1}
Note that an infinite tree can be rational. In fact, even if its set of nodes is infinite but $n$ subtrees linked to $n$ different nodes can be similar. Let us see this in the following example: 
\begin{example}
Let us consider the three trees presented in the beginning of Section \ref{ex}. Let us name them from left to righ by: $tr_1$, $tr_2$ and $tr_3$. The set of the subtrees of $tr_1$ is the following \emph{finite} set: 
\begin{center}
\[\left\{\begin{array}{l}
\{(\varepsilon,a)\},\\\{(\varepsilon,b)\},\\\{(\varepsilon,s),(1,a)\},\\\{(\varepsilon,f),(1,a),(2,b)\},\\\{(\varepsilon,f),(1,f),(2,s),(11,a),(12,b),(21,a)\}\\

\end{array}\right\}
\]
\end{center}
i.e. 

\includegraphics*[width=13cm]{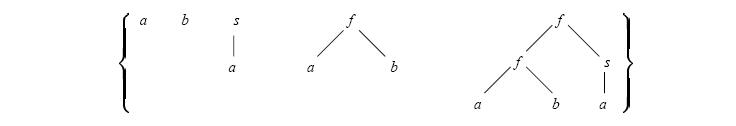}$\\$
The set of the subtrees of $tr_2$ is the following \emph{finite} set:
\begin{center}
\[\left\{\begin{array}{l}
\{(\varepsilon,a)\},\\\{(\varepsilon,b)\},\\\{(\varepsilon,f),(1,a),(2,f),(21,b),(22,f),(221,a),...\},\\\{(\varepsilon,f),(1,b),(2,f),(21,a),(22,f),(221,b),...\}\\

\end{array}\right\}
\]
\end{center}
i.e. 

\includegraphics*[width=13cm]{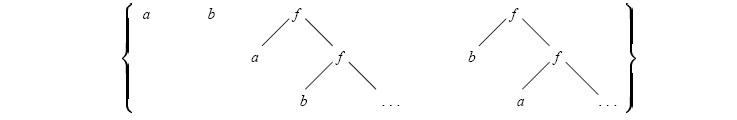}$\\$
The set of the subtrees of $tr_3$ is the following \emph{infinite} set:
\begin{center}
\[\left\{\begin{array}{l}
\{(\varepsilon,a)\},\\\{(\varepsilon,s),(1,a)\},\\\{(\varepsilon,s),(1,s),(11,a)\},\\\{(\varepsilon,s),(1,s),(11,s),(111,a)\},\\...\\\{(\varepsilon,f),(1,a),(2,f),(21,s),(22,f),...\},\\\{(\varepsilon,f),(1,s),(2,f),(11,a),(21,s),(22,f),...\}\\
...
\end{array}\right\}
\]
\end{center}
i.e. 

\includegraphics*[width=12cm]{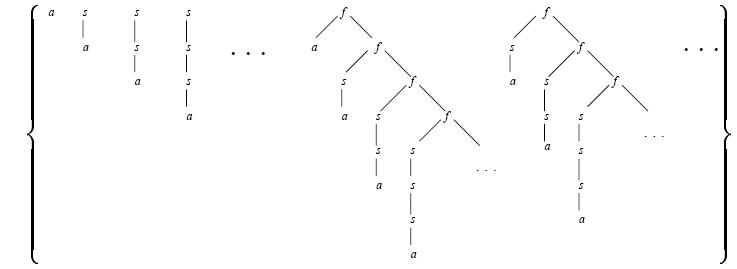}$\\$
Note  that the tree $tr_1$ has a finite set of nodes and a finite set of subtrees. Thus,  it is a finite rational  tree. The tree $tr_2$ has an infinite set of nodes but a finite set of subtrees. Thus, it is an infinite rational tree. The tree $tr_3$ has an infinite set of nodes and an infinite set of subtrees. Thus, it is an infinite non-rational tree. 

Note also that a rational tree can always be represented by a  \emph{finite directed graph}. For that, it is enough to merge all the nodes whose linked subtrees are similar. A non-rational tree cannot  be represented by a finite directed graph. In this case, only an \emph{infinite directed graph} representation will be possible. For example, the trees $tr_1$, $tr_2$ and $tr_3$ can be represented as follows: 

\includegraphics*[width=13cm]{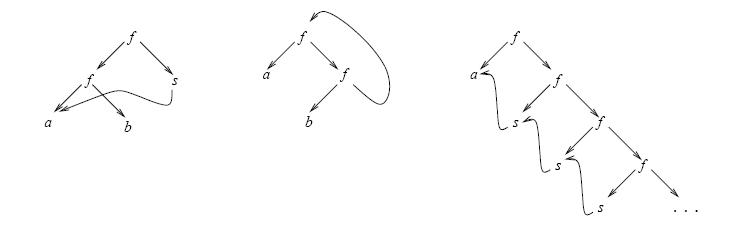}$\\$
Of course, two different directed graphs can represent the same tree. For example the trees $tr_2$ and $tr_3$ can also be represented as follows: 

\includegraphics*[width=13cm]{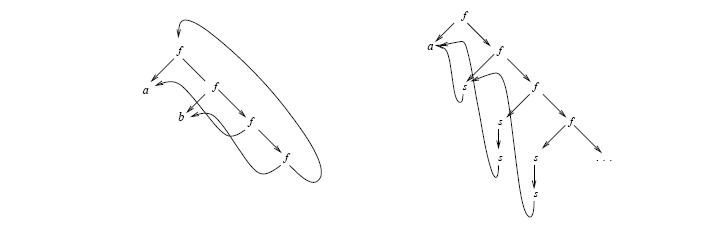}$\\$
\end{example}

\subsection{Construction operations}\label{alg}

We would like to provide the set $Tr$ of finite or infinite trees with a set of \emph{construction operations} ;  one for each label $l$ of $L$. These operations will be schematized  as follows: 

\includegraphics*[width=15cm]{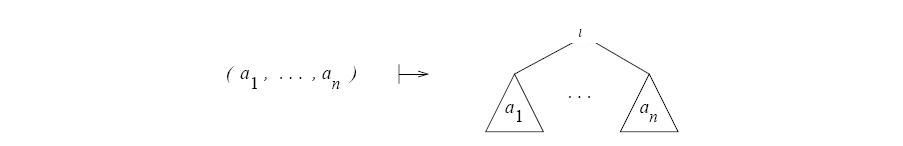}$\\$
with $n$ the arity of the label $l$. In order to formally define  these construction operations, we need first to define them  in the set $D$ of sets of nodes\footnote{In other words, each element of $D$ is a set of nodes, i.e. a subset of $N$.} of $N$. Let $i$ be a strictly positive integer. If $d=(j_1...j_k,l)$ is a node then we denote by $i.d$ the node $(ij_1...j_k,l)$. If $a$ is a set of nodes (i.e.  $a\in D$),  then we denote by $i.a$ the set of nodes $\{i.d\mid d\in a\}$.
$\\$

\begin{definition1}\label{cons}
In the set $D$, the construction operation linked to the $n$-ary label $l$ is the application $l^D:(a_1,...,a_n)\mapsto\{(\varepsilon,l)\}\cup 1.a_1\cup...\cup n.a_n$ with $a_1...a_n$ elements of $D$.
\end{definition1}

\begin{remarque}\label{25}
Let $a$ be an  element of $D$. Let us denote by $\nu_k(a)$ the set of  nodes of $a$ of depth $k$. Many remarks must be stated concerning any elements $a$, $a_i$ and $b$ of $D$:

\begin{enumerate}
\item
$a=b\leftrightarrow\bigwedge_{k=1}^{\infty} \nu_k(a)=\nu_k(b)$.
\item
$\nu_0(l^D(a_1,...,a_n))=\{(\varepsilon,l)\}$.
\item
For all $k\geq 0$, there exists a function $\varphi_{k+1}$ which is independent from all the $\nu_{k+1}(a_i)$, with $i\in\{1,...,n\}$, such that $\nu_{k+1}(l^D(a_1,...,a_n))=\varphi_{k+1}(\nu_k(a_1),...,\nu_k(a_n))$.
\item
The elements of $\nu_0(l^D(a_1,...,a_n))$ are arborescent in $l^D(a_1,...,a_n)$.
\item
For all $k\geq 0$, the elements of $\nu_{k+1}(l^D(a_1,...,a_n))$ are arborescent in $l^D(a_1,...,a_n)$ if and only if, for each $i\in\{1,...,n\}$, the elements of $\nu_k(a_i)$ are arborescent in $a_i$.
\item
If for all $k\geq 0$ the elements of $\nu_{k}(l^D(a_1,...,a_n))$ are arborescent in $l^D(a_1,...,a_n)$ then each element of $N$ is arborescent in $l^D(a_1,...,a_n)$.
\end{enumerate}
\end{remarque}

Let now $F$ be an infinite set of function symbols. Let us denote by:\begin{itemize}\item  $N$ the set of the nodes labeled by $F$,\item $D$ the set of sets of nodes of $N$, \item $Tr$ the set of the elements of $D$ which are trees,\item $Ra$ the set of the elements of $Tr$ which are rational, \item $Fi$ the set of the elements of $Tr$ which are finite.\end{itemize}
If $f$ is an $n$-ary function symbol taken from $F$ then the operation of construction $f^D$ associated to $f$ is an application of the form $D^n\rightarrow D$. Let $tr_1,...,tr_n$ be elements of $Tr$. From the fourth and fifth point of Remark \ref{25} we deduce that $f^D(tr_1,...,tr_n)$ is also a tree, i.e. an element of $Tr$. Thus, we can introduce the following  application:\begin{center}\begin{tabular}{l} $f^{Tr}:(tr_1,...,tr_n)\mapsto f^D(tr_1,...,tr_n)$ which is of type $Tr^n\rightarrow Tr$.\end{tabular}\end{center}
On the other hand, the set of the subtrees of the tree $f^D(tr_1,...,tr_n)$ is obtained by the union of the sets of the subtrees of all the $tr_i$ plus the tree $f^D(tr_1,...,tr_n)$. Thus, if all the $tr_i$'s are rational  trees then the tree $f^D(tr_1,...,tr_n)$ is rational. As a consequence, we can introduce the following application:  
\begin{center}\begin{tabular}{l} $f^{Ra}:(tr_1,...,tr_n)\mapsto f^D(tr_1,...,tr_n)$ which is of type $Ra^n\rightarrow Ra$.\end{tabular}\end{center}
Finally, if all the  $tr_i$'s are finite trees, then the tree $f^D(tr_1,...,tr_n)$ is  finite. Thus, we can introduce the following application: 
\begin{center}\begin{tabular}{l} $f^{Fi}:(tr_1,...,tr_n)\mapsto f^D(tr_1,...,tr_n)$ which is of type $Fi^n\rightarrow Fi$.\end{tabular}\end{center}
The pairs $<D,(f^{D})_{f\in F}>$, $<Tr,(f^{Tr})_{f\in F}>$,  $<Fi,(f^{Fi})_{f\in F}>$ and $<Ra,(f^{Ra})_{f\in F}>$ are known as the \emph{algebras} of sets  of nodes, of finite or infinite trees, of finite trees and of rational trees.

\section{The extended theory $T$ of finite or infinite trees}
\subsection{Formal preliminaries}
\subsubsection{Formulas}
We are given once and for all an infinite countable set $\Vv$ of {\em
variables} and the set $L$ of \emph{logical} symbols:
\[=, \vrai, \faux,\neg ,\wedge, \vee, \rightarrow, \leftrightarrow, \forall,\exists, (, ).\]

We are also given once and for all a \emph{signature} $S$, i.e. a set
of symbols partitioned into two subsets: the set of
\emph{function} symbols and the set of \emph{relation} symbols. To
each element $s$ of $S$ is linked a non-negative integer called
\emph{arity} of $s$. An $n$-ary symbol is a symbol of arity $n$.
A $0$-ary function symbol is called \emph{constant}.

As usual, an \emph{expression} is a word on $L\cup S\cup V$ which is
either a \emph{term}, i.e. of one of the two forms:
\begin{equation} \label{terme}
x,\;f(t_1,\ldots, t_n),
\end{equation}
or a \emph{formula}, i.e. of  one of the eleven forms:
\begin{equation}\label{formules}\begin{array}{@{}c@{}}
s= t,\; r(t_1,\ldots, t_n),\; \vrai,\; \faux,\;  \\
\neg \varphi,\;(\varphi \wedge \psi),\; (\varphi \vee
\psi),\;(\varphi \rightarrow \psi),\;(\varphi \leftrightarrow
\psi),\\(\forall x\,\varphi),\;(\exists x\,\varphi).
\end{array}
\end{equation}
In (\ref{terme}), $x$ is taken from $\Vv$, $f$ is an $n$-ary
function symbol taken from $S$ and the $t_i$'s are shorter terms.
In (\ref{formules}), $s,t$ and the $t_i$'s are terms, $r$ is an
$n$-ary relation symbol taken from $S$ and $\varphi$ and $\psi$
are shorter formulas. The set of the expressions forms   \emph{a first-order
language with equality}.

The formulas of the first line of (\ref{formules}) are known as \emph{atomic}, and
\emph{flat} if they are of one of the following forms:

\[\vrai,\; \faux,\; x_0=x_1, x_0=f(x_1,...,x_n),\; r(x_1,...,x_n),\]
where all the $x_i$'s are (possibly non-distinct)  variables taken from $\Vv$, $f$ is an $n$-ary function
symbol taken from $S$ and $r$ is an $n$-ary relation symbol taken
from $S$. An \emph{equation} is a formula of the form $s=t$ with $s$ and $t$ terms. 

An occurrence of a variable $x$ in a formula is
\emph{bound} if it occurs in a sub-formula of the form $(\forall
x\, \varphi)$ or $(\exists x\, \varphi)$. It is \emph{free} in the
contrary case. The \emph{free variables of a formula} are those
which have at least one free occurrence in this formula. A
\emph{proposition} or a \emph{sentence} is a formula without free
variables.  If $\varphi$ is a formula, then we denote by $var(\varphi)$ the set of the free variables of $\varphi$.

The syntax of the formulas being constraining, we allowed
ourselves to use infix notations for the binary symbols and to add
and remove brackets when there are no ambiguities.  Moreover, we do not distinguish two
formulas which can be made equal using the following
transformations of  sub-formulas:
\[\begin{array}{@{}c@{}}\varphi\wedge\varphi\Longrightarrow\varphi,\;\;\varphi\wedge\psi\Longrightarrow\psi\wedge\varphi,\;\;(\varphi\wedge\psi)\wedge\phi\Longrightarrow\varphi\wedge(\psi\wedge\phi),\\\varphi\wedge\vrai\Longrightarrow\varphi,\;\;\varphi\vee\faux\Longrightarrow\varphi.\end{array}\]

If $I$ is the set $\{i_1,...,i_n\}$, we call \emph{conjunction} of
formulas and write $\bigwedge_{i\in I}\varphi_i$, each formula of
the form
$\varphi_{i_1}\wedge\varphi_{i_2}\wedge...\wedge\varphi_{i_n}\wedge\vrai$.
In particular, for $I=\emptyset$, the conjunction $\bigwedge_{i\in
I}\varphi_i$ is reduced to $\vrai$.

\subsubsection{Model}
A \emph{model} is a tuple  $\mathcal{M} =\; < M, (f^M)_{{f}\in F},(R^M)_{{r}\in R}>$, where:
\begin{itemize}
\item $M$, the \emph{universe} or \emph{domain} of $\mathcal{M}$,
is a  nonempty set disjoint from $S$, its elements are called
\emph{individuals} of $\mathcal{M}$; \item $F$ and $R$ are  sets of $n$-ary functions and relations in the set $ M$, subscripted by
the elements of $S$ and such that: \begin{itemize}\item for every
$n$-ary function symbol $f$ taken from $S$, $f^M$ is an $n$-ary
operation in $M$, i.e. an application from ${
M}^n$ in ${ M}$. In particular, when $f$ is a constant,
$f^M$ belongs to ${ M}$; \item for every $n$-ary relation
symbol $r$ taken from $S$, $r^M$ is an $n$-ary relation in
${ M}$, i.e. a subset of ${M}^n$.
\end{itemize}\end{itemize}

Let $\mathcal{M} =\; < M, F,R>$ be a model. An
\emph{$\mathcal{M}$-expression} $\varphi$ is an expression built on the
signature $S\cup M$ instead of $S$, {by considering the
elements of $M$ as $0$-ary function symbols}. If for
each free variable $x$ of $\varphi$ we replace each free
occurrence of $x$ by a same element $m$ in $M$, we get an
$\mathcal{M}$-expression $\varphi'$ called \emph{instantiation}\footnote{We also say that the variable $x$ is instantiated by $m$ in $\varphi'$.}  or \emph{valuation} of $\varphi$ by
individuals of $\mathcal{M}$.

 If $\varphi$ is an $\mathcal{M}$-formula, we say that
$\varphi$ \emph{is true in $\mathcal{M}$} and we write
\begin{equation}\label{model}
{\mathcal M}\models\varphi,
\end{equation}
if for any instantiation $\varphi^\prime$ of $\varphi$ by
individuals of $\mathcal{M}$ the set $M$ has the property
expressed by $\varphi\prime$, when we interpret the function and
relation symbols of $\varphi\prime$ by the corresponding functions
and relations of $\mathcal{M}$ and when we give to the logical symbols their
usual meaning.
\begin{remarque} \label{modelecomplet} For every $\mathcal{M}$-formula $\varphi$ without free variables, one and only one of the following properties holds: $\mathcal{M}\models \varphi$, $\mathcal{M} \models \neg \varphi$.
\end{remarque}

Let us finish this sub-section by a convenient notation. Let
$\bar{x}=x_1...x_n$ be a word on $\Vv$ and let $\bar{i}=i_1...i_n$
be a word on $M$ or $\Vv$ of the same length as
$\bar{x}$. If $\varphi(\bar{x})$ and $\phi$ are two $\mathcal{M}$-formulas, then we denote
by $\varphi(\bar{i})$, respectively $\phi_{\bar{x}\leftarrow\bar{i}}$ , the $\mathcal{M}$-formula obtained by replacing in
$\varphi(\bar{x})$, respectively in $\phi$, each free occurrence of $x_j$ by $i_j$.

\subsubsection{Theory}
A \emph{theory} is a (possibly infinite) set of propositions called \emph{axioms}. 
We say
that the model $\mathcal{M}$ is a \emph{model of $T$}, if for each element
$\varphi$ of $T$, $\mathcal{M}\models\varphi$. If $\varphi$ is a formula, we
write
\[ T\models\varphi,\]
if for each model $\mathcal{M}$ of $T$, $\mathcal{M}\models\varphi$. We say that the
formulas $\varphi$ and $\psi$ are \emph{equivalent in} $T$ if $T
\models \varphi\leftrightarrow \psi$.

\begin{definition1}
A theory $T$ is \emph{complete} if for every proposition
$\varphi$, one and only one of the following properties holds:
$T\models \varphi$, $T \models \neg \varphi$.
\end{definition1}
Let $\phi$ be a formula and $\bar{x}=x_1...x_n$ be a word on $\Vv$ such that $var(\phi)=\bar{x}$. From the preceding  definition we deduce that  a decision procedure is sufficient  in the case where we want just to show the completeness of a theory $T$, as it was done in \cite{moitplp} for decomposable theories. In fact, the completeness of $T$ depends only  on the truth values of the propositions in $T$.  On the other  hand, finding for each model $\mathcal{M}$ of $T$  the  instantiations $\bar{i}$ of $\bar{x}$ such that $\mathcal{M}\models\phi_{\bar{x}\leftarrow\bar{i}}$ can be obtained only using a first-order constraint solver in $T$.  This kind of problem is generally known as \emph{first-order constraint satisfaction problem}.

\subsubsection{Vectorial quantifiers \label{besoins}}

Let $\mathcal{M}$ be a model and  $T$ a theory. Let $\bar{x}=x_1\ldots
x_n$ and $\bar{y}=y_1\ldots y_n$ be two words on $\Vv$ of the same
length. Let  $\phi$, $\varphi$ and $\varphi(\bar{x})$ be
$\mathcal{M}$-formulas. We write
\[\begin{tabular}{@{}l@{\,\,}l@{\,\,}l@{}}
$\exists\bar{x}\,\varphi$&for&$\exists x_1...\exists x_n\,\varphi$,\\
$\forall\bar{x}\,\varphi$&for&$\forall x_1...\forall x_n\,\varphi$,\\
$\exists?\bar{x}\,\varphi(\bar{x})$&for&$\forall\bar{x}\forall\bar{y}\,\varphi(\bar{x})\wedge\varphi(\bar{y})\rightarrow \bigwedge_{i\in\{1,...,n\}}x_i=y_i,$\\
$\exists!\bar{x}\,\varphi$&for&$(\exists\bar{x}\,\varphi)\wedge(\exists?\bar{x}\,\varphi).$
\end{tabular}\]
The word $\bar{x}$, which can be the empty word $\varepsilon$, is
called \emph{vector of variables}. Note that the formulas
$\exists?\varepsilon\varphi$ and $\exists!\varepsilon\varphi$ are
respectively equivalent to $\vrai$ and to $\varphi$ in any model
$\mathcal{M}$.

\begin{notation} Let $Q$ be a  quantifier taken from $\{\forall,\exists, \exists!, \exists?\}$. Let $\bar{x}$ be vector of  variables taken from $V$.  We write:   \[Q\bar{x}\,\varphi\wedge\phi\;\; for\;\; Q\bar{x}\,(\varphi\wedge\phi).\] 
\end{notation}
\begin{example}
Let $I=\{1,...,n\}$ be a finite  set. Let $\varphi$ and $\phi_i$ with $i\in I$ be formulas. Let $\bar{x}$ and $\bar{y}_i$ with $i\in I$ be vectors of variables. We write:
\[\begin{tabular}{@{}l@{\,\,}l@{\,\,}l@{}}
$\exists\bar{x}\,\varphi\wedge\neg\phi_1$&for&$\exists\bar{x}\,(\varphi\wedge\neg\phi_1)$,\\
$\forall\bar{x}\,\varphi\wedge\phi_1$&for&$\forall\bar{x}\,(\varphi\wedge\phi_1)$,\\
$\exists!\bar{x}\,\varphi\wedge\bigwedge_{i\in I}(\exists\bar{y}_i\phi_i)$&for&$\exists!\bar{x}\,(\varphi\wedge(\exists\bar{y}_1\phi_1)\wedge...\wedge(\exists\bar{y}_n\phi_n)\wedge\vrai),$\\
$\exists?\bar{x}\,\varphi\wedge\bigwedge_{i\in I}\neg(\exists\bar{y}_i\phi_i)$&for&$\exists?\bar{x}\,(\varphi\wedge(\neg(\exists\bar{y}_1\phi_1))\wedge...\wedge(\neg(\exists\bar{y}_n\phi_n))\wedge\vrai).$
\end{tabular}\]
\end{example}

\begin{notation}
If $\bar{x}$ is a vector of variables then we denote by $X$ the set of the variables of $\bar{x}$.
\end{notation}

Let $I$ be a (possible empty) finite set. The two following properties hold for any theory $T$: 
\begin{propriete}\label{r2}
If $T\models$\, $\exists?\bar{x}\,\varphi$ then  \[T\models\,
(\exists\bar{x}\,\varphi \wedge\bigwedge_{i\in I}
\neg\phi_i)\leftrightarrow((\exists\bar{x}\varphi)\wedge\bigwedge_{i\in I}\neg(\exists\bar{x}\,\varphi\wedge\phi_i)).\]
\end{propriete}

\begin{propriete}\label{unique1}
If $T\models\exists!\bar{x}\,\varphi$ then
\[T\models(\exists\bar{x}\,\varphi \wedge\bigwedge_{i\in I}
\neg\phi_i)\leftrightarrow\bigwedge_{i\in I}\neg(\exists\bar{x}\,\varphi\wedge\phi_i).\]
\end{propriete}
Full proofs of these two properties can be found in detail in \cite{moitplp}.

\subsection{The axioms of $T$}
Let $F$ be a set of function symbols containing infinitely many non-constant function symbols and at least one constant. Let $\fini$ be an $1$-ary relation symbol. The theory $T$ of finite or  infinite  trees built on the signature $S=F\cup\{\fini\}$  has as axioms the
infinite
set of propositions of one of the five following forms:
\[\begin{tabular}{lll}
$\forall\bar{x}\forall\bar{y}\hspace{5mm} $& $\neg (f(\bar{x})=g(\bar{y}))
$&\hspace{10mm} [1] \\
$\forall\bar{x}\forall\bar{y} $\hspace{5mm}& $f(\bar{x})=f(\bar{y})\rightarrow
\bigwedge_i x_i=y_i$  &\hspace{10mm} [2] \\
$\forall\bar{x}\exists!\bar{z}$\hspace{5mm}& $\bigwedge_i
z_i=t_i[\bar{x}\bar{z}]$ & \hspace{11mm}[3] \\
$\forall\bar{x}\forall u$\hspace{5mm}& $\neg(u=t[u,\bar{x}]\wedge\fini(u)) $ & \hspace{11mm}[4]\\
$\forall\bar{x}\forall u$\hspace{5mm}& $ (u=f(\bar{x})\wedge\fini(u))\leftrightarrow(u=f(\bar{x})\wedge\bigwedge_i \fini(x_i))$ & \hspace{11mm}[5]

\end{tabular}
\]
where $f$ and $g$ are distinct function symbols taken from $F$,
$\bar{x}$ is a vector of (possibly non-distinct)  variables $x_i$, $\bar{y}$ is a
vector of (possibly non-distinct)  variables $y_i$, $\bar{z}$ is a
vector of distinct variables $z_i$, $t_i[\bar{x}\bar{z}]$ is a
term which begins with an element of $F$ followed by variables
taken from $\bar{x}$ or $\bar{z}$, and $t[u,\bar{x}]$ is a term containing at least one occurrence of an element of  $F$ and the variable $u$  and possibly other variables taken from $\bar{x}$. For example, we have $T\models\forall x_1x_2\forall u\,\neg(u=f_1(x_1,f_2(u,x_2))\wedge\fini(u))$ and $T\models\forall u \neg(u=f_1(f_2(u,f_0),f_0)\wedge\fini(u))$ where $f_1$ and $f_2$ are $2$-ary function symbols and $f_0$ a constant of $F$.

The forms [1],..., [5] are also called \emph{schemas of axioms} of the theory
$T$. Proposition [1] called \emph{conflict of symbols} shows
that two distinct operations produce two distinct individuals. Proposition [2] called \emph{explosion} shows that the same
operation on two distinct individuals produces two distinct
individuals. Proposition [3] called $\emph{unique solution}$
shows that a certain form a conjunction of equations has a unique
set of solutions in $T$. In particular, the formula $\exists z\,z=f(z)$ has a unique solution which is the infinite tree $f(f(f(...)))$. Proposition [4]  means that a finite tree cannot  be a  strict subtree of itself. We emphasize strongly that $t[u,\bar{x}]$ should contain at least one occurrence of an element of $F$ and the variable $u$. In Axiom [5], if $\bar{x}$ is the empty vector and $f$ is a constant then we get $\forall u\, u=f\wedge\fini(u)\leftrightarrow u=f$, which means that the property $\fini(f)$ is true for each constant $f$   of $F$. 

This theory is an extension of the basic theory of finite or infinite trees given by M. Maher in~\cite{Maher} and built on a signature containing  an infinite set of function symbols. Maher's theory is composed of the three first axioms of $T$ and its completeness was shown using a decision procedure which transforms each proposition into a Boolean combination of 
existentially quantified conjunctions of atomic formulas. Note also  that both Maher's theory and the theory $T$ do not
accept full elimination of quantifiers, i.e. there exist some quantified formulas whose quantifiers cannot be eliminated. For example,  the formula $\exists
x\,y=f(x)$ is neither true nor false in $T$. It accepts in each model of $T$ a set of solutions and another set of non-solutions. As a consequence, we cannot  simplify it any further.  This non-full elimination of quantifiers makes the completeness of $T$ not evident.

\subsection{The models of $T$}
Let us  extend the algebras given at the end of section \ref{alg} by the relation $\fini$. More precisely, if $u_1$, $u_2$, $u_3$ and $u_4$ are respectively elements of $D$, $Tr$, $Fi$ and $Ra$ then the operations $\fini^{D}(u_1)$,  $\fini^{Tr}(u_2)$, $\fini^{Fi}(u_3)$ and $\fini^{Ra}(u_4)$ are true respectively in $D$, $Tr$, $Fi$ and $Ra$, if and only if  $u_1$, $u_2$, $u_3$ and $u_4$ have a finite set of nodes. 

Let us  now denote by: 
\begin{itemize}
\item
${\mathcal{D}}=<D,(f^{D})_{f\in F},\fini^{D}>$,  the model of sets  of nodes,
\item
 ${\mathcal{T}r}=<Tr,(f^{Tr})_{f\in F},\fini^{Tr}>$, the model of finite or infinite trees,
\item
${\mathcal{R}a}=<Ra,(f^{Ra})_{f\in F},\fini^{Ra}>$, the model of rational trees,
\item
${\mathcal{F}i}=<Fi,(f^{Fi})_{f\in F},\fini^{Fi}>$, the model of finite trees.
\end{itemize}
We have:
\begin{theoreme}\label{new}
The models ${\mathcal{D}}$, ${\mathcal{T}r}$ and ${\mathcal{R}a}$ are models of the theory $T$.
\end{theoreme}
This theorem is one of the essential contributions given in this paper and shows that our theory $T$ is in fact an axiomatization of the structures $D$, $Tr$ and $Ra$ together with an infinite set of construction operations and the $1$-ary relation $\fini$. It also shows that $T$ has at least one model and thus $T\models\neg(\vrai\leftrightarrow\faux)$.

\emph{Proof, first part: } Let us show first that the model ${\mathcal{D}}$ of  sets  of nodes is a model of $T$. In other words,  we must show that the following properties hold:
{\small
\[\begin{array}{ll}
[1^D]&(\forall a_1,...,a_m\in D)(\forall b_1,...,b_n\in D)\,\neg(f^D(a_1,...,a_m)=g^D(b_1,...,b_n))\\

[2^D]&(\forall a_1,...,a_n\in D)(\forall b_1,...,b_n\in D)\,(f^D(a_1,...,a_n)=f^D(b_1,...,b_n)\rightarrow\bigwedge_{i=1}^n a_i=b_i)\\

[3^D]&(\forall a_1,...,a_m\in D)(\exists! b_1,...,b_n\in D)\,(\bigwedge_{i=1}^{n} b_i=t_i^D[b_1,...,b_n,a_1,...,a_m])
\\

[4^D]&(\forall a_1,...,a_m\in D)(\forall u\in D)  \neg(u=t^D[u,a_1,...,a_n]\wedge\fini^D(u)) \\

[5^D]&(\forall a_1,...,a_n\in D) (\forall u\in D)  (u=f^D(a_1,...,a_n)\wedge\fini^D(u))\leftrightarrow\\
&(u=f^D(a_1,...,a_n)\wedge\bigwedge_{i=1}^n \fini^D(a_i)) 
\end{array}
\]
}
where $f$ and $g$ are distinct function symbols taken from $F$,
 $t_i^D[b_1,...,b_n,a_1,...,a_m]$ is a term which begins with an element of $F$ followed by variables  taken from 
$\{a_1,...,a_m,b_1,...,b_n\}$, and $t^D[u,a_1,...,a_n]$ is a term containing at least one occurrence of an element of  $F$ and the variable $u$  and possibly other variables taken from $\{a_1,...,a_n\}$. According to Definition \ref{cons} and the definition of the relation $\fini^D$, the  properties $[1^D]$,  $[2^D]$, $[4^D]$ and $[5^D]$ hold. On the other hand, property $[3^D]$ is much less obvious and deserves to be proved.

Let $a_1,...,a_m$ and $b_1,...,b_n$ be elements of $D$. According to the first point of Remark \ref{25}, the $\mathcal D$-formula \begin{equation}\label{e0}\bigwedge_{i=1}^{n} b_i=t_i^D[b_1,...,b_n,a_1,...,a_m],\end{equation}is equivalent in $\mathcal D$ to
\begin{equation}\label{e1} \bigwedge_{k=0}^{\infty}\bigwedge_{i=1}^n\nu_k(b_i)=\nu_k(t_i^D[b_1,...,b_n,a_1,...,a_m]).\end{equation}
Let $i\in\{1,...,n\}$. Let us denote by $f_i$ respectively $[b_1,...,b_n,a_1,...,a_m]_i$ the function symbol respectively the set of the variables which occur in the term $t_i^D[b_1,...,b_n,a_1,...,a_m]$. According to the second and third point of Remark \ref{25} we have:
\begin{itemize}
\item
For each $i\in\{1,...,n\}$ there exists one node $\varphi_0^i=(\varepsilon,f_i)$, such that \[\nu_0(t_i^D[b_1,...,b_n,a_1,...,a_m])=\{\varphi_0^i\}.\]
\item
For each $i\in\{1,...,n\}$ and each $k\geq 0$ there exists a function $\varphi_{k+1}^i$,  which is independent from  all the $\nu_{k+1}(x)$, with $x\in[b_1,...,b_n,a_1,...,a_m]_i$, such that
\[\nu_{k+1}(t_i^D[b_1,...,b_n,a_1,...,a_m])=\varphi_{k+1}^i([\nu_k(b_1),...,\nu_k(b_n),\nu_k(a_1),...,\nu_k(a_m)]_i),\]  where $[\nu_k(b_1),...,\nu_k(b_n),\nu_k(a_1),...,\nu_k(a_m)]_i$ is a tuple of elements of the form  $\nu_k(x)$ for all $x\in[b_1,...,b_n,a_1,...,a_m]_i$.

\end{itemize}
Thus, the $\mathcal D$-formula (\ref{e1}) is equivalent in $\mathcal D$ to

\[(\bigwedge_{i=1}^{n}\nu_0(b_i)=\{\varphi_0^i\})\wedge(\bigwedge_{k=0}^{\infty}\bigwedge_{i=1}^n\nu_{k+1}(b_i)=\varphi_{k+1}^i([\nu_k(b_1),...,\nu_k(b_n),\nu_k(a_1),...,\nu_k(a_m)]_i)),\]
from which we deduce that:
\begin{itemize}
\item
 (i) For all $i\in\{1,...,n\}$,  $\nu_0(b_i)$ has a  constant value, which is equal to $(\varepsilon,f_i)$.\item
 (ii) Each $\nu_{k+1}(b_i)$ depends in the worst case on  $\nu_k(b_1),...,\nu_k(b_n),\nu_k(a_1),...,\nu_k(a_m)$, i.e. on  $\nu_k(b_1),...,\nu_k(b_n)$ and $a_1,...,a_m$. 
 \end{itemize}
Thus, by recurrence\footnote{If $k=0$ then according to (ii) each $\nu_{1}(b_i)$ depends in the worst case on   $\nu_0(b_1),...,\nu_0(b_n)$ and $a_1$,...,$a_m$. According to (i) all the $\nu_0(b_1),...,\nu_0(b_n)$ have constant values  and thus each $\nu_{1}(b_i)$ depends  only on $a_1,...,a_m$. Let us now assume  that  each $\nu_{k}(b_i)$ depends  only on   $a_1,...,a_m$ and let us show that this hypothesis is true for $\nu_{k+1}(b_i)$. According to (ii), each $\nu_{k+1}(b_i)$ depends in the worst case on   $\nu_k(b_1),...,\nu_k(b_n)$ and $a_1,...,a_m$, which according to our hypothesis depend only on $a_1,...,a_m$. Thus, the recurrence is true for all $k\geq 0$.}
on $k$, we deduce that (iii)  each $\nu_{k+1}(b_i)$ with $k\geq 0$ and $i\in\{1,...,n\}$,  depends only on  $a_1,...,a_m$.  From (i) and  (iii) we deduce that all the $b_i$'s depend  only on  $a_1,...,a_m$ and thus property $[3^D]$ holds. In other words,  for each instantiation of $a_1,...,a_m$ by elements of $D$ we can deduce the values of $\nu_k(b_i)$ for all $i\in\{1,...,n\}$ and $k\geq 0$. 

We have shown that the model  $\mathcal D$ satisfies the five axioms of $T$ and thus it is a model of $T$.
$\\[2mm]$
\emph{Proof, second part: } Let us now show  that the model ${\mathcal{T}r}$  of finite or infinite trees is a model of $T$. For that, it is enough to show the validity of the following properties 
{\small 
\[\begin{array}{ll}
[1^{Tr}]&(\forall a_1,...,a_m\in Tr)(\forall b_1,...,b_n\in Tr)\,\neg(f^{Tr}(a_1,...,a_m)=g^{Tr}(b_1,...,b_n))\\

[2^{Tr}]&(\forall a_1,...,a_n\in Tr)(\forall b_1,...,b_n\in Tr)\,(f^{Tr}(a_1,...,a_n)=f^{Tr}(b_1,...,b_n)\rightarrow\bigwedge_{i=1}^n a_i=b_i)\\

[3^{Tr}]&(\forall a_1,...,a_m\in Tr)(\exists! b_1,...,b_n\in Tr)\,(\bigwedge_{i=1}^{n} b_i=t_i^{Tr}[b_1,...,b_n,a_1,...,a_m])\\

[4^{Tr}]&(\forall a_1,...,a_m\in Tr)(\forall u\in Tr)  \neg(u=t^{Tr}[u,a_1,...,a_n]\wedge\fini^{Tr}(u)) \\

[5^{Tr}]&(\forall a_1,...,a_n\in Tr) (\forall u\in Tr)  (u=f^{Tr}(a_1,...,a_n)\wedge\fini^{Tr}(u))\leftrightarrow\\

&(u=f^{Tr}(a_1,...,a_n)\wedge\bigwedge_{i=1}^{n} \fini^{Tr}(a_i))

\end{array}
\]
}
where $f$ and $g$ are distinct function symbols taken from $F$,
 $t_i^{Tr}[b_1,...,b_n,a_1,...,a_m]$ is a term which begins with an element of $F$ followed by variables  taken from 
$\{a_1,...,a_m,b_1,...,b_n\}$, and $t^{Tr}[u,a_1,...,a_n]$ is a term containing at least one occurrence of an element of  $F$ and the variable $u$  and possibly other variables taken from $\{a_1,...,a_n\}$. Since $Tr$ is a subset of $D$, then according to the definition of $f^{Tr}, f^D,\fini^{Tr}$ and $\fini^D$, the properties $[1^{D}]$, $[2^{D}]$, $[4^{D}]$ and $[5^{D}]$ imply $[1^{Tr}]$, $[2^{Tr}]$, $[4^{Tr}]$ and $[5^{Tr}]$. On the other hand, to show  property $[3^{Tr}]$, it is enough to show the following implication: 
{\small \begin{equation}\label{e4}
(\forall a_1,...,a_m,b_1,...b_n\in D)(((\bigwedge_{i=1}^n b_i=t_i^D[b_1,...,b_n,a_1,...,a_m])\wedge(\bigwedge_{i=1}^m a_i\in Tr))\rightarrow (\bigwedge_{i=1}^n b_i\in Tr))
\end{equation}
}
Let $a$, $b$, $a_1$,...,$a_m$, $b_1$,...,$b_n$ be elements of $D$. Let us consider the following notation: 
\begin{center}\begin{tabular}{l} $Arb(a,b)\leftrightarrow$ each element of $a$ is arborescent in $b$.
\end{tabular}
\end{center}
According to Definition \ref{arbre}, the ${\mathcal T}r$-formula  \[(\bigwedge_{i=1}^{n} b_i=t_i^D[b_1,...,b_n,a_1,...,a_m])\wedge(\bigwedge_{i=1}^m a_i\in Tr),\] is equivalent  in ${\mathcal T}r$ to
\[(\bigwedge_{i=1}^{n} b_i=t_i^D[b_1,...,b_n,a_1,...,a_m])\wedge(\bigwedge_{i=1}^m Arb(N,a_i)),\] which is equivalent  to
\begin{equation}\label{e5}(\bigwedge_{i=1}^{n} b_i=t_i^D[b_1,...,b_n,a_1,...,a_m])\wedge(\bigwedge_{k=0}^{\infty}\bigwedge_{i=1}^m Arb(\nu_k(N),a_i)),\end{equation} which for each $j\geq 0$ is equivalent  in ${\mathcal T}r$ to
\begin{equation}\label{e6}(\bigwedge_{i=1}^{n} b_i=t_i^D[b_1,...,b_n,a_1,...,a_m])\wedge(\bigwedge_{k=0}^{\infty}\bigwedge_{i=1}^m Arb(\nu_k(N),a_i))\wedge(\bigwedge_{i=1}^n Arb(\nu_j(b_i),b_i)).\end{equation} 
The equivalence $(\ref{e5}\leftrightarrow\ref{e6})$ holds for $j=0$ according to the fourth point of Remark \ref{25}, and if we  assume that this  equivalence holds for an integer $j$ with $j\geq 0$ then according to the fifth point of Remark \ref{25}, we deduce that  it holds also for $j+1$. Thus, since the  equivalence $(\ref{e5}\leftrightarrow\ref{e6})$ holds for  any $j\geq 0$ then according to the sixth point of Remark \ref{25} and Definition \ref{arbre} we deduce that (\ref{e6}) implies 
\[\bigwedge_{i=1}^n Arb(N,b_i),\]
which, according to Definition \ref{arbre}, implies 
\[\bigwedge_{i=1}^n b_i\in Tr.\]
Thus, the implication (\ref{e4}) holds and $Tr$ is a model of $T$.
$\\[2mm]$
\emph{Proof, third part: } Finally, let us show that the model ${\mathcal R}a$ is a model of $T$. For that, it is enough to show the validity of the following properties:
{\small
\[\begin{array}{ll}
[1^{Ra}]&(\forall a_1,...,a_m\in Ra)(\forall b_1,...,b_n\in Ra)\,\neg(f^{Ra}(a_1,...,a_m)=g^{Ra}(b_1,...,b_n))\\

[2^{Ra}]&(\forall a_1,...,a_n\in Ra)(\forall b_1,...,b_n\in Ra)\,(f^{Ra}(a_1,...,a_n)=f^{Ra}(b_1,...,b_n)\rightarrow\bigwedge_{i=1}^n a_i=b_i)\\

[3^{Ra}]&(\forall a_1,...,a_m\in Ra)(\exists! b_1,...,b_n\in Ra)\,(\bigwedge_{i=1}^{n} b_i=t_i^{Ra}[b_1,...,b_n,a_1,...,a_m])\\

[4^{Ra}]&(\forall a_1,...,a_m\in Ra)(\forall u\in Ra)  \neg(u=t^{Ra}[u,a_1,...,a_n]\wedge\fini^{Ra}(u)) \\

[5^{Ra}]& (\forall a_1,...,a_n\in Ra) (\forall u\in Ra)  (u=f^{Ra}(a_1,...,a_n)\wedge\fini^{Ra}(u))\leftrightarrow\\

& (u=f^{Ra}(a_1,...,a_n)\wedge\bigwedge_{i=1}^n \fini^{Ra}(a_i)) 

\end{array}
\]
}
where $f$ and $g$ are distinct function symbols taken from $F$,
 $t_i^{Ra}[b_1,...,b_n,a_1,...,a_m]$ is a term which begins with an element of $F$ followed by variables  taken from 
$\{a_1,...,a_m,b_1,...,b_n\}$, and $t^{Ra}[u,a_1,...,a_n]$ is a term containing at least one occurrence of an element of  $F$ and the variable $u$  and possibly other variables taken from $\{a_1,...,a_n\}$. Since $Ra$ is a subset of $Tr$ and according to the definitions of $f^{Tr}$, $f^{Ra}$, $\fini^{Tr}$ and $\fini^{Ra}$ then the properties $[1^{Tr}]$, $[2^{Tr}]$, $[4^{Tr}]$ and $[5^{Tr}]$ imply $[1^{Ra}]$, $[2^{Ra}]$, $[4^{Ra}]$ and $[5^{Ra}]$. On the other hand, in property $[3^{Tr}]$, (in the preceding proof), a subtree   of depth $k$ of any  $b_i$ is either one of the trees $b_1$,...$,b_n$ or a subtree of one of the $a_{j}$'s with $i\in\{1,...,n\}$ and $j\in\{1,...,m\}$. This is true for $k=0$ and if we assume that it is true for $k$ then  we deduce that it is true for $k+1$. Thus, if the $a_{j}$'s are rational then the $b_i$'s in $[3^{Tr}]$ are also rational  and thus we get $[3^{Ra}]$.

We have shown that the models $\mathcal D$, ${\mathcal T}r$ and ${\mathcal R}a$ are models of $T$. What about the model ${\mathcal F}i$ of finite trees? Since $F$ contains at least one function symbol $f$ which is not a constant then according to Axiom [3] of $T$ we have \[T\models\exists! x\,x=f(x,...,x).\]  It is obvious that this property cannot  be true in ${\mathcal F}i$, i.e. there exists no $x\in Fi$ such that $x=f^{Fi}(x,...,x)$. Thus, the model ${\mathcal F}i$ of finite trees is not a model of $T$.

Let us end this section by a property concerning the cardinality of any model of $T$:
\begin{propriete}\label{inff}
Let ${\mathcal M}=<M, (f^M)_{f\in F}, \fini^M>$ be a model of $T$. The model $\mathcal M$ has an infinity of individuals $i$ such that ${\mathcal M}\models \fini^M(i)$.
\end{propriete}
\begin{proof}
Since the set $F$ contains at least one function symbol $f$ which is a constant then according to Axiom [5], with $\bar{x}=\varepsilon$, we have \begin{equation}\label{int1} {\mathcal M}\models\fini^M(f^M).\end{equation} On the other hand, according to the definition of the signature of $T$, the set  $F$ contains an infinity of  distinct function symbols which are not  constants. Let $f_1$ one of these symbols.  According to (\ref{int1}) and Axiom [5] we have \[{\mathcal M}\models\fini^M(f_1^M(f^M,...,f^M)),\] thus the individual $f_1^M(f^M,...,f^M)$ is finite in $\mathcal M$. Since the set $F$ contains an infinity of distinct  function symbols $f_1,f_2,f_3,...$ which are not constants  then  we can create by following the same preceding  steps an infinity of finite individuals $f_1^M(f^M,...,f^M), f_2^M(f^M,...,f^M),f_3^M(f^M,...,f^M),...$ which start by distinct function symbols. According to Axiom [1], all these individuals are distinct. According to (9) and Axiom [5] all these individuals are finite in $\mathcal M$.   \end{proof}

\begin{corollaire}\label{inff2}
Each model of $T$ has an infinite domain, i.e. an infinite set of individuals. 
\end{corollaire}

\section{Solving first-order  constraints in $T$}\label{algo}

\subsection{Discipline of the formulas in $T$}\label{rest}
Let us assume that  the infinite set $\Vv$ is
 ordered by a strict linear dense order relation without endpoints denoted by
$\succ$. Starting from this section, we impose the following discipline to  every formula $\varphi$ in $T$: the quantified variables of $\varphi$ are renamed so that:
\begin{itemize}
\item
(i) The quantified variables of $\varphi$ have  distinct names and different from those of the free variables.
\item
(ii) For all variables $x$, $y$ and all sub-formulas\footnote{By considering  that each formula is also a sub-formula of itself.}  $\varphi_i$ of $\varphi$, if $y$ has a free occurrence in $\varphi_i$ and  $x$  has a bound occurrence in $\varphi_i$ then $x\succ y$.
\end{itemize}
\begin{example}
Let $x,y,z,v$ be variables of $V$ such that $x\succ y\succ z\succ v$. Let $\varphi$ be the formula \begin{equation}\label{e13}\exists x\,x=fy\wedge\left[\begin{array}{l}\neg(\exists z\, z=x)\wedge\\\neg(\exists z\, z=v)\end{array}\right].\end{equation}
The quantified variables of $\varphi$ have no distinct names.  Since the order $\succ$ is dense and without endpoints, there exists a variable $w$ in $V$ such that $x\succ y\succ z\succ v\succ w$, and thus $\varphi$ is equivalent in $T$ to 
\[\exists x\,x=fy\wedge\left[\begin{array}{l}\neg(\exists z\, z=x)\wedge\\\neg(\exists w\, w=v)\end{array}\right].\]
In the preceding formula, the variables  $z$ and $w$ have bound occurrences while the variables $y$ and $v$ have free occurrences. Since $x\succ y\succ z\succ v\succ w$ then $z$ and $w$ must be renamed. On the other hand, since the order $\succ$ is dense and without endpoints, there exist two variables $u$ and $d$ in $V$ such that  $x\succ u\succ d\succ y\succ z\succ v\succ w$. Thus, the preceding formula is equivalent in $T$ to 
\[\exists x\,x=fy\wedge\left[\begin{array}{l}\neg(\exists u\, u=x)\wedge\\\neg(\exists d\, d=v)\end{array}\right].\]
In the sub-formula $(\exists u\, u=x)$ the variable $x$ has a free occurrence while the variable $u$ has a bound occurrence. Since $x\succ u$ then $u$ must be renamed. On the other hand, since the order $\succ$ is dense and without endpoints, there exists a variable $n$  in $V$ such that  $n\succ x\succ u\succ d\succ y\succ z\succ v\succ w$. Thus, the preceding formula is equivalent in $T$ to 
\begin{equation}\label{e12}\exists x\,x=fy\wedge\left[\begin{array}{l}\neg(\exists n\, n=x)\wedge\\\neg(\exists d\, d=v)\end{array}\right].\end{equation}
This formula satisfies our conditions. Of course, the equivalence between (\ref{e12}) and (\ref{e13}) holds because in each step we   renamed only the quantified variables. It is obvious that we can always transform any formula $\varphi$ into an equivalent formula $\phi$, which respects the discipline of the formulas in $T$, only by renaming the quantified variables of $\varphi$. It is enough for that to rename the quantified variables by distinct names and different from those of the free variables and then check each sub-formula and rename the quantified variables if the condition (ii) does not hold.

\end{example}

We emphasize strongly  that all the formulas which will be used starting from now satisfy the discipline of the formulas in $T$.

\subsection{Basic formula}
In this sub-section we  introduce particular conjunctions of atomic formulas that we call \emph{basic formulas} and show some of their properties. All of them will be used to show the correctness of our rewriting rules given in section \ref{algoo}.
\begin{definition1}\label{solved1}
Let $v_1,...,v_n,u_1,...,u_m$ be variables. A \emph{basic formula} is a formula of the form \begin{equation}\label{basic} (\bigwedge_{i=1}^{n} v_i=t_i)\wedge(\bigwedge_{i=1}^{m}\fini(u_i))\end{equation} in which all the equations $v_i=t_i$ are flat. Note that if $n=m=0$ then (\ref{basic}) is reduced to $\vrai$. The basic formula (\ref{basic}) is called 
\emph{solved} if all the variables $v_1,...,v_n,u_1,...,u_m$ are distinct and for each equation of the form $x=y$ we have $x\succ y$.  If $\alpha$ is a basic formula then we denote by \begin{itemize}\item $Lhs(\alpha)$ the set of the variables which occur in the left hand sides of the equations of $\alpha$. 
\item $FINI(\alpha)$ the set of the variables which occur in a sub-formula of $\alpha$ of the form $\fini(x)$.
\end{itemize}
Note that if $\alpha$ is a solved basic formula then for all variables $x$ of $\alpha$ we have $x\in Lhs(\alpha)\rightarrow x\not\in FINI(\alpha)$.

\end{definition1}

\begin{example}
The basic formula $x=x\wedge\fini(y)$ is not solved because $x\not\succ x$. The basic formula  $x=f(y)\wedge z=f(y)\wedge\fini(x)$ is also not solved because $x$ is a left hand side of an equation and  occurs also in $\fini(x)$. The basic formulas $\vrai$ (empty conjunction) and  $x=f(y)\wedge z=f(y)\wedge\fini(y)$ are solved. 
  
\end{example}

 According to the axiom [3] of $T$ we deduce the following property:
\begin{propriete}\label{unique} Let  $\alpha$ be a solved basic formula containing only equations. Let $\bar{x}$ be the vector of the variables of $Lhs(\alpha)$.
We have: $T\models\exists!\bar{x}\,\alpha$.
\end{propriete}

\begin{propriete}\label{mg} Let $\alpha$ and $\beta$ be two solved  basic formulas containing only equations. If $Lhs(\alpha)=Lhs(\beta)$ and
$T\models \alpha\rightarrow\beta$ then $T\models \alpha\leftrightarrow\beta$.
\end{propriete}
\begin{proof}
Let $\alpha$ and $\beta$ be two solved basic formulas containing only equations 
 such that $Lhs(\alpha)=Lhs(\beta)$ and $T\models\alpha\rightarrow\beta$. Let us show that we have also  $T\models\beta\rightarrow\alpha$. Let $\bar{x}$ be the vector of the variables of $Lhs(\alpha)$ and let $\bar{y}$ be the vector of the variables which occur in $\alpha\rightarrow\beta$ and do not occur in $\bar{x}$.   Since $\alpha$ and $\beta$ are two solved  basic formulas such that 
 $Lhs(\alpha)=Lhs(\beta)$ then (i) $\bar{x}$ is also the vector of the left hand sides of the equations of $\beta$. Moreover, the following equivalences are true in $T$: 

\begin{tabular}{lll}
&$\alpha\rightarrow\beta$&\\
$\leftrightarrow$&$\forall\bar{x}\forall\bar{y}\, \alpha\rightarrow\beta$&  \\
$\leftrightarrow$&$\forall\bar{y}\forall\bar{x}\,\neg\alpha\vee\beta$&\\
$\leftrightarrow$&$\forall\bar{y}(\neg(\exists\bar{x}\,\alpha\wedge\neg\beta))$&\\
$\leftrightarrow$&$\forall\bar{y}(\neg(\neg(\exists\bar{x}\,\alpha\wedge\beta)))$& according to the properties \ref{unique} and \ref{unique1} \\
$\leftrightarrow$&$\forall\bar{y}(\neg(\neg(\exists\bar{x}\,\beta\wedge\alpha)))$& \\
$\leftrightarrow$&$\forall\bar{y}(\neg(\exists\bar{x}\,\beta\wedge\neg\alpha))$& according to: (i) and  Property \ref{unique} and using the other  \\ && sense (right to left) of the equivalence of Property \ref{unique1} \\
$\leftrightarrow$&$\forall\bar{y}\forall\bar{x}\,\neg\beta\vee\alpha$&\\
$\leftrightarrow$&$\forall\bar{y}\forall\bar{x}\,\beta\rightarrow\alpha$&\\
$\leftrightarrow$&$\beta\rightarrow\alpha$&\\
\end{tabular}
\end{proof}

\begin{propriete}\label{finiss}
Let $\alpha$ be a basic formula containing only equations and $\beta$ and $\delta$ two conjunctions of constraints of the form $\fini(x)$ such that $\alpha\wedge\beta$ and $\alpha\wedge\delta$ are solved basic formulas. We have $T\models (\alpha\wedge\beta)\leftrightarrow(\alpha\wedge\delta)$ if and only if $\beta$ and $\delta$ have exactly the same contraints. 
\end{propriete}
\begin{proof}
If $\beta$ and $\delta$ have the same constraints then it is evident that we have $T\models (\alpha\wedge\beta)\leftrightarrow(\alpha\wedge\delta)$. Let us now show  that if we have $T\models (\alpha\wedge\beta)\leftrightarrow(\alpha\wedge\delta)$ then $\beta$ and $\delta$ have  the same constraints. Suppose that we have (*) $T\models (\alpha\wedge\beta)\leftrightarrow(\alpha\wedge\delta)$ and let us show that if $\fini(u)$ occurs in $\beta$ then it occurs also in $\delta$ and vice versa. If $\fini(u)$ occurs in $\beta$ then $T\models (\alpha\wedge\beta)\rightarrow\fini(u)$, thus from (*) we have  (i) $T\models (\alpha\wedge \delta)\rightarrow \fini(u)$. Since $\alpha\wedge\beta$ is solved then $u$ is not the left hand side of an equation of $\alpha$. Thus, (ii) the conjunction $\alpha\wedge\delta$ does not contain sub-formulas of the form $u=t[\bar{x}]\wedge\bigwedge_i\fini(x_i)$. Since $\alpha\wedge\delta$ is solved then $\delta$ does not contain formulas of the form $\fini(v)$ where $v$ is the left hand side of an equation of $\alpha$. Thus, (iii) the conjunction $\alpha\wedge\delta$ does not contain also sub-formulas of the form $v=t[\bar{x},u]\wedge\fini(v)$. From (i), (ii)  and (iii), $\fini(u)$ should occur in $\delta$. By the same reasoning (we replace $\beta$ by $\delta$ and vice versa), we show that if $\fini(u)$ occurs in $\delta$ then it occurs in $\beta$.
\end{proof}

Let us now introduce the notion of \emph{reachable variable}: \begin{definition1}\label{acc}
Let $\alpha$ be a basic formula and $\bar{x}$ a vector of variables. The reachable variables and equations of $\alpha$ from the variable $x_0$ are those which occur in a sub-formula of $\alpha$ of the form:
\[x_0=t_0(x_1)\wedge x_1=t_1(x_2)\wedge...\wedge
x_{n-1}=t_{n-1}(x_n),\]
where $x_{i+1}$ occurs in the term $t_i(x_{i+1})$.  The reachable variables and equations of $\exists\bar{x}\,\alpha$ are those which are reachable in $\alpha$ from the free variables of $\exists\bar{x}\,\alpha$. A sub-formula of $\alpha$ of the form $\fini(u)$ is called reachable in $\exists\bar{x}\,\alpha$ if $u\not\in\bar{x}$ or $u$ is a reachable variable of $\exists\bar{x}\,\alpha$. 
\end{definition1}

\begin{example}
In the formula: $\exists uvw\, z = f(u,v) \wedge v = g(v,u) \wedge w
= f(u,v)\wedge\fini(u)\wedge\fini(x),$ the equations $z = f(u,v)$ and $v = g(v,u)$,  the variables 
$z$, $u$ and $v$ and the formulas $\fini(u)$ and $\fini(x)$  are reachable. On the other hand the equation $w =f(u,v)$ and the variable $w$ are not reachable.
\end{example}

\begin{remarque}\label{rem}
Let $\alpha$ be a solved basic formula. Let $\bar{x}$ be a vector of variables. We have: 
\begin{itemize}
\item
If all the variables of $\bar{x}$ are reachable in $\exists\bar{x}\,\alpha$ then all the equations and relations  of $\alpha$ are reachable in $\exists\bar{x}\,\alpha$.
\item
If $v=t[y]$ is a reachable equation in $\exists\bar{x}\,\alpha$, then $\alpha$  contains  a sub-formula  of the form \begin{equation}\label{e25}\bigwedge_{j=1}^k v_j=t_j[v_{j+1}]\end{equation} with $k\geq 1$ and  (i) $v_1\not\in X$, (ii) for all $j\in\{1,...,k\}$ the variable $v_{j+1}$ occurs in the term $t_j[v_{j+1}]$, (iii) $v_{k}$ is the variable $v$, (iv) $v_{k+1}$ is the variable $y$ and $t_{k}[v_{k+1}]$ is the term $t[y]$. 
\end{itemize}
\end{remarque}

According to the first point of Remark \ref{rem} and Definition \ref{acc} we have the following property:
\begin{propriete}\label{astuce}
Let $\alpha$ be a solved basic formula. If the formula $\exists\bar{x}\,\alpha$ has no free variables and if all the variables of $\bar{x}$ are reachable in $\exists\bar{x}\,\alpha$ then $\bar{x}$ is the empty vector $\varepsilon$ and $\alpha$ is the formula $\vrai$.
\end{propriete}

According to the axioms [1] and [2] of $T$ we have the following property:
\begin{propriete}\label{acc1}
Let $\alpha$ be a basic formula. If all the
variables of $\bar{x}$ are reachable in $\exists\bar{x}\,\alpha$ then
\[T\models\exists?\bar{x}\,\alpha.\]
\end{propriete}

\begin{propriete}\label{tech}
Let $\bar{x}$ be a vector of variables and $\alpha$ a solved basic formula. We have: \[T\models(\exists\bar{x}\,\alpha)\leftrightarrow(\exists\bar{x}'\,\alpha'),\]
where:\begin{itemize}\item $\bar{x}'$ is the vector of the variable of $\bar{x}$ which are reachable in $\exists\bar{x}\,\alpha$, \item $\alpha'$ is the conjunction of the equations and the formulas of the form $\fini(x)$ which are reachable in  $\exists\bar{x}\,\alpha$.
\end{itemize}
\end{propriete}

\begin{proof}
Let us decompose $\bar{x}$ into three vectors $\bar{x}', \bar{x}''$ and $\bar{x}'''$ such that:
\begin{itemize}
\item
$\bar{x}'$ is the vector of the variables of $\bar{x}$ which are reachable in  $\exists\bar{x}\,\alpha$.
\item
$\bar{x}''$ is the vector of the variables of $\bar{x}$ which are non-reachable in $\exists\bar{x}\,\alpha$ and do not occur in the left hand sides of the equations of $\alpha$.
\item
$\bar{x}'''$ is the vector of the variables of $\bar{x}$ which are non-reachable in $\exists\bar{x}\,\alpha$ and occur in a left hand side of an equation of $\alpha$.

\end{itemize}
Let us now decompose  $\alpha$ into three  formulas $\alpha'$, $\alpha''$ and $\alpha'''$ such that:
\begin{itemize}
\item
$\alpha'$ is the conjunction of the equations and the formulas of the form $\fini(x)$ which are reachable in  $\exists\bar{x}\,\alpha$.
\item
$\alpha''$ is the conjunction of the formulas of the form $\fini(x)$  which are non-reachable in  $\exists\bar{x}\,\alpha$.
\item
$\alpha'''$ is the conjunction of the equations which are non-reachable in  $\exists\bar{x}\,\alpha$.

\end{itemize}
According to Definition \ref{acc}, all the variables of $\bar{x}''$ and $\bar{x}'''$ do not occur in $\alpha'$ (otherwise they will be reachable)  and since $\alpha$ is solved then $\bar{x}'''$ is the vector of the left hand sides of the equations of $\alpha'''$ and its variables do not occur in $\alpha''$. Thus the formula $\exists\bar{x}\,\alpha$ is equivalent in $T$ to \[(\exists\bar{x}'\,\alpha'\wedge(\exists\bar{x}''\,\alpha''\wedge(\exists\bar{x}'''\,\alpha'''))).\]
According to  Property \ref{unique} we have $T\models\exists!\bar{x}'''\,\alpha'''$. According to Corollary   \ref{inff2} we have $T\models\exists\bar{x}''\,\alpha''$. Thus, the preceding formula is equivalent in $T$ to 
 $(\exists\bar{x}'\,\alpha')$.
\end{proof}

\begin{example}
The formula $\exists xyzw\,v=f(x,x)\wedge w=g(y,z,x)\wedge\fini(x)\wedge \fini(y)$ is equivalent in $T$ to
\[\exists x\,v=f(x,x)\wedge\fini(x)\wedge(\exists yz\,\fini(y)\wedge(\exists w\,w=g(y,z,x))),\]
which, since $T\models\exists!w\,w=g(y,z,x)$ and $T\models\exists yz\,\fini(y)$, is equivalent in $T$ to 
\[\exists x\,v=f(x,x)\wedge\fini(x).\]
Property \ref{tech} confirms the fact that the theory $T$ does not accept full elimination of quantifiers and shows that we can  eliminate only non-reachable quantified variables. On the other hand, reachable variables cannot  be removed since their values depend on the instantiations of the free variables. In fact,  the formula $\exists x\,v=f(x,x)\wedge\fini(x)$ is neither true nor false in $T$ since for each model  $\mathcal M$ of $T$ there exist instantiations of the free variable $v$ which make it false in $\mathcal M$ and others which make it true in $\mathcal M$, and thus the reachable quantified variable $x$ cannot  be eliminated and the formula $\exists x\,v=f(x,x)\wedge\fini(x)$ cannot  be simplified anymore. On the other hand,  the formula $\exists w\,w=g(y,z,x)$ is true in any model of $T$ and for any instantiation of $z$. The quantified non-reachable variable $w$ can then be eliminated and the formula is replaced by $\vrai$. As we will see in section \ref{algoo}, reachability, has a crucial role while solving first-order  constraints in $T$. It shows which quantifications can be eliminated and enables to simplify complex quantified basic formulas. 
\end{example}

According to the axioms [1] and [2] and since the set $\Ff$ is
infinite we have the following property:

\begin{propriete}\label{infini}
Let $I=\{1,...,n\}$ be a finite (possibly empty) set and $\bar{x}$ and $\bar{x}'$ two disjoint vectors of variables. Let  $\bar{y}_{1}$,...,$\bar{y}_n$ be vectors of variables and 
$\alpha_1$,...,$\alpha_n$  solved basic formulas such that for all $i\in I$
all the variables of $\bar{y}_i$ are reachable in
$\exists\bar{y}_i\,\alpha_i$. If each conjunction $\alpha_i$ contains at least (1) one sub-formula of the form $\fini(x)$ with $x\in X$, {\bfseries or} (2) one equation which contains at least one occurrence of a variable $x\in X\cup X'$, then: 
\begin{equation}\label{e10}T\models \exists \bar{x}\bar{x}'(\bigwedge_{x\in X'}\fini(x))\wedge(\bigwedge_{i\in I}
\neg(\exists\bar{y}_i\,\alpha_i)).\end{equation}
\end{propriete}
\begin{proof}
Let ${\mathcal M}=<M, (f^M)_{f\in F}, \fini^M>$ be a model of $T$. To show  the validity of (\ref{e10}) it is enough to show that:
\begin{equation}\label{e11}{\mathcal M}\models \exists \bar{x}\bar{x}'(\bigwedge_{x\in X'}\fini^M(x))\wedge(\bigwedge_{i\in I}
\neg(\exists\bar{y}_i\,\alpha_i)).\end{equation}
Since the basic formulas $\alpha_i$ are solved, they do not contain equations of the form $x=x$. Suppose now that one of the $\alpha_i$ contains one equation of the form  $x=v$ with $x\in X\cup X'$ and $v\in Y_i$. Since $\alpha_i$ is solved then $x\succ v$ but according to the discipline of the formulas in $T$ we have $v\succ x$\footnote{In fact,  the variable $x$ has a free occurrence in $\exists\bar{y}_i\,\alpha_i$ and the variable $v$ has a bound occurrence in $\exists\bar{y}_i\,\alpha_i$ (because $v$ is a quantified reachable variable in $\exists\bar{y}_i\,\alpha_i$) and thus according to the discipline of our formulas we  have $v\succ x$.}. Since the order $\succ$ is strict then $x=v$ cannot  be a sub-formula of  $\alpha_i$. Thus, according to the conditions of Property \ref{infini}, each conjunction $\alpha_i$ contains at least (1) one sub-formula of the form $\fini(x)$ with $x\in X$, {\bf or} (2) one equation of the one of the following  forms: \begin{itemize}\item   (*) $x=f(v_1,...v_n)$ with $x\in X\cup X'$, \item  (**)  $x=v$ with $x$ and $v$ two distinct variables such that $x\in X\cup X'$ and $v\not\in Y$, \item  (***) $v=t[x]$ where $x$ is a variable of $X\cup X'$ which occurs in the term $t[x]$. According to the first point of Remark \ref{rem} and since for all $i\in\{1,...,n\}$ the variables of $\bar{y}_i$ are reachable in $\exists\bar{y_i}\,\alpha_i$, then the equation $v=t[x]$ is reachable in $\exists\bar{y_i}\,\alpha_i$ and thus according to the second point of Remark \ref{rem}  the conjunction $\alpha_i$  contains  a  sub-formula  of the form $(\bigwedge_{j=1}^k v_j=t_j[v_{j+1}])$ with $v_1\not\in Y_i$, for all $j\in\{1,...,k\}$ the variable $v_{j+1}$ occurs in the term $t_j[v_{j+1}]$ and $v_{k+1}$ is the variable $x$. But, since the case $v_1\in X\cup X'$  is  already treated in (*) and (**), then we can restrict ourself without loosing generality  to the case where $v_1\not\in Y_i\cup X\cup X'$,  i.e.  $v_1$ is free in (\ref{e11}).
\end{itemize}
 Let \begin{equation}\label{faitchier} \exists \bar{x}\bar{x}'(\bigwedge_{x\in X'}\fini^M(x))\wedge(\bigwedge_{i\in I}
\neg(\exists\bar{y}_i\,\alpha_i^*))\end{equation} be an any instantiation of $\exists \bar{x}\bar{x}'(\bigwedge_{x\in X'}\fini^M(x))\wedge(\bigwedge_{i\in I}
\neg(\exists\bar{y}_i\,\alpha_i))$ by individuals of $\mathcal M$. Let us show that there exists an instantiation for the variables of $X$ and $X'$ which satisfies the preceding formula. For that, let us chose an instantiation which respects the following conditions: 
\begin{itemize}
\item
(i) For each $x\in X'$, the instantiation $x^*$ of $x$ satisfies ${\mathcal M}\models\fini^M(x^*)$.
\item
(ii) If a conjunction $\alpha^*_i$  contains a sub-formula of the form $\fini^M(x)$ with $x\in X$ then the instantiation $x^*$ of $x$ satisfies ${\mathcal M}\models x^*=f^M(x^*,...,x^*)$ with $f$ an $n$-ary function symbol of strictly positive arity which does not occur in any $\alpha_i$ with $i\in I$.
\item
(iii) If a conjunction $\alpha^*_i$  contains a sub-formula of the form $x=f^M(v_1,...v_n)$ with $x\in X\cup X'$,  then the instantiation of $x$ starts with a different function symbol than $f$.
\item
(iv) If a conjunction $\alpha^*_i$  contains a sub-formula of the form $x=v$ with $x$ and $v$ two distinct variables such that $x\in X\cup X'$ and $v\not\in Y$, then the instantiation of  $x$ is different from those of  $v$.
\item
(v) If a conjunction $\alpha^*_i$ contains a sub-formula of the form $(\bigwedge_{j=1}^k v_j=t_j[v_{j+1}])$ with $v_1\not\in (X\cup X'\cup Y)$, for all $j\in\{1,...,k\}$ the variable $v_{j+1}$ occurs in the term $t_j[v_{j+1}]$, and $v_{k+1}\in X\cup X'$, then the instantiation of $v_{k+1}$ is different from $v^*$, where $v^*$ is the instantiation of $v_{k+1}$ obtained from those of $v_1$ in\footnote{Recall that  $v_1\not\in (X\cup X'\cup Y)$ and thus $v_1$ is a free variable in (\ref{e11}). As a consequence, it is already instantiated in (\ref{faitchier}).}  (\ref{faitchier}) so that ${\mathcal M}\models \bigwedge_{j=1}^k v_j=t_j[v_{j+1}]$.  \end{itemize}

A such instantiation of the variables of $X$ and $X'$ is always possible since : (1) there exists an infinity of function symbols in $F$ which are not constants (2) the set of the individuals $i$ of $\mathcal M$ such that ${\mathcal M}\models\fini^M(i)$ is infinite (see Property \ref{inff}). As a consequence, according to axioms [1] and [4], this instantiation implies a conflict inside each sub-instantiated-formula $\exists\bar{y}_i\,\alpha^*_i$, with $i\in\{1,...,n\}$ and thus  \[{\mathcal M}\models\exists \bar{x}\bar{x}'(\bigwedge_{i\in I}
\neg(\exists\bar{y}_i\,\alpha_i^*)).\]  Since this instantiation satisfies the first condition (i) of the preceding list of conditions then (\ref{faitchier}) holds and thus (\ref{e11}) holds.
 \end{proof}
We emphasize strongly  that this property holds only if the formula (\ref{e10}) satisfies the discipline of the formulas in $T$. This property is vital for solving first-order constraint over finite or infinite trees. In fact, since the variables of each $\bar{y}_i$ with $i\in\{1,...,n\}$ are reachable in $\exists\bar{y}_i\,\alpha_i$ then we cannot   eliminate or remove the quantification $\exists\bar{y}_i$ form $\exists\bar{y}_i\,\alpha_i$, and thus solving a constraint containing such formulas is not evident. Property \ref{infini} enables us to surmount this problem by reducing to $\vrai$ particular formulas containing sub-formulas which does not accept full elimination of quantifiers. 
\begin{example}
Let $x$, $y$, $z$ and $v$ be variables such that $y\succ x\succ z\succ w$. Let us consider the following formula $\varphi$:

\begin{equation}\label{e14}\exists x\left[\begin{array}{l}\neg(\exists y\,z=f(y)\wedge y=g(x))\wedge\\\neg(\exists\varepsilon\,x=w)\wedge\\\neg(\exists\varepsilon\,x=g(x))\end{array}\right].\end{equation}
This formula satisfies the  discipline of the formulas in $T$. Let $\mathcal M=<M,(f^M)_{f\in F},\fini^M>$ be a model of $T$. Note that we cannot eliminate the quantifier $\exists y$ in the sub-formula $\exists y\,z=f(y)\wedge y=g(x)$. In fact, this sub-formula is neither true nor false in $T$ because there exist  instantiations of the free variable $z$ in $\mathcal M$ which satisfy this sub-formula in $\mathcal M$ and others which do not satisfy it. On the other hand,  Property \ref{infini}, states that  formula (\ref{e14}) is true in $T$ for all instantiations of $z$ even if the sub-formula $\exists y\,z=f(y)\wedge y=g(x)$ is neither true nor false in $T$.  Let us check this strange result. For that, let us show that for each instantiation of the free variables $z$ and $w$ by two individuals $z^*$ and $w^*$ of $\mathcal M$,  there exists an instantiation $x^*$ of $x$ which makes false the three $\mathcal M$-formulas $(\exists y\,z^*=f^M(y)\wedge y=g(x^*))$, $(\exists\varepsilon\,x^*=w^*)$ and $(\exists\varepsilon\,x^*=g(x^*))$. We have:

\begin{itemize}
\item
In the formula $(\exists y\,z=f(y)\wedge y=g(x))$, the variable $x$ is reachable. Thus, its value is determined by the value of $z$ (because $z=f(g(x))$). Two cases arise: \begin{itemize}\item
If $z^*$ is  of the form $f(g(i))$ with $i\in M$ then it is enough to instantiate $x$ by an individual $x^*\in M$ which is different from\footnote{For example, we can take $x^*=f^M(i)$.} $i$, in order to make false $(\exists y\,z^*=f^M(y)\wedge y=g^M(x^*))$ in $\mathcal M$.
\item if $z^*$ is not of the form $f(g(i))$ with $i\in M$ then the $\mathcal M$-formula $(\exists y\,z^*=f^M(y)\wedge y=g^M(x))$ is  false in $\mathcal M$ for all the instantiations of $x$.
\end{itemize}

\item
In the $\mathcal M$-formula $(\exists\varepsilon\,x=w^*)$, it is enough to instantiate  $x$ by an element $x^*$ of $\mathcal M$ which is different from $w^*$ in order to  make false the $\mathcal M$-formula $(\exists\varepsilon\,x^*=w^*)$.

\item
In the $\mathcal M$-formula  $(\exists\varepsilon\,x=g^M(x))$, it is enough to instantiate $x$ by an individual which starts by a distinct function symbol than $g$ in order to make false $(\exists\varepsilon\,x=g^M(x))$ in $\mathcal M$.
\end{itemize}
Since the set of the functions symbols which are not constants is infinite then there exists an infinity of instantiations of $x$ which satisfy the three preceding conditions. Each of these instantiations  $x^*$  makes false the three $\mathcal M$-formulas $(\exists y\,z^*=f^M(y)\wedge y=g^M(x^*))$, $(\exists\varepsilon\,x^*=w^*)$ and $(\exists\varepsilon\,x^*=g^M(x^*))$ and thus  (\ref{e14}) holds. 
\end{example}

\subsection{Normalized  formula}

\begin{definition1} A normalized formula $\varphi$ of depth $d\geq 1$ is a formula of the form
\begin{equation}\label{norm}
\neg(\exists\bar{x}\,\alpha\wedge\bigwedge_{i\in I} \varphi_i),
\end{equation}
 with $I$ a finite (possibly empty) set, $\alpha$ a basic formula 
 and the $\varphi_i's$ are normalized formulas of depth $d_i$ with
 $d=1+\max\{0,d_1,...,d_n\}$.
 \end{definition1}

\begin{example}
 Let $f$ and $g$ be two $1$-ary function symbols which belong to $F$. The formula 
 \[\neg\left[\exists\varepsilon\,\fini(u)\wedge\left[\begin{array}{l} \neg(\exists x\,y=f(x)\wedge x=g(y)\wedge\neg(\exists\varepsilon\,y=g(x)\wedge\fini(x)))\wedge\\\neg(\exists\varepsilon\,x=f(z)\wedge\fini(z))\end{array}\right]\right]\]
is a  normalized formula of depth equals to three. The formula $\neg(\exists\varepsilon\,\vrai)$ is a   normalized formula of depth 1.  The smallest value of a depth of a normalized formula is 1. Normalized formulas of depth 0 are not defined and do not exist.
 \end{example}
 
 We will use now the
abbreviation wnfv for \emph{``without new free variables}". A
formula $\varphi$ is equivalent to a wnfv formula $\psi$ in $T$
means that $T\models\varphi\leftrightarrow\psi$ and $\psi$ does
not contain other free variables than those of $\varphi$.

\begin{propriete}\label{norm11} 
Every formula $\varphi$ is equivalent in $T$ to a wnfv normalized formula of depth $d\geq 1$.
\end{propriete}
\begin{proof}
 It is easy to transform any formula into a normalized formula, it is enough for example to follow the followings steps: \begin{enumerate}\item 
Introduce a supplement of equations and existentially quantified
variables to transform the conjunctions of atomic formulas into
conjunctions of flat formulas. \item Replace each sub-formula of the form $\faux$ by $\neg\vrai$ then  express all the
quantifiers and logical connectors using only the logical symbols $\neg$, $\wedge$
and $\exists$. This can be done using the following transformations\footnote{These equivalences are true in the empty theory and thus in any theory $T$. } of sub-formulas:
\[\begin{array}{lll}
(\varphi\vee\phi)&\Longrightarrow& \neg(\neg\varphi\wedge\neg\phi),\\

(\varphi\rightarrow\phi)&\Longrightarrow& \neg(\varphi\wedge\neg\phi),\\

(\varphi\leftrightarrow\phi)&\Longrightarrow& (\neg(\varphi\wedge\neg\phi)\wedge\neg(\phi\wedge\neg\varphi)),\\

(\forall x\,\varphi)&\Longrightarrow& \neg(\exists x\,\neg\varphi).\\

\end{array}
\]

 \item If the formula $\varphi$ obtained
does not start with the logical symbol $\neg$, then  replace it by $\neg(\exists\varepsilon\,\vrai\wedge\neg \varphi)$.
\item  Rename the quantified  variables so that the obtained formula satisfies the imposed discipline in $T$ (see Section \ref{rest}).
\item Lift the
quantifier before the conjunction, i.e.
$\varphi\wedge(\exists\bar{x}\,\psi)$ or  $(\exists\bar{x}\,\psi)\wedge\varphi$, becomes
$\exists\bar{x}\,\varphi\wedge \psi$ because the free variables
of $\varphi$ are distinct from those of $\bar{x}$.
\item Group the quantified variables into a vectorial quantifier, i.e. $\exists\bar{x}(\exists\bar{y}\,\varphi)$ or $\exists\bar{x}\exists\bar{y}\,\varphi$ becomes $\exists\overline{xy}\,\varphi$.
\item Insert empty vectors and formulas of the form $\vrai$ to get the  normalized form using the following transformations of sub-formulas: 
\begin{equation}\label{regle1}
\neg(\bigwedge_{i\in I}\neg\varphi_i)\Longrightarrow \neg(\exists\varepsilon\,\vrai\wedge\bigwedge_{i\in I}\neg\varphi_i),\end{equation}\begin{equation}\label{regle2} \neg(\alpha\wedge\bigwedge_{i\in I}\neg\varphi_i)\Longrightarrow \neg(\exists\varepsilon\,\alpha\wedge\bigwedge_{i\in I}\neg\varphi_i),\end{equation}
\begin{equation}\label{regle3}
\neg(\exists\bar{x}\,\bigwedge_{j\in J}\neg\varphi_j) \Longrightarrow \neg(\exists\bar{x}\,\vrai\wedge\bigwedge_{j\in J}\neg\varphi_j).
\end{equation}
with $\alpha$ a conjunction of elementary equations, $I$ a finite (possibly empty) set and $J$ a finite non-empty set.
\item Rename the quantified variables so that  the obtained normalized formula satisfies the discipline of the formulas  in $T$.
\end{enumerate}  If the starting
formula does not contain the logical symbol $\leftrightarrow$ then this
transformation will be linear, i.e. there exists a constant $k$
such that $n_2\leq kn_1$, where $n_1$ is the size of the starting
formula and $n_2$ the size of the normalized formula. We show easily by contradiction that the final formula obtained after application of these steps is normalized. 
\end{proof}

\begin{example}
Let $x$, $v$, $w$, $u$ be variables such that $x\succ v\succ w\succ u$. Let $f$ be a $2$-ary function symbol which  belongs to $F$. Let us apply the preceding steps to transform the following formula into a normalized formula: \[(f(u,v)=f(w,u)\wedge(\exists x\, u=x))\vee(\exists u\,\forall w\, u=f(v,w)).\] 
Note that the  formula does not start with $\neg$ and the variables $u$ and $w$ are free in $f(u,v)=f(w,u)\wedge(\exists x\, u=x)$ and bound in $\exists u\,\forall w\, u=f(v,w)$. Note also that this formula does not respect the discipline of the formulas  in $T$. $\\$Step 1: Let us first transform the equations into flat equations. The preceding formula is equivalent in $T$ to 
 \begin{equation}\label{ayay1}(\exists u_1\, u_1=f(u,v)\wedge u_1=f(w,u)\wedge(\exists x\, u=x))\vee(\exists u\,\forall w\, u=f(v,w)),\end{equation}where $u_1$ is a variable of $V$ such that $u_1\succ x\succ v\succ w\succ u$.$\\$
Step 2: Let us now express the quantifier $\forall$ using $\neg$, $\wedge$ and $\exists$. Thus, the formula  (\ref{ayay1}) is equivalent in $T$ to

 \[(\exists u_1\, u_1=f(u,v)\wedge u_1=f(w,u)\wedge(\exists x\, u=x))\vee(\exists u\,\neg(\exists w\, \neg(u=f(v,w)))).\]
Let us also express the logical symbol $\vee$ using $\neg$, $\wedge$ and $\exists$. Thus, the preceding formula is equivalent in $T$ to 
\begin{equation}\label{ayay2} \neg (\neg(\exists u_1\, u_1=f(u,v)\wedge u_1=f(w,u)\wedge(\exists x \,u=x))\wedge\\\neg(\exists u\,\neg(\exists w\, \neg(u=f(v,w))))).\end{equation}
Step 3: As the formula starts with $\neg$,  we move to Step 4.$\\$
Step 4: The occurrences of the quantified variables $u$ and $w$ in $(\exists u\,\neg(\exists w\, \neg(u=f(v,w))))$ must be renamed. Thus, the formula (\ref{ayay2}) is equivalent in $T$ to 
\[\neg (\neg(\exists u_1\, u_1=f(u,v)\wedge u_1=f(w,u)\wedge(\exists x \,u=x))\wedge\neg(\exists u_2\,\neg(\exists w_1\, \neg(u_2=f(v,w_1))))),\]
where $u_2$ and $w_1$ are variables of $V$ such that $w_1\succ u_2\succ u_1\succ x\succ v\succ w\succ u.$$\\$
Step 5: By lifting the existential quantifier  $\exists x$, the preceding formula is equivalent in $T$ to 
\[\neg (\neg(\exists u_1\,\exists x\, u_1=f(u,v)\wedge u_1=f(w,u)\wedge u=x)\wedge\neg(\exists u_2\,\neg(\exists w_1\, \neg(u_2=f(v,w_1))))).\]
Step 6: Let us group the two quantified variables $x$ and $u_1$ into a vectorial quantifier. Thus, the preceding formula is equivalent in $T$ to 

\[\neg (\neg(\exists u_1x\, u_1=f(u,v)\wedge u_1=f(w,u)\wedge u=x)\wedge\neg(\exists u_2\,\neg(\exists w_1\, \neg(u_2=f(v,w_1))))).\]
Step 7: Let us introduce empty vectors of variables and formulas of the form $\vrai$ to get the  normalized formula. According to the rule (\ref{regle1}), the preceding formula is equivalent in $T$ to 
\[\neg \left[\exists\varepsilon\,\vrai\wedge\left[\begin{array}{l} \neg(\exists u_1x\, u_1=f(u,v)\wedge u_1=f(w,u)\wedge u=x)\wedge\\\neg(\exists u_2\,\neg(\exists w_1\, \neg(u_2=f(v,w_1))))\end{array}\right]\right],\]
which  using the rule  (\ref{regle2}) with $I=\emptyset$   is equivalent in $T$ to 
\[\neg \left[\exists\varepsilon\,\vrai\wedge\left[\begin{array}{l} \neg(\exists u_1x\, u_1=f(u,v)\wedge u_1=f(w,u)\wedge u=x)\wedge\\\neg(\exists u_2\,\neg(\exists w_1\, \neg(\exists\varepsilon\, u_2=f(v,w_1))))\end{array}\right]\right],\]
which   using the rule (\ref{regle3})  is equivalent in $T$ to  
\[\neg \left[\exists\varepsilon\,\vrai\wedge\left[\begin{array}{l} \neg(\exists u_1x\, u_1=f(u,v)\wedge u_1=f(w,u)\wedge u=x)\wedge\\\neg(\exists u_2\,\vrai\wedge\neg(\exists w_1\,\vrai\wedge \neg(\exists\varepsilon\, u_2=f(v,w_1))))\end{array}\right]\right].\]
Step 8: This is a  normalized formula of depth 4 which respects the discipline of the formulas in $T$ since $w_1\succ u_2\succ u_1\succ x\succ v\succ w\succ u$.

\end{example}

\subsection{General solved formula}
\begin{definition1}\label{sol}
A \emph{general solved} formula is a normalized formula of the form
\[
\neg(\exists\bar{x}\,\alpha\wedge\bigwedge_{i=1}^{n}\neg(\exists\bar{y}_i\,\beta_i)),
\] 
with $n\geq 0$ and such that: 
 \begin{enumerate}\item $\alpha$ and all the $\beta_i$, with $i\in\{1,...,n\}$, are solved basic formulas. \item If $\alpha'$ is the conjunction of the equations of $\alpha$ then all the conjunctions $\alpha'\wedge\beta_i$, with $i\in\{1,...,n\}$, are solved basic formulas. \item All the variables
of $\bar{x}$  are reachable in
 $\exists\bar{x}\,\alpha$. \item For all $i\in\{1,...,n\}$, all the
variables of $\bar{y}_i$ are reachable
in  $\exists\bar{y}_i\,\beta_i$.
\item
If $\fini(u)$ is a sub-formula of $\alpha$ then for all $i\in\{1,...,n\}$, the formula $\beta_i$ contains either $\fini(u)$, or $\fini(v)$ where $v$  is a reachable variable from $u$ in $\alpha\wedge\beta_i$  and does not occur in a left hand side of an equation of $\alpha\wedge\beta_i$.

 \item For all $i\in\{1,...,n\}$, the formula  $\beta_i$ contains at least one atomic formula which does not occur in $\alpha$.
\end{enumerate} 
\end{definition1}

\begin{example}
Let $w$, $v$, $u_1$, $u_2$, $u_3$ be variables such that $w\succ v\succ u_1\succ u_2\succ u_3$.  The following formula is not a general  solved formula
\begin{equation}\label{merdee}\neg(\exists\varepsilon\,\fini(w)\wedge\neg(\exists v\, w=v\wedge\fini(v))).\end{equation}
This formula satisfies all the conditions of Definition \ref{sol} but it does not satisfy the discipline of the formulas in $T$. In fact, the variable $v$ is bound in $(\exists v\, w=v\wedge\fini(v))$ and the variable $w$ is free in $(\exists v\, w=v\wedge\fini(v))$ and thus we should have $v\succ w$ and not $w\succ v$. Let $u_4$ be a variable such that $u_4\succ w\succ v\succ u_1\succ u_2\succ u_3$. The formula (\ref{merdee}) is equivalent in $T$ to 
\[\neg(\exists\varepsilon\,\fini(w)\wedge\neg(\exists u_4\, w=u_4\wedge\fini(v))).\] This formula respects the discipline of the formulas of $T$ but is not a general solved formula since it does not satisfy the first  condition of Definition \ref{sol}. In fact, $w=u_4\wedge\fini(v)$ is not a solved basic formula since we have $u_4\succ w$.

The following formula is a general  solved formula
\[\neg(\exists v\, u_1=f(v)\wedge v=u_2\wedge\fini(u_2)\wedge \neg(\exists w\, u_2=f(w)\wedge \fini(w)\wedge\fini(u_3))).\]
\end{example}

\begin{propriete}\label{vrai} Let $\varphi$ be a general solved formula. If $\varphi$
  has no free variables then $\varphi$ is the formula
  $\neg(\exists\varepsilon\,\vrai)$ else neither $T\models\neg\varphi$ nor
  $T\models\varphi$.  \end{propriete} 
 
 \begin{proof}
 Let $\varphi$ be a general solved formula of the form \begin{equation}\label{eqq3}
\neg(\exists\bar{x}\,\alpha\wedge\bigwedge_{i\in I}\neg(\exists\bar{y}_{i}\,\beta_{i})),\end{equation}
two cases arise:

(1) If $\varphi$ does not contain free variables, then according to the first and third condition of Definition \ref{sol} and using Property \ref{astuce} we get  $\bar{x}=\varepsilon$ and $\alpha=\vrai$. As a consequence, the formula (\ref{eqq3}) is  equivalent in $T$ to \begin{equation}\label{ez}
\neg(\exists\varepsilon\,\vrai\wedge\bigwedge_{i\in I}\neg(\exists\bar{y}_{i}\,\beta_{i})),\end{equation}
 Since (\ref{ez}) has no free variables then  each $\exists\bar{y}_i\,\beta_i$  has no free variables.  According to the first and fourth condition of Definition \ref{sol}, and using Property \ref{astuce} we get: for all $i\in I$: $\bar{y}_i=\varepsilon$ and $\beta_i=\vrai$. But according to the last condition of Definition \ref{sol}  all the formulas $\beta_i$ should be different from $\vrai$ (since we do not distinguish between $\alpha$ and $\alpha\wedge\vrai$). Thus, the set $I$ must be empty. As a consequence, $\varphi$ is the formula   $\neg(\exists\varepsilon\,\vrai)$.
 
 (2) If $\varphi$ contains free variables then it is enough to show that there exist two distinct instantiations $\varphi'$ and $\varphi''$ of $\varphi$ by individuals of ${\mathcal T}r$\footnote{ Recall that ${\mathcal T}r$ is the model of finite or infinite trees.}  such that \[{\mathcal T}r\models\varphi'\,\,  and \,\,{\mathcal T}r\models\neg\varphi''.\] Note first that if $I\neq\emptyset$  then each  $(\exists\bar{y}_i\,\beta_i)$, with $i\in I$,  should contain at least one free variable. In fact, if $(\exists\bar{y}_i\,\beta_i)$, with $i\in I$,  does not contain free variables then this formula  is of the form $(\exists\varepsilon\,\vrai)$ according to the first and fourth point of Definition \ref{sol} and Property \ref{astuce}, which contradicts the last condition of Definition \ref{sol} (since we do not distinguish between $\alpha$ and $\alpha\wedge\vrai$).  Thus each $(\exists\bar{y}_i\,\beta_i)$, with $i\in I$, contains at least one free variable that can be instantiated. On the other hand: 
 
 \emph{Case 1}: If $\exists\bar{x}\,\alpha$ contains free variables then  we can easily  find an instantiation of the free variables of $\exists\bar{x}\,\alpha$ which contradicts the constraints of $\alpha$. In fact, let $z$ be a free variable. Four cases arise:

\begin{itemize}
 \item
 If $z=w$ is a sub-formula of $\alpha$ then according to Definition \ref{sol} $\alpha$ is a solved basic formula and thus  $z\succ w$. As a consequence,  $w$ cannot be a quantified variable otherwise the formula $\varphi$ does not respect the discipline of the formulas in $T$. Thus is enough to instantiate $z$ and $w$ by two distinct values. 
 \item
 If $z=f(\bar{w})$ is a sub-formula of $\alpha$ then it is enough to instantiate $z$ by a tree which starts by a function symbol which is different from $f$.
 \item
 If  $w=z$ or $w=t[z]$ is a sub-formula of $\alpha$ then according to Definition \ref{sol} all the variables of $\bar{x}$ are reachable in $\exists\bar{x}\,\alpha$ and thus according to the first point of Remark \ref{rem} the equations $w=z$ and $w=t[z]$ are reachable. According to the second point of Remark \ref{rem} the value of $z$ is linked to another free variable $v$ which occurs in a left hand side of an equation of $\alpha$. This case is already treated in two preceding cases. 
 \item
 If  $\fini(z)$ is a sub-formula of $\alpha$ then it is enough to instantiate $z$ by an infinite tree.
 \end{itemize}
 As a consequence, the instantiated formula of $\exists\bar{x}\,\alpha$ will be false in ${\mathcal T}r$ and thus ${\mathcal T}r\models\varphi'$. On the other hand, by following the same preceding steps and since:
 
  (i) the set  $F$ contains an infinity of function symbols which are not constants, 
  
  (ii) ${\mathcal T}r$ contains an infinity of individuals $u$ of ${\mathcal T}r$ such that ${\mathcal T}r\models\fini^{{\mathcal T}r}( u)$,
  
   (iii) $\varphi$ is a general solved formula, 
   
   \noindent then we show that there exists at least one  instantiation which satisfies all the constraints of $\alpha$ and contradicts the constraints of each $\beta_i$, with $i\in I$. In fact, (iv) in order to contradicts each constraint $\beta_i$, it is enough to follow the preceding discussion (by replacing $\alpha$ by $\beta_i$ ) and use (i) and (ii). On the other hand, according to Definition \ref{sol} all the variables of $\bar{x}$ are reachable in $\exists\bar{x}\,\alpha$, thus according to the first point of remark \ref{rem} all the equations and relations of $\alpha$ are reachable in $\exists\bar{x}\,\alpha$. According to the second point of remark \ref{rem} the values of the free variables which occur in these formulas are mainly linked to those of free variables which occur in left hand side of equations of $\alpha$. According to the two first conditions of Definition \ref{sol},  the variables of $Lhs(\alpha)$ are distinct and do not occur in $FINI(\alpha)$, $Lhs(\beta_i)$ and $FINI(\beta_i)$ for all $i\in\{1,...,n\}$. As a consequence,  from (iv) and using  (i), (ii) and (iii)  there exists at least one instantiation which satisfies $\exists\bar{x}\,\alpha$ and contradicts each $\exists\bar{y}_i\,\beta_i$ in ${\mathcal T}r$, with $i\in I$ and thus ${\mathcal T}r\models\neg\varphi''$.  Note that if $I=\emptyset$ then we have also  ${\mathcal T}r\models\neg\varphi''$ and ${\mathcal T}r\models\varphi'$ using the preceding instantiations. 
  
 \emph{Case 2}: If $\exists\bar{x}\,\alpha$ does not contain free variables then according to the first and third  condition of Definition \ref{sol} and Property \ref{astuce} we have $\bar{x}=\varepsilon$ and $\alpha=\vrai$. Since $\varphi$ contains at least one free variable then  $I\neq \emptyset$.  Let $k\in I$. Since:
 
  (i) the set  $F$ contains an infinity of function symbols which are not constants, 
  
  (ii) ${\mathcal T}r$ contains an infinity of individuals $u$ of ${\mathcal T}r$ such that ${\mathcal T}r\models\fini^{{\mathcal T}r}( u)$, 
  
  (iii) $\varphi$ is a general solved formula,
  
  \noindent  then we can easily find an instantiation of the free variables of $\exists\bar{y}_k\,\beta_k$ which satisfies the constraints of $\beta_k$ (similar to the second part of Case 1 by replacing $\alpha$ by $\beta_k$).  Such an instantiation makes false the instantiated formula $\neg(\exists\bar{y}_k\,\beta_k)$ in ${\mathcal T}r$ and thus ${\mathcal T}r\models\varphi'$. On the other hand, according to (i), (ii) and (iii), we show that  there exists at least one instantiation which contradicts the constraints of each $\beta_i$, with $i\in I$ (similar to the second part of Case 1 with $\alpha=\vrai$ and $\bar{x}=\varepsilon$). As a consequence, this instantiation satisfies all the $\neg(\exists\bar{y}_i\,\beta_i)$ in ${\mathcal T}r$, with $i\in I$ and thus ${\mathcal T}r\models\neg\varphi''$. 
 
 From Case 1 and Case 2, we have  ${\mathcal T}r\models\varphi'$  and ${\mathcal T}r\models\neg\varphi'',$  and thus  neither ${ T}\models\varphi$  nor ${T}\models\neg\varphi$.
   
 \end{proof}
 
 \begin{example}
Let $v_1$, $v_2$, $v$, $u$ and $w$ be variables such that $v_1\succ v_2\succ v\succ u\succ w$. Let $\varphi$ be the following general solved formula
\begin{equation}\label{dredi}
\neg(\exists v\, u=g(v,w)\wedge \neg(\exists v_1\,v=g(v,v_1)\wedge v_1=f(v))\wedge\neg(\exists v_2\,w=g(w,v_2)\wedge v_2=f(w))
\end{equation}
Let us consider for example the model ${\mathcal T}r$ of finite or infinite trees. If we instantiate the free variable $u$ by the finite tree $1$ where $1$ is a constant in $F$ which is distinct from  $g$ then according to axiom [1] of conflict of symbols, the instantiated formula of (\ref{dredi}) is true in ${\mathcal T}r$. On the other hand, if $u$ is instantiated by a tree  of the form $g(v^*,w^*)$ with $v^*\neq g(v^*,f(v^*))$ (for example $v^*=1$) and $w^*\neq g(w^*,f(w^*))$ (for example $w^*=1$) then the instantiated formula of (\ref{dredi}) is false in ${\mathcal T}r$. As a consequence (\ref{dredi}) is neither true nor false in the theory $T$. The reader should not think that the fact that we have neither $T\models\neg\varphi$ nor $T\models\varphi$ means that $\varphi$ is unsatisfiable in $T$. This  is of course false. In fact, since neither $T\models\neg\varphi$ nor $T\models\varphi$ then $\varphi$ has in each model $\mathcal M$ of $T$ a set of solutions which make it true in $\mathcal M$ and  another set of non-solutions which make it false in $\mathcal M$.  We also remind the reader that all the properties given after Section \ref{rest} hold only for formulas that respect the  discipline of the formulas of $T$.    \end{example}

   A similar property has been shown for the finite trees of J. Lassez \cite{33} and  the rational trees of M. Maher \cite{37}.  M. Maher in \cite{37} has also shown that if the set $F$ is finite and contains at least one $n$-ary function symbol  with $n\geq 2$, then the problem of deciding if a formula containing equations and the logical symbols $\wedge$, $\vee$, $\neg$ is equivalent to a disjunction of conjunctions of equations is a co-NP-complete problem, and  the problem of deciding if an expression represents a nonempty set of rational trees is NP-complete. Note also that in all our proofs we have not used the famous independence of inequations \cite{col84,32,com,34} but only the condition that the signature of $T$ is infinite and contains an infinity of function symbols which are not constants and at least one symbol which is a constant, which implies in this case the independence of the inequations.

\begin{propriete}\label{comb} Every general solved formula of the form $\neg(\exists\bar{x}\,\alpha\wedge\bigwedge_{i=1}^{n}\neg(\exists\bar{y}_i\,\beta_i))$
is equivalent in $T$ to the following Boolean combination of existentially quantified 
basic formulas: 

\[(\neg(\exists\bar{x}\,\alpha))\vee\bigvee_{i=1}^{n}(\exists\bar{x}\bar{y}_i\,\alpha\wedge
\beta_i).\]
\end{propriete}

\begin{proof}
Let \begin{equation}\label{e15}
\neg(\exists\bar{x}\,\alpha\wedge\bigwedge_{i=1}^{n}\neg(\exists\bar{y}_i\,\beta_i)),
\end{equation}
 be a general solved formula. According to the third point of Definition \ref{sol},  all the variables of $\bar{x}$ are reachable in $\exists\bar{x}\,\alpha$. Thus, according to Property \ref{acc1}, we have $T\models\exists?\bar{x}\,\alpha$. According to Property \ref{r2}, the formula (\ref{e15}) is equivalent in $T$ to
\[\neg((\exists\bar{x}\,\alpha)\wedge\bigwedge_{i=1}^n\neg(\exists\bar{x}\,\alpha\wedge(\exists\bar{y}_i\,\beta_i))),\] 
i.e. to
\[(\neg(\exists\bar{x}\,\alpha))\vee\bigvee_{i=1}^n(\exists\bar{x}\,\alpha\wedge(\exists\bar{y}_i\,\beta_i)),\] 
which, since the quantified variables have distinct names and different from those of the free variables, is equivalent in $T$ to 
\[(\neg(\exists\bar{x}\,\alpha))\vee\bigvee_{i=1}^n(\exists\bar{x}\bar{y}\,\alpha\wedge \beta_i),\] 
which is a Boolean combination of existentially quantified basic formulas.
\end{proof}

\begin{definition1}\label{kiki}
Let $\varphi$ be a formula of the form \begin{equation}\label{zsx1}
\exists\bar{x}\,\alpha\wedge\bigwedge_{i=1}^{n}\neg(\exists\bar{y}_i\,\beta_i),
\end{equation} 
with $\bar{x}$ and $\bar{y}$ two vectors of variables, $n\geq 0$ and $\alpha$ and the $\beta_i$, with $i\in\{1,...,n\}$,  basic formulas. We say that $\varphi$ is written in an \emph{explicit solved form} if and only if the formula $\neg\varphi$, i.e. \begin{equation}\label{zsx2}
\neg(\exists\bar{x}\,\alpha\wedge\bigwedge_{i=1}^{n}\neg(\exists\bar{y}_i\,\beta_i)),
\end{equation} is a general solved formula.
\end{definition1}
This definition shows how to easily extract from a general solved formula, a simple formula $\varphi$ which has only one level of negation and where the solutions of the free variables are given in clear and explicit way, i.e. for each model ${\cal M}$ of $T$, it is easy to find all the possible instantiations of the free variables of $\varphi$ which make it true in ${\cal M}$. In fact, according to Definition \ref{sol}, we warrant among other things that the left hand sides of the equations of $\alpha$ are distinct and do not occur in those of each $\beta_{i}$, the left hand sides of the equations of each $\beta_{i}$ are distinct and we cannot eliminate any quantification since all the variables are reachable. 
\begin{example}
Let $w$, $v$, $u_1$, $u_2$, $u_3$ be variables such that $w\succ v\succ u_1\succ u_2\succ u_3$.  Let $\varphi$ be the following general  solved formula
\[\neg(\exists v\, u_1=f(v)\wedge v=u_2\wedge\fini(u_2)\wedge \neg(\exists w\, u_2=f(w)\wedge \fini(w)\wedge\fini(u_3))).\]
According to Definition~\ref{kiki}, the following formula $\phi$ is written in an explicit solved form:
\begin{equation}\label{aqw}
\exists v\, u_1=f(v)\wedge v=u_2\wedge\fini(u_2)\wedge \neg(\exists w\, u_2=f(w)\wedge \fini(w)\wedge\fini(u_3)).
\end{equation}
 Let us chose the model ${{\mathcal T}r}$ of finite or infinite trees and let us give all the possible instantiations  $u_1^*,u_2^*,u_3^*$ of the free variables $u_1,u_2,u_3$ so that the instantiated formula of $\phi$ is true in the model ${{\mathcal T}r}$. From  (\ref{aqw}) it is clear that we have two possibilities: 
 
 \begin{itemize}
 \item \underline{Solution 1}:
 \begin{itemize}
 \item
 $u_3^*$ is any infinite tree.
 \item
 $u_2^*$ is any finite tree.
 \item
 $u_1^*$ is the tree $f(u_2^*)$.
 \end{itemize}
   \item \underline{Solution 2}:
\begin{itemize}
 \item
 $u_3^*$ is any finite tree.
  \item
 $u_2^*$ is any finite tree which starts by a function symbol which is different from $f$. 
 \item
 $u_1^*$ is the tree $f(u_2^*)$.
 \end{itemize}

 \end{itemize}

\end{example}

\subsection{Working formula}
\begin{definition1}\label{wor}
A working formula is a normalized formula in which all the
occurrences of $\neg$ are replaced by $\neg^k$ with
$k\in\{0,...,5\}$ and such that each occurrence of a sub-formula
of the form \begin{equation}\label{e17} p=\neg^k(\exists\bar{x}\,\alpha\wedge q),\:\:\:\:\:\:
with\:\:\: k>0,\end{equation} {\bfseries satisfies the $k$ first conditions} of the
 condition list bellow. In (\ref{e17}) $\alpha$ is a basic formula, $q$ is a conjunction of working formulas of the form $\bigwedge_{i=1}^{n}\neg^{k_i}(\exists\bar{y}_i\,\beta_i\wedge
q_i),$ with $n\geq 0$, $\beta_i$ a basic formula, $q_i$ a conjunction of working formulas,  and in the below condition list  $\alpha'$ is the basic formula of the immediate top-working formula\footnote{In other words, $p'$ is of the form $\neg^{k'}(\exists\bar{x}'\,\alpha'\wedge p^*\wedge p)$ where $p^*$ is a conjunction of working formulas and $p$ is the formula (\ref{e17}).}  $p'$ of $p$ if it
exists.
\begin{enumerate}
\item If $p'$ exists then $T\models \alpha\rightarrow \alpha'$ and $T\models \alpha_{eq}\rightarrow \alpha'_{eq}$ where $\alpha_{eq}$ and $\alpha'_{eq}$ are the conjunctions of the equations of $\alpha$ respectively $\alpha'$. Moreover, the set of the variables of $Lhs(\alpha')\cup FINI(\alpha')$ is included in those of $Lhs(\alpha)\cup FINI(\alpha)$.\item
The left hand sides of the equations of  $\alpha$ are distinct and for all equations of the form $u=v$ we have $u\succ v$.
\item
$\alpha$ is a basic solved formula.
 \item If $p'$ exists then the set of the equations of $\alpha'$ is
included in those of $\alpha$. \item The
variables of $\bar{x}$, the equations of $\alpha$ and  the constraints of the form $\fini(x)$ of $\alpha$ are reachable in $\exists\bar{x}\,\alpha$. Moreover,  if $n>0$ then for all $i\in\{1,...,n\}$ the conjunction $\beta_i$ contains at least one atomic formula which does not occur in $\alpha$.
\end{enumerate}
The intuitions behind these working formulas come from an  aim to have a  full control on the execution of
our rewriting rules  by adding semantic informations on a syntactic form of formulas. We emphasize strongly  that $\neg^k$ does not mean  that the normalized formula satisfies only the $k^{th}$ condition but all the conditions $i$ with $1\leq  i \leq k$.
\end{definition1}

\begin{example}\label{ahoi}
Let $w_1$, $w_2$, $w_3$, $v_1$, $u$ be variables such that $w_1\succ w_2\succ w_3\succ v_1\succ u$. This is a working formula of depth 2: 
\[\neg^2\left[\exists v_1\,u=f(v_1)\wedge\fini(u)\wedge\left[\begin{array}{l}\neg^2(\exists w_1\,u=f(w_1)\wedge w_1=v_1\wedge\fini(u))\wedge\\\neg^3(\exists w_2\,u=f(v_1)\wedge w_2=f(v_1)\wedge\fini(v_1))\wedge\\\neg^4(\exists w_3\,u=f(v_1)\wedge v_1=f(w_3)\wedge\fini(w_3))\end{array}\right]\right]\]
\end{example}

\begin{definition1} An \emph{initial}  working formula is a working formula which begins with 
$\neg^4$ and such that $k=0$ for all the other occurrences of
$\neg^k$. A \emph{final} working formula is a working formula of depth less
or equal to 2 with $k=5$ for all the occurrences of $\neg^k$. 
\end{definition1}

The
relation between the final working formulas and the general solved formulas is
expressed in the following property: \begin{propriete}\label{fina}
Let $p$ be the following final working formula
$\neg^5(\exists\bar{x}\,\alpha\wedge\bigwedge_{i=1}^{n}\neg^5(\exists\bar{y}_i\,
\beta_i)).$  The formula
$\neg(\exists\bar{x}\,\alpha\wedge\bigwedge_{i=1}^{n}\neg(\exists\bar{y}_i\,
\beta^*_i)),$ is a general solved formula equivalent to $p$ in $T$ where $\beta^*_i$ is the basic formula $\beta_i$ from which we have removed all the equations which occur also in $\alpha$. 
\end{propriete}

\begin{example}\label{ancien}
Let $w_2$, $v$, $u$ and $u_1$ be variables such that $w_2\succ v\succ u\succ u_1$. Let $\varphi$ be the following final working formula
\[\neg^5\left[\begin{array}{l}\exists \varepsilon\,v=u\wedge\fini(u)\wedge\\
\neg^5(\exists\varepsilon\, v=u\wedge u=u_1\wedge\fini(u_1))\wedge\\
\neg^5(\exists w_2\, v=u\wedge u=s(w_2)\wedge\fini(w_2))
\end{array}\right].
\]
The formula 
\[\neg\left[\begin{array}{l}\exists \varepsilon\,v=u\wedge\fini(u)\wedge\\
\neg(\exists\varepsilon\,u=u_1\wedge\fini(u_1))\wedge\\
\neg(\exists w_2\, u=s(w_2)\wedge\fini(w_2))
\end{array}\right].
\]
is a general solved formula equivalent to $\varphi$ in $T$. 
\end{example}

\subsection{Rewriting rules}\label{algoo}

We now present  the rewriting rules which transform an initial
working formula of any depth $d$ into an equivalent  conjunction of final working 
formulas. To apply the rule $p_1\Longrightarrow
p_2$ to the working formula $p$ means to replace in $p$ a
sub-formula $p_1$ by the formula $p_2$,  by considering that the
connector $\wedge$ is
 associative and commutative. In the following, the letters $u$, $v$ and $w$
represent
 variables, the letters
 $\bar{x}$, $\bar{y}$ and $\bar{z}$ represent vectors of variables, the letters  
$a$, $b$ and $c$ represent basic formulas, the letter  $q$ 
represents a conjunction of working
 formulas, the letter  $r$ represents a conjunction of flat equations, formulas of the form $\fini(x)$ and working
formulas. All these letters can be subscripted or have primes.

{\small
\[\begin{array}{cccc}
(1)&\neg^1(\exists\bar{x}\,u=u\wedge r)&\Longrightarrow&
\neg^1(\exists\bar{x}\,r)\\[1,5mm]
 (2)&\neg^1(\exists\bar{x}\,v=u\wedge
r)&\Longrightarrow&
\neg^1(\exists\bar{x}\,u=v\wedge r)\\[1,5mm]
(3)&\neg^1(\exists\bar{x}\,u=v\wedge u=t\wedge r)&\Longrightarrow&
\neg^1(\exists\bar{x}\,u=v\wedge v=t \wedge r)\\[1,5mm]
(4)&\neg^1(\exists\bar{x}\,u=fv_1...v_n\wedge u=gw_1...w_m\wedge
r)&\Longrightarrow& \vrai\\[1,5mm]
(5)&\neg^1(\exists\bar{x}\,u=fv_1...v_n\wedge u=fw_1...w_n\wedge
r)&\Longrightarrow& \neg^1(\exists\bar{x}\,
u=fv_1...v_n\wedge\bigwedge_{i=1}^{n} v_i=w_i\wedge r)
\\[1,5mm]
(6)&\neg^1(\exists\bar{x}\,a\wedge q)&\Longrightarrow &
\neg^2(\exists\bar{x}\,a\wedge q)\\[1,5mm]
(7)&\neg^2(\exists\bar{x}\,\fini(u)\wedge\fini(u)\wedge r)&\Longrightarrow & \neg^2(\exists\bar{x}\,\fini(u)\wedge r)
\\[1,5mm]
(8)&\neg^2(\exists\bar{x}\,u=v\wedge\fini(u)\wedge  r)&\Longrightarrow & \neg^2(\exists\bar{x}\,u=v\wedge \fini(v)\wedge r)
\\[1,5mm]
(9)&\neg^2(\exists\bar{x}\,\fini(u)\wedge a\wedge q )&\Longrightarrow & \vrai
\\[1,5mm]
(10)&\neg^2(\exists\bar{x}\,u=f(v_1,...,v_n)\wedge\fini(u)\wedge  r)&\Longrightarrow & \neg^2(\exists\bar{x}\,u=f(v_1,...,v_n)\wedge\bigwedge_{i=1}^{n} \fini(v_i) \wedge r)
\\[1,5mm]
(11)&\neg^2(\exists\bar{x}\,a\wedge q)&\Longrightarrow & \neg^3(\exists\bar{x}\,a\wedge q)
\\[1,5mm]
(12)&\neg^4(\exists\bar{x}\,a\wedge q\wedge\neg^0(\exists
\bar{y}\,r))&\Longrightarrow & \neg^4(\exists\bar{x}\,a\wedge
q\wedge\neg^1(\exists \bar{y}\,a\wedge r))
\\[1,5mm]
(13)&\neg^4(\exists\bar{x}\,a\wedge a'\wedge q\wedge\neg^3(\exists
\bar{y}\,a''\wedge r))&\Longrightarrow &
\neg^4(\exists\bar{x}\,a\wedge a'\wedge 
q\wedge\neg^4(\exists\bar{y}\,a\wedge r))\\[1,5mm]
(14)&\neg^4(\exists\bar{x}\,a\wedge q\wedge\neg^5(\exists
\bar{y}\,a))&\Longrightarrow & \vrai\\[1,5mm]
(15)& \neg^4(
\exists\bar{x}\,a\wedge
\bigwedge_{i=1}^{n}\neg^5(\exists\bar{y}_i\,
b_i))
&\Longrightarrow &\neg^5
(\exists\bar{x}'\,a'\wedge
\bigwedge_{i\in K}\neg^5(\exists\bar{y}'_i\,
b'_i)^*
)\\
(16)& \neg^4\left[\begin{array}{l} \exists\bar{x}\,a\wedge
q\wedge\\[2mm] \neg^5\left[\begin{array}{l} \exists\bar{y}\,
b\wedge\\[2mm]\bigwedge_{i=1}^{n}\neg^5(\exists\bar{z}_i\,c_i)
\end{array}\right]\end{array}\right]&\Longrightarrow &\left[\begin{array}{l}
\neg^4(\exists\bar{x}\,a\wedge q\wedge\neg^5(\exists\bar{y}\,b))\wedge\\[2mm]\bigwedge_{i=1}^{n}\neg^4(\exists\bar{x}\bar{y}\bar{z}_i\, c_i\wedge q_0)^*\\
\end{array}\right]\\[2mm]
\end{array}\]}
with  $u\succ v$, $f$ and $g$ two distinct function symbols taken from $F$. In  rule (3), $t$ is a flat term, i.e. either a variable or a term of the form $f(x_1,...,x_n)$ with $f$ an $n$-ary function symbol taken from $F$. In  rule (6), the equations of  $a$ have distinct left hand sides and for each equation of the form $u=v$ we have $u\succ v$. In
 rule (9), the variable $u$ is reachable from $u$ in  $a$. In  rule (10), the variable $u$ is non-reachable from $u$ in  $a$. Moreover, if $f$ is a constant then $n=0$. In  rule (11), $a$ is a  solved basic formula. In  rule (13), $a$ and $a''$ are conjunctions of equations having the same left hand sides and  $a'$ is a conjunction of formulas of the form $\fini(u)$.  In  rule (15), 
$n\geq 0$ and for all $i\in\{1,...,n\}$  the formula $b_i$ is different from the formula $a$. The pairs  
$(\bar{x}',a')$ and $(\bar{y}'_i,b'_i)$ are obtained by a decomposition of $\bar{x}$ and $a$ into $\bar{x}'\bar{x}''\bar{x}'''$ and $a'\wedge a''\wedge a'''$ as follows: \begin{itemize}\item
$a'$ is the conjunction of the equations and the formulas of the form $\fini(x)$  which are reachable in    $\exists\bar{x}\,a$.
\item
$\bar{x}'$ is the vector the variables of $\bar{x}$ which are reachable in $\exists\bar{x}\,a$.
\item
$a''$ is the conjunction of the formulas of the form $\fini(x)$  which are non-reachable in $\exists\bar{x}\,a$. 
\item
$\bar{x}''$ is the vector the variables of $\bar{x}$ which are non-reachable in $\exists\bar{x}\,a$ and do not occur in the left hand sides of the equations of $a$.
\item
$a'''$ is the conjunction of the equations  which are non-reachable in $\exists\bar{x}\,a$.
\item
$\bar{x}'''$ is the vector the variables of $\bar{x}$ which are non-reachable in $\exists\bar{x}\,a$ and occur in the left hand sides of the equations of $a$.
\item
$b^*_i$ is the formula obtained  by removing from $b_i$ the formulas of the form $\fini(u)$ which occur also  in $a''$
\item
$\bar{y}'_i$ is the vector of the variables of $\bar{y}_i\bar{x}'''$ which are reachable in $\exists\bar{y}_i\bar{x}'''\,b^*_i$.
\item
$b'_i$  is the conjunction of the equations and the formulas of the form $\fini(x)$  which are reachable in $\exists\bar{y}_i\bar{x}'''\,b^*_i$.

\item $K\subseteq\{1,...,n\}$ is the set of the indices $i$ such that $i\in K$ if and only if no variable of $\bar{x}''$ occurs in $b'_i$.
\item
The formula $\bigwedge_{i\in K}\neg^5(\exists\bar{y}'_i\,
b'_i)^*$ is the formula $\bigwedge_{i\in K}\neg^5(\exists\bar{y}'_i\,
b'_i)$ in which we have renamed the quantified variables  so that they satisfy the discipline of the formulas in $T$.
\end{itemize}
 In  rule (16), $n>0$  and $q_0$ is the formula $q$ in which all the
occurrences of $\neg^k$ have been replaced by
$\neg^0$. The formula $\bigwedge_{i=1}^{n}\neg^4(\exists\bar{x}\bar{y}\bar{z}_i\, c_i\wedge q_0)^*$ is the formula $\bigwedge_{i=1}^{n}\neg^4(\exists\bar{x}\bar{y}\bar{z}_i\, c_i\wedge q_0)$ in which we have renamed the quantified variables so that they satisfy the discipline of the formulas of $T$.

 The use of  indices on the negations of the working formulas enables us to force the application of the rules to follow a clear  strategy until reaching a conjunction of final working formulas. In fact, the algorithm follows two main steps while solving any first-order  constraint in $T$: \begin{itemize}\item (i) A top-down propagation of basic formulas following the tree structure of the working formulas and using the rules (1),...,(13). In this step,  basic  formulas are solved and copied in all sub-working formulas. Finiteness is also check and inconsistent basic formulas are removed by the rules (4) and (9). \item (ii) A bottom-up elimination of quantifiers and depth reducing of the  working formulas using the rules (14),...,(16).  Inconsistent working formulas are also removed in this step.  \end{itemize} More precisely, starting from an initial working formula $\varphi$ of the form $\neg^4(\exists\bar{x}\,a\wedge\bigwedge_{i\in I}  q_i)$,  where all the $q_i$ are working formulas whose negations are of the form  $\neg^0$, rule (12) propagates the atomic formulas  of $a$  into a sub-formula  $q_i$, with $i\in I$, and changes the first negation of $q_i$  into $\neg^1$. The rules (1),...,(5) can now be applied until the equations of  $a$ have distinct left hand sides and  for each equation of the form $u=v$ we have $u\succ v$. Rule (6) is then  applied and changes the first negation of $q_i$ into $\neg^2$. The algorithm starts now a new phase which consists in solving the basic formulas using  the rules (7),...,(10). In particular finiteness is checked by rule (9). When a solved basic formula is obtained, rule (11) is applied and changes the negation into $\neg^3$. Note that if a working formula starts by $\neg^3$ then its top working formula starts by $\neg^4$. Rule (13) is then applied. It restores some equations  and changes the first negation into $\neg^4$. Rule (12) can now be applied again since all the nested negations are of the form $\neg^0$ and so on. This is the first step of our algorithm. Once the sub-working formulas of depth 1 are of the form $\neg^4(\exists\bar{y}_i\,b_i)$, the second step starts using rule (15)  with $n=0$ on all these sub-working-formulas of depth 1 and transforms their negations into $\neg^5$. Inconsistent working formulas of the form $\neg^4(\exists\bar{x}\,\alpha\wedge\neg^5(\exists\bar{y}\,\alpha)\wedge q)$ are then removed by rule (14). When all the inconsistent working formulas have been removed, rule (15) with $n\neq 0$ can  be applied on the sub-working-formulas of depth 2 of the form $\neg^4(\exists\bar{x}\,a\wedge\bigwedge_{i\in I}\neg^5(\exists\bar{y}_i\,b_i))$ and produces working formulas of the form $\neg^5(\exists\bar{x}\,a\wedge\bigwedge_{i\in I}\neg^5(\exists\bar{y}_i\,b_i))$. Rule (16) can now be applied  on the working formulas of depth $d>2$ of the form $\neg^4(\exists\bar{x}\,a\wedge
q\wedge\neg^5( \exists\bar{y}\,
b\wedge\bigwedge_{i=1}^{n}\neg^5(\exists\bar{z}_i\,c_i)))$. After each application of this rule, new working formulas containing negations of the form $\neg^0$ are created which implies the execution of the rules of the first step of our algorithm, starting by rule (12) and so on. After several applications of our rules, we  get a conjunction of working formulas whose depth is less or equal to 2. The rules are then applied again until all the negations of  these working formulas are of the form $\neg^5$. It is a conjunction of final working formulas.

\begin{example}\label{exem}
Let $f$ and $g$ be two function symbols taken from $F$ of respective arities $2,1$. Let $w_1$, $w_2$, $v_1$, $u_1$, $u_2$, $u_3$ be variables such that $w_1\succ w_2\succ v_1\succ u_1\succ u_2\succ u_3$. Let us run our rules on the following initial working formula
\begin{equation}\label{zob1} \neg^4\left[\begin{array}{l}\exists v_1\, v_1 = f(u_1,u_2)\wedge u_2 =
g(u_1)\wedge\\\neg^0( \exists w_1\, v_1 =
g(w_1))\wedge\\\neg^0(\exists w_2\, u_2 = g(w_2)\wedge w_2 =
g(u_3)\wedge\fini(w_2))\end{array}\right].\end{equation} According to rule (12), the preceding formula is equivalent in $T$ to \[ \neg^4\left[\begin{array}{l} \exists v_1\, v_1
= f(u_1,u_2)\wedge u_2 = g(u_1)\wedge\\\neg^1( \exists w_1\, v_1 =
g(w_1)\wedge v_1 = f(u_1,u_2)\wedge u_2 = g(u_1))\wedge\\\neg^0(\exists
w_2\, u_2 = g(w_2)\wedge w_2 = g(u_3)\wedge\fini(w_2))\end{array}\right].\] The
application of rule (4)  on the sub formula $\neg^1( \exists
w_1\, v_1 = g(w_1)\wedge v_1 = f(u_1,u_2)\wedge u_2 = g(u_1)\wedge\fini(w_2))$ simplifies
this sub formula into the formula $\vrai$. Thus, the preceding formula is equivalent in $T$ to \[\neg^4\left[\begin{array}{l}
\exists v_1\, v_1 = f(u_1,u_2)\wedge u_2 = g(u_1)\wedge\\\neg^0(\exists
w_2\, u_2 = g(w_2)\wedge w_2 = g(u_3)\wedge\fini(w_2))\end{array}\right],\] 
which according to rule (12)  is equivalent in $T$ to 
 \[\neg^4\left[\begin{array}{l}
\exists v_1\, v_1 = f(u_1,u_2)\wedge u_2 = g(u_1)\wedge\\\neg^0(\exists
w_2\, v_1 = f(u_1,u_2)\wedge u_2 = g(u_1)\wedge u_2 = g(w_2)\wedge w_2 = g(u_3)\wedge\fini(w_2))\end{array}\right].\] 
Rule (5) can now be applied. Thus, the preceding formula is equivalent in $T$ to 
\[\neg^4\left[\begin{array}{l}\exists v_1\, v_1 = f(u_1,u_2) \wedge
u_2 = g(u_1)\wedge\\\neg^1(\exists w_2\, v_1 = f(u_1,u_2) \wedge u_2 =
g(w_2)\wedge w_2 = u_1\wedge w_2 = g(u_3)\wedge\fini(w_2))\end{array}\right],\] 
which according to rule (3)  is equivalent in $T$ to 
\[\neg^4\left[\begin{array}{l}\exists v_1\, v_1 = f(u_1,u_2)\wedge
u_2 = g(u_1) \wedge\\\neg^1(\exists w_2\, v_1 = f(u_1,u_2)\wedge u_2 =
g(w_2) \wedge w_2 = u_1\wedge u_1 = g(u_3)\wedge\fini(w_2))\end{array}\right].\] Since the 
conjunction of equations of the sub-formula which starts by $\neg^1$ has distinct left hand sides and $w_2\succ u_1$, then rule (6) can be applied. Thus, the preceding formula is equivalent in $T$ to
\[\neg^4\left[\begin{array}{l}\exists v_1\, v_1 = f(u_1,u_2)\wedge
u_2 = g(u_1)\wedge\\ \neg^2(\exists w_2\, v_1 = f(u_1,u_2)\wedge u_2 =
g(w_2)\wedge w_2 = u_1\wedge u_1 = g(u_3)\wedge\fini(w_2))\end{array}\right],\]
which according to rule (8) is equivalent in $T$ to 
\[\neg^4\left[\begin{array}{l}\exists v_1\, v_1 = f(u_1,u_2)\wedge
u_2 = g(u_1)\wedge\\ \neg^2(\exists w_2\, v_1 = f(u_1,u_2)\wedge u_2 =
g(w_2)\wedge w_2 = u_1\wedge u_1 = g(u_3)\wedge\fini(u_1))\end{array}\right],\]
which according to  rule (10) is equivalent in $T$ to 
\[\neg^4\left[\begin{array}{l}\exists v_1\, v_1 = f(u_1,u_2)\wedge
u_2 = g(u_1)\wedge\\ \neg^2(\exists w_2\, v_1 = f(u_1,u_2)\wedge u_2 =
g(w_2)\wedge w_2 = u_1\wedge u_1 = g(u_3)\wedge\fini(u_3))\end{array}\right].\]
Since the basic formulas are solved then rule (11) can be applied. Thus, the preceding formula is equivalent in $T$ to 
\[\neg^4\left[\begin{array}{l}\exists v_1\, v_1 = f(u_1,u_2)\wedge
u_2 = g(u_1)\wedge\\ \neg^3(\exists w_2\, v_1 = f(u_1,u_2)\wedge u_2 =
g(w_2)\wedge w_2 = u_1\wedge u_1 = g(u_3)\wedge\fini(u_3))\end{array}\right],\]
which according to rule (13)  is equivalent in $T$ to 
\[\neg^4\left[\begin{array}{l}\exists v_1\, v_1 = f(u_1,u_2)\wedge
u_2 = g(u_1)\wedge\\ \neg^4(\exists w_2\, v_1 = f(u_1,u_2)\wedge u_2 =
g(u_1)\wedge w_2 = u_1\wedge u_1 = g(u_3)\wedge\fini(u_3))\end{array}\right].\]
Rule (15) can now be applied  with $n=0$. Thus,  the preceding formula is equivalent in $T$ to 
\[\neg^4\left[\begin{array}{l}\exists v_1\, v_1 = f(u_1,u_2)\wedge
u_2 = g(u_1)\wedge\\ \neg^5(\exists\varepsilon\, v_1 = f(u_1,u_2)\wedge u_2 =
g(u_1)\wedge  u_1 = g(u_3)\wedge\fini(u_3))\end{array}\right].\]
 Once again rule (15) can be applied, with $n\neq 0$ and we get the following final working formula 
\[\neg^5\left[\begin{array}{l}\exists \varepsilon\, u_2 = g(u_1)\wedge\\ \neg^5(\exists\varepsilon\, u_2 =
g(u_1)\wedge  u_1 = g(u_3)\wedge\fini(u_3))\end{array}\right],\]
which according to Property \ref{fina} is equivalent in $T$ to the following general 
solved formula
\[\neg\left[\begin{array}{l}u_2 = g(u_1)
\wedge\\\neg(u_1 = g(u_3)\wedge\fini(u_3))\end{array}\right].\]
\end{example}
We have seen in the preceding example how the rules (1),...,(15) can be applied. Let us now  see  how rule (16) is applied.
$\\$

\begin{example}
Let $s$ and $0$ be two function symbols taken from $F$ of respective arities $1,0$. Let $w_1$, $w_2$, $u$, $v$ be variables such that $w_1\succ w_2\succ v\succ u$. Let us apply our rules on the following working formula of depth 3: 
\[
\neg^4\left[\exists\varepsilon\,\vrai\wedge\left[\begin{array}{l}\neg^5(\exists\varepsilon\,u=s(v))\wedge\\\neg^5(\exists w_1\, u=s(w_1)\wedge w_1=s(v))\wedge\\\neg^5(\exists\varepsilon\,v=u\wedge\neg^5(\exists\varepsilon\,v=u\wedge u=0)\wedge\neg^5(\exists w_2\, v=u\wedge u=s(w_2)))\end{array}\right]\right].\]
By considering that \begin{itemize}
\item
$(\exists\bar{x}\,a)=(\exists\varepsilon\vrai)$
\item
$q=\left[\begin{array}{l}\neg^5(\exists\varepsilon\,u=s(v))\wedge\\\neg^5(\exists w_1\, u=s(w_1)\wedge w_1=s(v))\end{array}\right]$
\item
$(\exists\bar{y}\,b)=(\exists\varepsilon\,v=u)$
\item
$\bigwedge_{i=1}^n\neg^5(\exists\bar{z}_i\,c_i)=\left[\begin{array}{l}\neg^5(\exists\varepsilon\,v=u\wedge u=0)\wedge\\\neg^5(\exists w_2\, v=u\wedge u=s(w_2))\end{array}\right]$
\end{itemize}
rule (16) can be applied and produces the following formula 
\[
\left[\begin{array}{l}\neg^4(\exists\varepsilon\,\vrai\wedge\neg^5(\exists\varepsilon\,u=s(v))\wedge\neg^5(\exists w_1\, u=s(w_1)\wedge w_1=s(v))\wedge\neg^5(\exists\varepsilon\,v=u))\wedge\\
\neg^4(\exists\varepsilon\,v=u\wedge u=0\wedge\neg^0(u=s(v))\wedge\neg^0(\exists w_{11}\, u=s(w_{11})\wedge w_{11}=s(v)))\wedge\\
\neg^4(\exists w_2\, v=u\wedge u=s(w_2)\wedge\neg^0(\exists\varepsilon\,u=s(v))\wedge\neg^0(\exists w_{12}\, u=s(w_{12})\wedge w_{12}=s(v)))\end{array}\right],\]
where $w_{11}$ and $w_{12}$ are variables such that  $w_{11}\succ w_{12}\succ w_1\succ w_2\succ v\succ u$. Now, only the rules (1),...,(15) will be applied until all the negations are of the form $\neg^5$. Rule (16) will not be applied anymore since there exists no working formulas of depth greater or equal to 3 and the rules (1),...,(15) never increase the depth of the working formulas.  
\end{example}

\begin{propriete}\label{resol}
Every repeated application of the preceding rewriting rules on an
initial working formula $p$ is terminating and producing a wnfv 
conjunction of final working formulas equivalent to $p$ in $T$.
\end{propriete}

\begin{proof}
 $\\$ {\emph{Proof, first
part:}} The application of the rewriting rules terminates. Let us introduce the  function $\alpha: q\rightarrow n$, where $q$ is a conjunction of working formulas, $n$ an integer and such that  \begin{itemize} \item $\alpha(\vrai)=0$, \item
$\alpha(\neg(\exists\bar{x}\,a\wedge\varphi))=2^{\alpha(\varphi)}$,
\item $\alpha(\bigwedge_{i\in I}\varphi_i)=\sum_{i\in
I}\alpha(\varphi_i),$
\end{itemize}
with $a$ a basic formula, $\varphi$ a conjunction of working formulas and
the $\varphi_i$'s working formulas. Note that if $\alpha(p_2) <
\alpha(p_1)$ then $\alpha(p[p_2]) < \alpha(p)$ where $p[p_2]$ is
the formula obtained from $p$ when we replace the occurrence of
the formula $p_1$ in $p$ by $p_2$. This function has been introduced in \cite{vo14} and \cite{dao2} to show the non-elementary  complexity of all algorithms solving propositions in the theory of finite or infinite trees. It has also  the property to decrease if the depth of the working formula decreases after application of  distributions as it is done in our rule (16). 

Let us  introduce also the function $\lambda: (u,a)\rightarrow n$, where $u$ is a variable, $a$ a basic formula, $n$ an integer and such that 

\[\lambda(u,a)=\left[\begin{array}{ll}
 \,0, & $if the conjunction of the equations of $ a $ has $\\
&  $ not distinct left hand sides or contains a $\\
& $sub-formula of the form $ x=y $ with $y\succ x,\, ${\bf else}$\\[1mm]
 \,1, & $ if $ u $ does not occur in a left hand side of an equation$ \\
& $ of $a $, or $ u $ is reachable from $ u $  in $ a, \, ${\bf else}$\\[1mm]
\,1+\lambda(v,a), & $if the equation $u=v$ is in $a,\, ${\bf else}$\\[1mm]
 \, 2+\sum_{i=1}^{n} \lambda(v_i,a), & $if the equation $ u=f(v_1,...,v_n) $ is in $ a.\\
\end{array}\right]\]
Since
the variables which occur in our formulas are ordered by the order
relation $``\succ"$, we can number them by positive integers
such that
\[x\succ y \leftrightarrow no(x)>no(y),\] where $no(x)$ is the
number associated  to the variable $x$. Let us consider the
10-tuple $(n_1,n_2,n_3,n_4,n_5,n_6,n_7,n_8,n_9,n_{10})$ where the $n_i$'s are
the following positive integers:
\begin{itemize}
\item $n_1=\alpha(p)$,\item $n_2$ is the number of $\neg^0$, \item $n_3$ is the number of 
$\neg^1$, \item $n_4$ is the number of  occurrences of function
symbols in  sub-formulas of the form $\neg^1(...)$. For example, if we have $\neg^1(\exists x\, x=f(y)\wedge y=f(x)\wedge x=g(x,w)\wedge y=f(y))$ then $n_4=4$. \item $n_5$
is the sum of all the $no(x)$ for each occurrence of a variable $x$  in a basic formula of 
a sub-formula of the form $\neg^1(...)$. For example, if we have $\neg^1(\exists w\,x=f(x,z)\wedge y=x\wedge\fini(z)\wedge...)$ then $n_5=no(x)+no(x)+no(z)+no(y)+no(x)+no(z)+...$. \item $n_6$ is the number
 of formulas of the form $v=u$ with $u\succ v$ in
 sub-formulas of the form $\neg^1(...)$,\item $n_7$ is the
number  of $\neg^2$, 
\item $n_8$ is the sum of all the $\lambda(u,a)$ for each occurrence of a sub-formula  $\fini(u)$  in a basic-formula $a$ of a working formula of the form $\neg^2(\exists\bar{x}\, a\wedge q)$.  For example, if we have $\neg^2(\exists z\,x=f(x,z)\wedge z=f(y,y)\wedge\fini(x)\wedge\fini(x)\wedge\fini(z))$ then $n_8=\lambda(x,a)+\lambda(x,a)+\lambda(z,a)=1+1+(2+1+1)$ where $a$ is the basic formula $x=f(x,z)\wedge z=f(y,y)\wedge\fini(x)\wedge\fini(x)\wedge\fini(z)$.

\item
$n_9$ is the number  of $\neg^3$
\item
$n_{10}$ is the number  of $\neg^4$.
\end{itemize}
For each rule, there exists a positive integer $i$ such that the application of
this rule decreases or does not change the values of the $n_j$'s, with $1\leq
j<i$, and decreases the value of $n_i$. These $i$ are equal to: 1
for the rules (4), (9), (14) and (16), 2 for rule (12), 3 for 
rule (6), 4 for rule (5), 5 for  the rules (1), (3), (7) and (8) , 6 for
 rule (2), 7 for  rule (11), 8 for  rule (10), 9 for  rule (13), and 10 for rule (15). To
each sequence of formulas obtained by a finite application of the
preceding rewriting rules, we can associate a series of 10-tuples
$(n_1,n_2,n_3,n_4,n_5,n_6,n_7,n_8,n_9,n_{10})$ which is strictly decreasing
in the lexicographic order. Since the $n_i$'s are
positive integers, they cannot be negative, thus, this series of
10-tuples is a finite
series and the application of the rewriting rules terminates.%
$\\[3mm]${\emph{Proof, second part:}} Let us now show  that for each rule of the form
$p\Longrightarrow p'$ we have $T\models p\leftrightarrow p'$ and
the formula $p'$ remains a conjunction of working formula.
\subsubsection*{Correctness of  the rules (1),...,(14)}
The rules (1),...(5) are correct according to the axioms [1] and [2] of $T$. Rules (6) and (11) are evident.  The rules (7) and (8) are true in the empty theory and thus true in $T$. In rule (9), the variable $u$ is reachable from itself in $a$, i.e. the basic formula $a$ contains a sub-formula of the form 
\begin{equation}\label{ho} u=t_1\wedge u_2=t_2\wedge...\wedge u_n=t_n\end{equation}
where $u_i$ occurs in the term $t_{i-1}$ for all $i\in\{2,...,n\}$ and $u$ occurs in $t_n$. According to Definition \ref{wor}, since our working formula starts with $\neg^2$ then all the equations of $a$ have distinct lef hand sides and for all equations of the form $x=y$ we have $x\succ y$. Thus, there exists at least one equation in (\ref{ho}) which contains a function symbol which is not a constant, otherwise (\ref{ho}) is of the form $u=u_2\wedge u_2=u_3\wedge...\wedge u_n=u$ which implies $u\succ u_2\succ ...\succ u$, i.e. $u\succ u$ which is false since the order $\succ$ is strict. Thus, according to the fourth axiom of $T$ we have $T\models a\rightarrow\neg\fini(u)$. As a consequence, rule (9) is correct in $T$. Rule (10) is correct according to the last axiom of $T$. Rule (13) is correct according to Property \ref{mg} and  Definition \ref{wor}. The rules (12) and (14) are true in the empty theory and thus true in $T$. Note that according to Property \ref{finiss},  two solved basic formulas having the same equations are equivalent if and only if  they have the same relations $\fini(x)$. This is why in Definition \ref{wor} of the working formulas (more precisely in  condition 4) we force only the equations to be included in the sub-forworking formulas and use the elementary  rule (14) to remove inconsistent working formulas of depth 2.

\subsubsection*{Correctness of rule (15)} 

\[ \neg^4(
\exists\bar{x}\,a\wedge
\bigwedge_{i=1}^{n}\neg^5(\exists\bar{y}_i\,
b_i))
\Longrightarrow \neg^5
(\exists\bar{x}'\,a'\wedge
\bigwedge_{i\in K}\neg^5(\exists\bar{y}'_i\,
b'_i)^*
)
\]
with $n\geq 0$, and for all $i\in\{1,...,n\}$  the formula $b_i$ is different from the formula $a$. The pairs  
$(\bar{x}',a')$ and $(\bar{y}'_i,b'_i)$ are obtained by a decomposition of $\bar{x}$ and $a$ into $\bar{x}'\bar{x}''\bar{x}'''$ and $a'\wedge a''\wedge a'''$ as follows: \begin{itemize}\item
$a'$ is the conjunction of the equations and the formulas of the form $\fini(x)$  which are reachable in    $\exists\bar{x}\,a$.
\item
$\bar{x}'$ is the vector the variables of $\bar{x}$ which are reachable in $\exists\bar{x}\,a$.
\item
$a''$ is the conjunction of the formulas of the form $\fini(x)$  which are non-reachable in $\exists\bar{x}\,a$. 
\item
$\bar{x}''$ is the vector the variables of $\bar{x}$ which are non-reachable in $\exists\bar{x}\,a$ and do not occur in the left hand sides of the equations of $a$.
\item
$a'''$ is the conjunction of the equations which are non-reachable in $\exists\bar{x}\,a$.
\item
$\bar{x}'''$ is the vector the variables of $\bar{x}$ which are non-reachable in $\exists\bar{x}\,a$ and occur in the left hand sides of the equations of $a$.
\item
$b^*_i$ is the formula obtained  by removing from $b_i$ the formulas of the form $\fini(u)$ which occur also  in $a''$
\item
$\bar{y}'_i$ is the vector of the variables of $\bar{y}_i\bar{x}'''$ which are reachable in $\exists\bar{y}_i\bar{x}'''\,b^*_i$.
\item
$b'_i$  is the conjunction of the equations and the formulas of the form $\fini(x)$  which are reachable in $\exists\bar{y}_i\bar{x}'''\,b^*_i$.

\item $K\subseteq\{1,...,n\}$ is the set of the indices $i$ such that $i\in K$ if and only if no variable of $\bar{x}''$ occurs in $b'_i$.
\item
The formula $\bigwedge_{i\in K}\neg^5(\exists\bar{y}'_i\,
b'_i)^*$ is the formula $\bigwedge_{i\in K}\neg^5(\exists\bar{y}'_i\,
b'_i)$ in which we have renamed the quantified variables  so that they satisfy the discipline of the formulas in $T$.
\end{itemize}

Let $\bar{x}',\bar{x}'',\bar{x}''',\bar{y}'$ and $a',a'',a''',b^*_i,b'_i$ be the vector of variables and the basic formulas defined above. According to Definition \ref{acc}, (i) all the variables of $\bar{x}''$ and $\bar{x}'''$ do not occur in $a'$, otherwise they are reachable in $\exists\bar{x}\,a$. On the other hand, since the first negation in the left hand side of rule (15)  is of the form $\neg^4$ then according to Definition \ref{wor} (ii)  $a$ is a solved basic formula and thus  $\bar{x}'''$ is the vector of the left hand sides of the equations of $a'''$ and its variables do not occur in $a''$. Thus, according to (i) and (ii) the left hand side of rule (15) is equivalent in $T$ to 

\[\neg(\exists\bar{x}'\, a'\wedge(\exists\bar{x}''\, a''\wedge(\exists\bar{x}'''\, a'''\wedge\bigwedge_{i=1}^{n}\neg(\exists\bar{y}_i\,b_i)))).\]
Since $a$ is a solved basic formula then $a'''$ is a solved basic formula which contains only equations and thus according  to Property \ref{unique} we have $T\models\exists!\bar{x}'''\,a'''$. Thus, according to Property \ref{unique1} the preceding formula is equivalent in $T$ to 
\[\neg(\exists\bar{x}'\, a'\wedge(\exists\bar{x}''\, a''\wedge\bigwedge_{i=1}^{n}\neg(\exists\bar{x}'''\,a'''\wedge(\exists\bar{y}_i\,b_i)))),\]
which, according to the discipline of the formulas in $T$ (the quantified variables have distinct names and different from those of the free variables ), is equivalent in $T$ to 
\begin{equation}\label{e20}\neg(\exists\bar{x}'\,a'\wedge(\exists\bar{x}''\, a''\wedge\bigwedge_{i=1}^{n}\neg(\exists\bar{x}'''\bar{y}_i\,a'''\wedge b_i))).\end{equation}
 Since all the nested negations in the left hand side of rule (15) are of the form $\neg^5$  then according to Definition \ref{wor}, for all $i\in\{1,...,n\}$, the set of the equations of $a$ is included in those of  $b_i$. As a consequence, the formula (\ref{e20}) is equivalent in $T$ to 
 \[\neg(\exists\bar{x}'\,a'\wedge(\exists\bar{x}''\, a''\wedge\bigwedge_{i=1}^{n}\neg(\exists\bar{x}'''\bar{y}_i\,b_i))),\]
i.e. to 
 \[\neg(\exists\bar{x}'\,a'\wedge(\exists\bar{x}''\, a''\wedge\bigwedge_{i=1}^{n}\neg(\exists\bar{x}'''\bar{y}_i\,b^*_i))).\]
Since all the nested negations in the left hand side of rule (15) are of the form $\neg^5$, then according to Definition \ref{wor}, for all $i\in\{1,...,n\}$,  $b^*_i$ is a solved basic formula. Thus, according to Property  \ref{tech}, the preceding formula is equivalent in $T$ to  
\[ \neg(
\exists\bar{x}'\,a'\wedge(\exists\bar{x}''\,a''\wedge 
\bigwedge_{i=1}^{n}\neg(\exists\bar{y}'_i\,b'_i))),
\]
which is equivalent in $T$ to 
\[ \neg(\exists\bar{x}'\, a'\wedge(\bigwedge_{i\in K}\neg(\exists\bar{y}'_i\,b'_i))\wedge(
\exists\bar{x}''\, a''\wedge 
\bigwedge_{i\in \{1,...,n\}-K}\neg(\exists\bar{y}'_i\,b'_i))),
\]
where $K\subseteq\{1,...,n\}$ is the set of the indices $i$ such that $i\in K$ if and only if no variable of $\bar{x}''$ occurs in $b'_i$.  Since all the nested negations in the left hand side of rule (15) are of the form $\neg^5$  then according to Definition \ref{wor}, for all $i\in\{1,...,n\}-K$, the variables of $\bar{y}'_i$ are reachable in $\exists\bar{y}'_i\,b'_i$ and  the formula $b'_i$ is a solved basic formula. Moreover, since each $b'_i$ does not contain sub-formulas of the form  $\fini(x)$ which occur also in $a''$ (see the construction of $b^*_i$), then the  formula $\exists\bar{x}''\, a''\wedge 
\bigwedge_{i\in \{1,...,n\}-K}\neg(\exists\bar{y}'\,b'_i)$ satisfies  the conditions of  Property \ref{infini}. As a consequence, according to Property \ref{infini} the preceding formula is equivalent in $T$ to 

\[ \neg(\exists\bar{x}'\, a'\wedge\bigwedge_{i\in K}\neg(\exists\bar{y}'_i\,b'_i)),
\]
i.e. to 
\[ \neg(\exists\bar{x}'\, a'\wedge\bigwedge_{i\in K}\neg(\exists\bar{y}'_i\,b'_i)^*),
\]
where $\bigwedge_{i\in K}\neg^5(\exists\bar{y}'_i\,
b'_i)^*$ is the formula $\bigwedge_{i\in K}\neg^5(\exists\bar{y}'_i\,
b'_i)$ in which we have renamed the quantified variables  so that they satisfy the discipline of the formulas in $T$. According to the conditions of application of rule (15) and the form of the negations in the left hand side of this rule, we check easily that we can fix the negations of the preceding formula as follows 
\[ \neg^5(\exists\bar{x}'\, a'\wedge\bigwedge_{i\in K}\neg^5(\exists\bar{y}'_i\,b'_i)^*).
\]
Thus, rule (15) is correct in $T$.

\subsubsection*{Correctness of rule (16)}
\[
\neg^4\left[\begin{array}{l} \exists\bar{x}\,a\wedge
q\wedge\\[2mm] \neg^5\left[\begin{array}{l} \exists\bar{y}\,
b\wedge\\[2mm]\bigwedge_{i=1}^{n}\neg^5(\exists\bar{z}_i\,c_i)
\end{array}\right]\end{array}\right]\Longrightarrow \left[\begin{array}{l}
\neg^4(\exists\bar{x}\,a\wedge q\wedge\neg^5(\exists\bar{y}\,b))\wedge\\[2mm]\bigwedge_{i=1}^{n}\neg^4(\exists\bar{x}\bar{y}\bar{z}_i\, c_i\wedge q_0)^*\\
\end{array}\right]\]
with $n>0$,  and $q_0$ is the formula $q$ in which all the
occurrences of $\neg^k$ have been replaced by
$\neg^0$. The formula $\bigwedge_{i=1}^{n}\neg^4(\exists\bar{x}\bar{y}\bar{z}_i\, c_i\wedge q_0)^*$ is the formula $\bigwedge_{i=1}^{n}\neg^4(\exists\bar{x}\bar{y}\bar{z}_i\, c_i\wedge q_0)$ in which we have renamed the quantified variables so that they satisfy the discipline of the formulas of $T$.

The left hand side of rule (16)  is equivalent in
 $T$ to
 \[
\neg(\exists\bar{x}\,a\wedge q\wedge \neg( \exists\bar{y}\,
b\wedge\neg\bigvee_{i=1}^{n}(\exists\bar{z}_i\,c_i))).
\]
Since the first negation of $\neg(\exists\bar{y}\,b...$ in the left hand side of rule (16) is of the form $\neg^5$ then according to  Definition \ref{wor}, all the variables of $\bar{y}$ are
reachable in $\exists\bar{y}\,b$, and thus according to Property
\ref{acc1} we have $T\models\exists?\bar{y}\,b$. According to
Property \ref{r2}, the precedent formula is equivalent in $T$ to
 \[
\neg(\exists\bar{x}\,a\wedge q\wedge \neg((\exists\bar{y}\,
b)\wedge\neg(\exists\bar{y}\,b\wedge\bigvee_{i=1}^{n}(\exists\bar{z}_i\,c_i)))).
\]
By distributing the $\wedge$ on the $\vee$ and the $\exists$ on
the $\vee$ and since the quantified variables have distinct names and different from those of the free variables then the preceding formula is equivalent in $T$ to

\[
\neg(\exists\bar{x}\,a\wedge q\wedge \neg((\exists\bar{y}\,
b)\wedge\neg\bigvee_{i=1}^{n}(\exists\bar{z}_i\bar{y}\,b\wedge c_i))),
\]
i.e. to 
\[
\neg(\exists\bar{x}\,a\wedge q\wedge ((\neg(\exists\bar{y}\,
b))\vee\bigvee_{i=1}^{n}(\exists\bar{z}_i\bar{y}\,b\wedge c_i))),
\]
i.e. to 
\[
\neg(\exists\bar{x}\,(a\wedge q\wedge \neg(\exists\bar{y}\,
b))\vee\bigvee_{i=1}^{n}(a\wedge q\wedge(\exists\bar{z}_i\bar{y}\,b\wedge c_i))),
\]
which, according to the discipline of the formulas in $T$ (the quantified variables have distinct names and different from those of the free variables), is equivalent in $T$ to 
\[
\neg(\exists\bar{x}\,(a\wedge q\wedge \neg(\exists\bar{y}\,
b))\vee\bigvee_{i=1}^{n}(\exists\bar{z}_i\bar{y}\,a\wedge q\wedge b\wedge c_i)),
\]
i.e. to 
\[
\neg((\exists\bar{x}\,a\wedge q\wedge \neg(\exists\bar{y}\,
b))\vee\bigvee_{i=1}^{n}(\exists\bar{x}\bar{z}_i\bar{y}\,a\wedge q\wedge b\wedge c_i)),
\]
i.e. to 

\[
\neg(\exists\bar{x}\,a\wedge q\wedge \neg(\exists\bar{y}\,
b))\wedge\bigwedge_{i=1}^{n}\neg(\exists\bar{x}\bar{y}\bar{z}_i\,a\wedge q\wedge b \wedge
 c_i).
\]
Since we have  $\neg^5(\exists\bar{y}\,b...$  in the left hand side of rule (16) then according to Definition \ref{wor}, we have (i) $T\models b\rightarrow a$.  But since we have also $\neg^5(\exists\bar{z}_i\,c_i)$ for all $i\in\{1,...,n\}$, then according to Definition \ref{wor} we have (ii) $T\models c_i\rightarrow b$. From (i) and (ii) we have $T\models c_i\rightarrow (a\wedge b)$. Thus the preceding formula is equivalent in $T$ to 
\[
\neg(\exists\bar{x}\,a\wedge q\wedge \neg(\exists\bar{y}\,
b))\wedge\bigwedge_{i=1}^{n}\neg(\exists\bar{x}\bar{y}\bar{z}_i\,c_i\wedge
 q),
\]
i.e. to 
\[
\neg(\exists\bar{x}\,a\wedge q\wedge \neg(\exists\bar{y}\,
b))\wedge\bigwedge_{i=1}^{n}\neg(\exists\bar{x}\bar{y}\bar{z}_i\,c_i\wedge
 q)^*,
\]
where $\bigwedge_{i=1}^{n}\neg^4(\exists\bar{x}\bar{y}\bar{z}_i\, c_i\wedge q)^*$ is the formula $\bigwedge_{i=1}^{n}\neg^4(\exists\bar{x}\bar{y}\bar{z}_i\, c_i\wedge q)$ in which we have renamed the quantified variables so that they satisfy the discipline of the formulas of $T$. According to the conditions of application of rule (16) and the form of the negations in the left hand side of this rule, we check easily that we can fix the negations of the preceding formula as follows 
 \[
\neg^4(\exists\bar{x}\,a\wedge q\wedge \neg^5(\exists\bar{y}\,
b))\wedge\bigwedge_{i=1}^{n}\neg^4(\exists\bar{x}\bar{y}\bar{z}_i\,c_i\wedge
 q_0)^*,
\]
where $q_0$ is the formula $q$ in which all the
occurrences of $\neg^k$ have been replaced by
$\neg^0$. Thus  rule (16) is correct in $T$.
$\\[3mm]${\emph{Proof, third part:}} Every repeated application until termination of the rewriting rules on an initial working formula 
produces a conjunction of final working formulas. Recall that  we
write $\bigwedge_{i\in I}\varphi_i$, and call \emph{conjunction}
each formula of the form
$\varphi_{i_1}\wedge\varphi_{i_2}\wedge...\wedge\varphi_{i_n}\wedge\vrai$.
In particular, for $I=\emptyset$, the conjunction $\bigwedge_{i\in
I}\varphi_i$ is reduced to $\vrai$. Moreover, we do not
distinguish two formulas which can be made equal using the
following transformations of  sub-formulas:
\[\begin{array}{@{}c@{}}\varphi\wedge\varphi\Longrightarrow\varphi,\;\;\varphi\wedge\psi\Longrightarrow\psi\wedge\varphi,\;\;(\varphi\wedge\psi)\wedge\phi\Longrightarrow\varphi\wedge(\psi\wedge\phi),\\\varphi\wedge\vrai\Longrightarrow\varphi,\;\;\varphi\vee\faux\Longrightarrow\varphi.\end{array}\]

Let us show first that every substitution of a sub-working formula of a conjunction of working formulas by a conjunction of working formulas produces a conjunction of working formulas.  Let $\bigwedge_{i\in I}\varphi_i$ be a conjunction of working formulas. Let $\varphi_k$ with $k\in I$ be an element of this conjunction  of depth $d_k$. Two cases arise: \begin{enumerate} \item 
 We replace $\varphi_k$ by a conjunction of working formulas. Thus, let $\bigwedge_{j\in J_k}\phi_j$ be a conjunction of working formulas which is equivalent to $\varphi_k$ in $T$. The conjunction of working formulas $\bigwedge_{i\in I}\varphi_i$ is equivalent in $T$ to \[(\bigwedge_{i\in I-\{k\}}\varphi_i)\wedge(\bigwedge_{j\in J_k}\phi_j)\] which is clearly a conjunction of working formulas.
\item
 We replace a strict sub-working formula of $\varphi_k$ by a conjunction of working formulas. Thus, let $\phi$ be a sub-working formula of $\varphi_k$ of depth $d_{\phi}<d_k$ (thus $\phi$ is different from $\varphi_k$). Thus, $\varphi_k$ has a sub-working formula\footnote{By considering that the set of the sub-formulas of any formula $\varphi$ contains also the whole formula $\varphi$.} of the form 
\[\neg(\exists\bar{x}\alpha\wedge(\bigwedge_{l\in L} \psi_l)\wedge\phi),
\]
where $L$ is a finite (possibly empty) set and all the $\psi_l$ are working formulas. Let $\bigwedge_{j\in J}\phi_j$ be a conjunction of working formulas which is equivalent to $\phi$ in $T$.  Thus the preceding sub-working formula of $\varphi_k$ is equivalent in $T$ to 
\[\neg(\exists\bar{x}\alpha\wedge(\bigwedge_{l\in L} \psi_l)\wedge(\bigwedge_{j\in J}\phi_j)),
\]
which is clearly a sub-working formula and thus $\varphi_k$ is equivalent to a working formula and thus  $\bigwedge_{i\in I}\varphi_i$ is equivalent to a conjunction of working formulas.
\end{enumerate}
 From 1 and 2 we deduce that  (i) every substitution of a sub-working formula of a conjunction of working formulas by a conjunction of working formulas produces a conjunction of working formulas. 

 Since each rule transforms a working formula into a conjunction of working formulas, then according to the sub-section \emph{``proof: first part''} and (i) we deduce that every repeated application of the rewriting rules on an initial  working formula terminates and produces a conjunction of working formulas. Thus, since an initial working formula starts by $\neg^4$ and all its
other negations are of the form $\neg^0$ then all long the 
application of our rules and by going down along the nested negations of any working formula $\varphi$ obtained after any finite application of our rules, we can build many series of negations which represent the paths that we should follow from the top negation of $\varphi$ to reach one of the  sub-working formulas of $\varphi$ of depth equal to one. Each of these series is of the one of the following forms: 
\begin{itemize}
\item
a series of  $\neg^4$ followed by a possibly series of  $\neg^0$,
\item
a series of  $\neg^4$ followed by one $\neg^1$, followed by a possibly series of $\neg^0$, 
\item
a series of  $\neg^4$ followed by one $\neg^2$, followed by a possibly series of $\neg^0$, 
\item
a series of  $\neg^4$ followed by one $\neg^3$, followed by a possibly series of $\neg^0$, 
\item
a series of  $\neg^4$ followed by one or two $\neg^5$,
\item
one or two $\neg^5$.
\end{itemize}
While all the negations of these series are not of the form $\neg^5$ or their length is greater than 2 then one of the rules (1),...,(16) can still be  applied. As a consequence, when no rule can be
applied, we obtain a conjunctions  of formulas of depth less or equal
to 2 in which all the negations are of the form $\neg^5$. It is a conjunction of final working formulas. Since all the rules do not introduce new free variables then Property \ref{resol} holds. \end{proof}

\subsection{The Solving Algorithm}
Let $p$ be a formula. Solving  $p$ in $T$  proceeds  as follows:

(1) Transform the formula $\neg p$ (the negation of p) into a wnfv normalized formula $p_1$
equivalent to $\neg p$ in $T$. 

(2) Transform $p_1$ into the following initial 
working formula $p_2$
\[p_2=\neg^4(\exists\varepsilon\,\vrai\wedge\neg^0(\exists\varepsilon\,\vrai\wedge p_1)),\] where all the occurrences of $\neg$ in $p_1$ are replaced by $\neg^0$.  

(3) Apply the preceding rewriting rules on $p_2$ as many time as possible. According to Property \ref{resol} we obtain at the end 
 a wnfv conjunction $p_3$ of final working formulas of the form 
\[\bigwedge_{i=1}^n\neg^5(\exists\bar{x}_i\,\alpha_i\wedge\bigwedge_{j=1}^{n_i}\neg^5(\exists\bar{y}_{ij}\,
\beta_{ij})).\]  According to Property \ref{fina}, the formula $p_3$ is equivalent in $T$ to the following wnfv conjunction $p_4$ of general solved formulas 
\[\bigwedge_{i=1}^n\neg(\exists\bar{x}_i\,\alpha_i\wedge\bigwedge_{j=1}^{n_i}\neg(\exists\bar{y}_{ij}\,
\beta^*_{ij})),\]
where $\beta^*_{ij}$ is the formula $\beta_{ij}$ from which we have removed all the equations which occur also in $\alpha_i$. Since $p_4$ is equivalent to $\neg p$ in $T$, then $p$ is equivalent  in $T$ to 

 \[\neg\bigwedge_{i=1}^n\neg(\exists\bar{x}_i\,\alpha_i\wedge\bigwedge_{j=1}^{n_i}\neg(\exists\bar{y}_{ij}\,
\beta^*_{ij})),\] 
which is equivalent  to the following disjunction $p_5$
 \[\bigvee_{i=1}^n(\exists\bar{x}_i\,\alpha_i\wedge\bigwedge_{j=1}^{n_i}\neg(\exists\bar{y}_{ij}\,
\beta^*_{ij})).\] 
This is the final answer of our solver to the initial constraint $p$.  Note that the negations which were at the beginning of each general solved formula of $p_4$ have been removed and the top conjunction of $p_4$ has been replaced by a disjunction. As a consequence, the set of the solutions of the free variables of $p_5$ is nothing other than the union of the solutions of each formula of the form $\exists\bar{x}_i\,\alpha_i\wedge\bigwedge_{j=1}^{n_i}\neg(\exists\bar{y}_{ij}\,
\beta^*_{ij})$. According to Definition \ref{kiki}, each of these formulas is written in an explicit solved form which enables us to easily extract the solutions of its free variables. On the other hand, two cases arise: 
\begin{itemize}
\item
 If $p_4$ does not contain free variables then according to Property \ref{vrai} the formula $p_4$ is of the form $\bigwedge_{i=1}^n \neg(\exists\varepsilon\,\vrai)$ and thus $p_5$ is  of the form $\bigvee_{i=1}^n \exists\varepsilon\,\vrai$. Two cases arise:  if $n=0$ then $p_5$ is the empty disjunction (i.e. the formula $\faux$). Else, if $n\neq 0$ then since we do not distinguish between $\varphi\wedge\varphi$ and $\varphi$,  $p_5$ is the formula $\exists\varepsilon\,\vrai$.
 
\item   If $p_4$ contains at least one free variable then according to Property \ref{vrai} neither $T\models p_4$ nor $T\models\neg p_4$ and thus neither $T\models\neg p_5$ nor $T\models p_5$. 
\end{itemize}
Since $T$ has at least one model and since $p_5$ is equivalent to $p$ in $T$ and does not contain news free variables then we have  the following theorem: 
\begin{theoreme}\label{thm}
Every formula is equivalent in $T$ either to $\vrai$, or to $\faux$,
or to a wnfv formula which has at least one free variable, which is equivalent neither to $\vrai$ nor to $\faux$, and where the solutions of the free variables are expressed in a clear  and explicit way.

\end{theoreme}
The fact that $T$ accepts at least one model is vital in this theorem. In fact, if $T$ does not have models then the formula $\vrai$ can be equivalent to $\faux$ in $T$. In other words,  a formula can be equivalent to $\vrai$ in $T$ using a finite application of our rules and equivalent to $\faux$ using another different finite application of our rules.  Theorem \ref{new} prevents these kinds of conflicts and shows that $T$ has at least three models ${\mathcal{D}}$, ${\mathcal{T}r}$ and ${\mathcal{R}a}$ and thus $T\models\neg(\vrai\leftrightarrow\faux)$.

\begin{corollaire}
$T$ is a complete theory.
\end{corollaire}
Note that using Theorem \ref{thm} and the properties  \ref{comb} and \ref{tech}, we
get  Maher's decision procedure~\cite{Maher} for the basic theory of finite or infinite trees.

\section{ Implementation of our algorithm}

We have implemented our algorithm in C++ and CHR (Constraint Handling
Rules)~\cite{Fru98,book,chrsite}.  The C++ implementation is a straightforward
extension of those given in \cite{moi3}.  It uses records and pointers and
releases unused pointers after each rule application.  The CHR implementation
was done using Christian Holzbaur's CHR library of Sicstus Prolog 3.11.0. It
consists of 18 CHR constraints and 73 CHR rules -- most of them are needed for
the complicated rules (15) and (16) of our algorithm.  Even if our C++
implementation has given better performances, we think that it is interesting to
show how can we translate our rules into CHR rules. We will be able to quickly
prototype optimizations and variations of our algorithm and to parallelize
it. For CHR, the implementation of this complex solver helps to understand what
programming patterns and language features can be useful. 
The CHR code without
comments and examples, but pretty-printed, is about 250 lines, which is one
seventh of the size of our C++ implementation.
Indeed for code size
and degree of abstraction it seems only possible and interesting to describe the
CHR implementation, and we do so in the following.  
The reader can find our full CHR
implementation at \url{http://khalil.djelloul.free.fr/solver.txt} and can
experiment with it online using webchr at
\url{http://chr.informatik.uni-ulm.de/~webchr/}.

\subsection{Constraint Handling Rules (CHR) Implementation}


CHR manipulates conjunctions of constraints that reside in a constraint store. 
Let $H$, $C$ and $B$ denote conjunctions
of constraints.  A simplification rule $H \simp C \; \com \; B$ replaces
instances of the CHR constraints $H$ by $B$ provided the guard test $C$ holds.
A propagation rule $H \prop C \; \com \; B$ instead just adds $B$ to $H$
without removing anything. The hybrid simpagation rules will come handy in the
implementation: $H_1 \backslash H_2 \simp C \; \com \; B$ removes matched
constraints $H_2$ but keeps constraints $H_1$.

The constraints of the store comprise the state of an execution. 
Starting from an arbitrary initial store (called query), CHR rules
are applied exhaustively until a fixpoint is reached. 
%
%
Trivial non-termination of a propagation rule application is avoided by applying
it at most once to the same constraints.

Almost all CHR implementations execute queries from left to right
and apply rules top-down in the textual order of the program \cite{refined}.  A
CHR constraint in a query can be understood as a procedure that goes efficiently
through the rules of the program. When it
matches a head constraint of a rule, it will look for the other
constraints of the head in the constraint store and check the guard. 
On success, it will apply the rule. The rule application cannot be undone.
If the initial constraint has not been removed after trying all rules,
it will be put into the constraint store. Constraints from the store will be
reconsidered if newly added constraints constrain its variables.

\subsubsection{CHR Constraints}

The implementation consists of 18 constraints: two main constraints that
encode the tree data structure of the working formulas (nf/4) and the atomic
formulas (of/2),
9 auxiliary constraints that perform reachability analysis, variable renaming
and copying of formulas, and 7 constraints that encode execution control
information, mainly for rules (15) and (16).

In more detail, {\tt nf(ParentId,Id,K,ExVars)} describes a {\bf n}egated
quantified basic {\bf f}ormula with the identifier of its parent node, its own
identifier Id, the level K from $\neg^k$ and the list of existentially
quantified variables.  {\tt Var=FlatTerm of Id} denotes an equation between a
variable and a flat term (a variable or a function symbol applied to variables)
that belongs to the negated sub-formula with the identifier Id.  {\tt finite(U)
of Id} denotes the relation $\fini(U)$.


It is easy to represent any working formula $\varphi$ using conjunctions of nf/4
and of/2 constraints. It is enough to create one nf/4 constraint for
each quantified basic formula of $\varphi$ and to use a conjunction of of/2
constraints to enumerate the atomic formulas linked to each quantified basic
formula. 
\[\]

\begin{example}
Let $\varphi$ be the following working formula
\[
\neg^4\left[\begin{array}{l}\exists u\, u=1\wedge\\\left[\begin{array}{l}\neg^0(\exists\varepsilon\,u=s(v))\wedge\\\neg^0(\exists w_1\, u=s(w_1)\wedge w_1=s(v))\wedge\\\neg^5(\exists\varepsilon\,v=s(u)\wedge u=1\wedge\left[\begin{array}{l} \neg^5(\exists\varepsilon\,v=s(u)\wedge u=1\wedge\fini(w_1))\wedge\\\neg^5(\exists w_3\, v=s(u)\wedge u=1\wedge w_2=s(w_3)\wedge\fini(w_3))\end{array}\right])\end{array}\right]\end{array}\right].\]
$\varphi$ can be expressed using the following conjunction of constraints:
\[\begin{array}{l} {\tt nf(Q,P1,4,[U]),U=1 \,of\, P1,}\\
                             {\tt  nf(P1,P2,0,[\,]), U=S(V) \,of\, P2,}\\
                             {\tt  nf(P1,P3,0,[W1]), U=S(W1) \,of\, P3 , W1=S(V)\,of \, P3 ,}\\
                              {\tt nf(P1,P4,5,[\,]), V=S(U)\, of\, P4, U=1\,of\, P4}\\
                              {\tt nf(P4,P5,5,[\,]), V=S(U) \,of \,P5, U=1\, of\, P5, finite(W1)\,of\,P5 } \\
                              {\tt nf(P4,P6,5,[W3]), V=S(U) \,of\, P6, U=1 \,of \,P6,W2=S(W3)\,of\, P6, finite(W3)\,of\,P6    }
\\
\end{array}
\]

\end{example} 

\subsubsection{CHR Rules}

The rules (1) to (14) have a rather direct translation into CHR rules.
It seems hard to come up with a more concise implementation.
{\small
%
\begin{verbatim}
% 1 Locally simplify equations
(1) @ nf(Q,P,1,Xs) \ U=U of P <=> true.
(2) @ nf(Q,P,1,Xs) \ V=U of P <=> gt(U,V) | U=V of P.
(3) @ nf(Q,P,1,Xs), U=V of P \ U=G of P <=> gt(U,V) | V=G of P.
(4) @ nf(Q,P,1,Xs), U=F of P,  U=G of P <=> notsamefunctor(F,G) | true(P).
(5) @ nf(Q,P,1,Xs), U=F of P \ U=G of P <=>    samefunctor(F,G) | 
                                                          same_args(F,G,P).
(6) @ nf(Q,P,1,Xs) <=> nf(Q,P,2,Xs).

% 2 finiteness check
(7) @ nf(P0,P,2,Xs), finite(U) of P \ finite(U) of P <=> true.
(8) @ nf(P0,P,2,Xs), U=V of P \ finite(U) of P <=> var(V) | finite(V) of P.
(9+10)@nf(P0,P,2,Xs),U=T of P \ finite(U) of P <=> nonvar(T) |
                                     reach_args(U,T,P), finite_args(U,T,P).
(11) @ nf(Q,P,2,Xs) <=> nf(Q,P,3,Xs).

% 4/0-4/1 copy down before solving
(12) @ nf(Q,P,4,Xs), A of P, nf(P,P1,0,Ys) ==> A of P1.
       nf(Q,P,4,Xs)        \ nf(P,P1,0,Ys) <=> nf(P,P1,1,Ys).

% 4/3-4/4 replace down after solving
(13) @ nf(Q,P,4,Xs),U=V of P, nf(P,P1,3,Ys)\ U=G of P1 <=> V\==G | U=V of P1.
       nf(Q,P,4,Xs)         \ nf(P,P1,3,Ys)            <=> nf(P,P1,4,Ys).

% 4/5-true trivial satisfaction - each A of P1 also occurs as A of P
(14) @ nf(Q,P,4,Xs), nf(P,P1,5,Ys) <=>
       \+(findconstraint(P1,(A of P1),_), \+findconstraint(P,(A of P),_)) | 
                                                           true(P).
\end{verbatim}
} 
\noindent
Note that rules (1) to (5) are similar to the classical CHR equation solver for
flat rational trees \cite{book,marc-thom-rt}. By
applying results of \cite{marc-thom-rt}, we can show that the worst-case time
complexity of these rules of the algorithm is quadratic in the
size of the equations.

In the rules (2) and (3), the predicate {\tt gt(U,V)} checks if
${\tt U}\succ {\tt V}$. 
%
%
%
Note that the constraint {\tt true(P)} used in rule (4) removes all
constraints associated with {\tt P} using an auxiliary rule not shown.

In rule (9+10) {\tt reach\_args(U,T,P)} checks reachability of {\tt U} from
itself in {\tt P}. If so, {\tt true(P)} will be executed and thus 
{\tt P} will be removed, implementing rule (9). Otherwise, 
the subsequent {\tt finite\_args(U,T,P)} will propagate down the {\tt
finite} relation from {\tt U} to its arguments, implementing rule (10).

In the rules (12) and (13) we handle equations one by one (due to the chosen
granularity of the constraints), and thus we need auxiliary second CHR rules
that perform the update of the level {\tt K} afterwards.

For rule (14) the implementation is easy when nested
negation-as-absence~\cite{chr06-neg} is used to verify that there is no
constraint in the sub-formula that is not in the main formula.
Negation-as-absence can be directly encoded in CHR, but then it requires two
additional rules per negation. Instead, we have chosen to use in the guard of
the rule the CHR library built-in {\tt findconstraint(Var,Pattern,Match)} that
returns on backtracking all constraints {\tt Match} that match {\tt Pattern} and
that are indexed on variable {\tt Var} together with negation-as-failure
provided by the Prolog built-in {\tt $\backslash+$}.

 

The translation of the complex rules (15) and (16) of the
algorithm require 40 CHR rules, because several non-trivial new expressions have to be computed. Simpagation rules and auxiliary constraints collect the nested nf/4 constraints, compute the
reachable variables and atomic formulas, rename the quantified variables and
produce updated nf/4 and of/2 constraints. In order not to overburden the
reader with technical details, we omit the description of those 40 rules.

\subsection{Benchmarks: Two partner game}

Let us consider 
the following two partner game:  An ordered pair $(i,j)$ is given, with $i$ a non-negative (possibly null)  integer and $j\in\{0,1\}$. One after another, each player changes the values of $i$ and $j$ according to the following rules
\begin{itemize}
\item
If $j=0$ then the actual player should replace $i$ by $i-1$ in the pair $(i,j)$.
\item
If $j=1$ and $i$ is odd then the actual player can either replace $i$ by $i+1$ or replace $j$ by $j-1$, in the pair $(i,j)$.
\item
If $j=1$ and $i$ is even then the actual player can either replace  $i$ by $i+1$ and $j$ by $j-1$ in the pair  $(i,j)$ or replace only $i$ by $i+1$ in the pair let $(i,j)$
\end{itemize}
The first player who cannot  keep $i$ non negative has lost. This game can be represented by the following directed infinite graph: 

\includegraphics*[width=12cm]{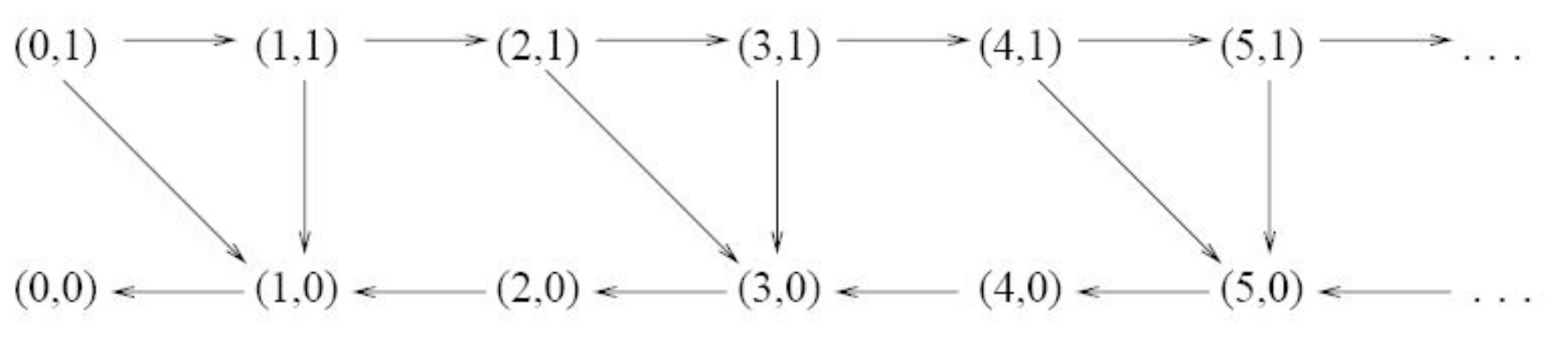}

It is clear that the player which is at the position $(0,0)$ and should play has lost. Suppose that it is the turn of player $A$ to play. A position $(n,m)$   is
called \emph{k-winning} if, no matter the way the other player $B$  plays, it
is always possible for $A$ to win, after having made at most $k$ moves. It is easy to show that
\[winning_k(x)=\left[\begin{array}{l}\exists
y\,move(x,y)\wedge\neg(\\\exists
x\,move(y,x)\wedge\neg(\\...\\\exists
y\,move(x,y)\wedge\neg(\\\exists
x\,move(y,x)\wedge\neg(\\false\hspace{1.4cm}{\underbrace{)...)}}_{2k}\end{array}\right]\]
where move$(x,y)$ means: ``starting from the position $x$ we
play one time and reach the position $y$". By moving down the
negations, we get an embedding of 2k alternated quantifiers.

Suppose that $\Ff$ contains the function symbols $0$, $1$, $ f$, $g$,
$c$ of respective arities $0$, $0$, $1$, $1$, $2$. We code the vertices
$(i,j)$ of the game graph by the trees $c(\bar{i},0)$ and $c(\bar{i},1)$ with 
$\bar{i} = (fg)^{i/2}(0)$ if $i$ is
even, and $\bar{i} = g(\overline{i-1})$ if $i$ is odd.%
\footnote{$(fg)^{0}(x)=x$ and $(fg)^{i+1}(x)=f(g({(fg)}^i(x)))$.}
The relation $move(x,y)$  is then defined as follows:
\[move(x,y)\stackrel{\mathrm{def}}{\leftrightarrow}
transition(x,y)\vee (\neg(\exists uv\, x=c(u,v))\wedge x=y)\] with
\[\begin{array}{lll}
transition(x,y)&\stackrel{\mathrm{def}}{\leftrightarrow}&\left[\begin{array}{l}\exists
u_1v_1u_2v_2\\
x=c(u_1,v_1)\wedge y=c(u_2,v_2)\wedge\\
\left[\begin{array}{l} (v_1=0\wedge v_2=v_1\wedge pred(u_1,u_2))\\
\vee\\ 
(v_1=1\wedge\left[\begin{array}{l}(\exists
w\,u_1=g(w)\wedge \left[\begin{array}{l}(u_2=f(u_1)\wedge v_2=v_1)\vee\\
(u_2=u_1\wedge v_2=0)
\end{array}\right])\vee\\
(\neg(\exists w\, u_1=g(w))\wedge u_2=g(u_1)\wedge (v_2=v_1\vee v_2=0))\end{array}\right])\\
\vee\\ 

(\neg(v_1=0)\wedge\neg(v_1=1)\wedge u_2=u_1\wedge v_2=v_1)
\end{array}\right]\end{array}\right]
\\ &&\\
pred(u_1,u_2)&\stackrel{\mathrm{def}}{\leftrightarrow}&\left[\begin{array}{l}(\exists
j\,u_1=f(j)\wedge\left[\begin{array}{l}(\exists k\,j=g(k)\wedge
u_2=j)\vee\\(\neg(\exists k\,j=g(k))\wedge
u_2=u_1)\end{array}\right])\vee\\(\exists j\,u_1=g(j)\wedge
\left[\begin{array}{l}(\exists k\,j=g(k)\wedge
u_2=u_1)\vee\\(\neg(\exists k\,j=g(k))\wedge
u_2=j)\end{array}\right])\vee\\
(\neg(\exists j\,u_1=f(j))\wedge\neg(\exists
j\,u_1=g(j))\wedge\neg(u_1=0)\wedge u_2=u_1)\end{array}\right]
\end{array}\]

If we take as input of our solver the formula $winning_k(x)$ then we
will get as output a disjunction of simple formulas where the solutions of the free variable $x$ represent all the $k$-winning
positions.

For $winning_1(x)$ our algorithm gives the following 
formula:
\[\exists u_1u_2\, x=c(u_1,u_2)\wedge u_1=g(u_2)\wedge u_2=0,\]
which corresponds to the solution $x=c(g(0),0)$. For $winning_2(x)$ our algorithm gives the following disjunction of simple  formulas
\[\left[\begin{array}{l}
(\exists u_1u_2\, x=c(u_1,u_2)\wedge u_1=g(u_2)\wedge u_2=0)\\
\vee\\
(\exists u_3u_4u_5u_6\, x=c(u_3,u_6)\wedge u_3=g(u_4)\wedge u_4=f(u_5)\wedge u_5=g(u_6)\wedge  u_6=0)
\end{array}\right],
\]
which corresponds to the solution $x=c(g(0),0)\vee x=c(g(f(g(0))) ,0)$. Note that $x$ is the only free variable in the two preceding disjunctions and its solutions represent the positions which are $k$-winning.

The times of execution (CPU time in milliseconds) of the formulas $winning_k(x)$
are given in the following table as well as a comparison with those obtained
using a decision procedure for decomposable theories \cite{moitplp} (even though
the later does not produce comprehensible results, i.e. explicit solved
forms). The benchmarks are performed on a 2.5Ghz Pentium IV
processor, with 1024Mb of RAM. The symbol ``-"  bellow means \emph{exhausting memory}.

\begin{tabular}{|l|c|c|c|c|c|c|c|c|c|}
 \hline k ($winning_k(x)$)  & 1 & 2 & 4 & 5 & 
7 & 10 & 20 & 40  \\\hline CHR  (our 16 rules) & 320 & 690 & 1750 & 2745 & 5390 & $ - $ &
$ - $ & $-$ \\\hline C++ \cite{moitplp}  & 28 & 50 & 115& 150 & 245 & 430 & 2115 &
$ - $   \\\hline C++  (our 16 rules)  & 25 & 40 & 90 & 115 & 175 & 315 & 1490
& 15910    \\\hline \end{tabular}

This decision procedure takes from $10\%$ to $40\%$ more time,
comparing with our C++ implementation to solve the $winning_k(x)$ formulas of our game and overflows the memory  for
$k>20$, i.e. 40 nested alternated quantifiers. Our C++ implementation has
better performance and is able to give all the $winning_k$ strategies in a
clear and explicit way until $k=40$, i.e. 80 nested alternated quantifiers.

The execution times of $winning_k(x)$ using our CHR implementation are 12-30
times slower than those obtained using our C++ implementation and the maximal
depth of working formula that can be solved is 14 ($k=7$). These results are in
line with the experience that the overhead of using declarative CHR without
optimisations induces an overhead of about an order of magnitude over
implementations in procedural languages. As discussed in the conclusions,
switching to a more recent optimizing CHR compiler may close the gap to a small
constant factor.

The algorithm given in \cite{moitplp} is a decision procedure in the form of
five rewriting rules which for every decomposable theory $T$ transforms a
first-order formula $\varphi$ into a conjunction $\phi$ of final formulas easily
transformable into a Boolean combination of existentially quantified
conjunctions of atomic formulas. This decision procedure does not warrant that the solutions of the free variables are expressed in a clear and explicit way and can even produce formulas having free variables but being always true or false in $T$. In fact, for our two player game, we got
conjunctions of final formulas where the solutions of the free variable $x$ was
incomprehensible, especially from $k=5$.

 We also tried to use Remark 4.4.2 of \cite{moitplp} which gives a way to get a
disjunction of the form
 \begin{equation}\label{trop}
 \bigvee_{i\in I}(\exists\bar{x}'_i\,\alpha'_i\wedge\bigwedge_{j\in
J_i}\neg(\exists\bar{y}'_{ij}\,\beta'_{ij}))\end{equation} as output of the
decision procedure. As the author of \cite{moitplp} wrote: \emph{"it is more
easy to understand the solutions of the free variables of this disjunction of
solved formulas than those of a conjunction of solved formulas"}. That is of
course true, but this does not mean that the solutions of the free variables of
this formula are expressed in a clear and explicit way. In fact, we got a
disjunction of the form (\ref{trop}) where many variables which occurred in left
hand sides of equations of $\alpha'_i$ occurred also in left hand sides of
equations of some $\beta'_{ij}$. Moreover,  many formulas of the preceding disjunction contained occurrences of the free variable $x$ but after a hard and complex manual checking we found them equivalent to $\faux$.  As a consequence, the solutions of $x$  
 was completely not evident to understand and we could not
extract  clear and understandable $winning_k(x)$ strategies for all $k\geq 5$. In order to simplify the formula
(\ref{trop}) we finally used our solving algorithm on it and have got a
disjunction of simple formulas equivalent to (\ref{trop}) in $T$ in which: (1) all the formulas having free occurrences of $x$  but being always  false in $T$ have been removed,  (2) the solutions of the free variable $x$ were expressed in a clear and explicit way.

We now discuss why our  solver  is faster
than the decision procedure of K. Djelloul. The latter uses many times a
particular distribution (rule (5) in \cite{moitplp}) which decreases the depth
of the working formulas but increases exponentially the number of conjunctions
of the working formulas until overflowing the memory. Our solving algorithm uses
a similar distribution (rule (16)) but only after a necessary propagation step
which copies the basic formulas into the sub-working formulas and checks if
there exists no working formulas which contradict their top-working formula. This
step enables us to remove  the inconsistent working formulas and to not lose time with solving a huge working formulas (i.e. of big depth) which contradicts their top-working formulas. It also prevents us
from making exponential distributions between huge inconsistent working formulas
which finally are all equivalent to $\faux$. Unfortunately, we cannot add this
propagation step to the decision procedure of \cite{moitplp} since it uses many
properties which hold only for the theory of finite or infinite trees and not
for any decomposable theory $T$. 

The game introduced in this paper was inspired from those given in \cite{moitplp} but is different.  Solving a $winning_k(x)$ formula in this game generates many huge working formulas which contradict their top-working formulas. Our algorithm removes directly these huge working formulas  after the first propagation step (rules (1),...,(13)). The decision procedure cannot detect this inconsistency and is obliged to apply a  costly rule (rule (5) in \cite{moitplp})  to decrease the size of these inconsistent working formulas until  finding basic inconsistent formulas of the form $\neg(a\wedge\neg(\exists\varepsilon\,\vrai))$ or $\neg(\exists\varepsilon\,\faux\wedge\varphi)$. At each application of this rule, the depth of the working formulas decreases but the number of conjunctions increase exponentially until overflowing the memory. This explains why for this game the decision procedure overflows the memory for $k>20$ while our solver can compute the $winning_k(x)$ strategies until $k=40$. 

\subsection{Benchmarks: Random normalized formulas}

We have also tested our 16 rules on   randomly generated normalized formulas such that in each sub-normalized formula of the form $\neg(\exists\bar{x}\,\alpha\wedge\bigwedge_{i=1}^n\varphi_i)$, with the $\varphi_i$'s normalized formulas and $n\geq 0$,  we have: 
\begin{itemize}
\item $n$ is a positive integer randomly chosen between 0 and 4.
\item The number of the atomic formulas in the basic formula $\alpha$  is randomly chosen between 1 and 8. Moreover, the atomic formula $\vrai$ occurs at most once in $\alpha$. 
\item The vector of variables and the atomic formulas of $\exists\bar{x}\,\alpha$ are randomly generated starting from a set containing 10 variables, the relation $\fini$ and 6 function symbols: $f_0, f_1,f_2, g_0,g_1,g_2$. Each function symbol $f_j$ or $g_j$ is of arity $j$ with  $0\geq j\geq 2$.
\end{itemize}

The benchmarks were realized on a 2.5Ghz Pentium IV processor  with 1024Mb of RAM as follows: For each integer $1\geq d\geq 42$ we generated 10 random normalized formulas\footnote{We  of course renamed the quantified variables of each randomly generated normalized formula so that it respects the discipline of the formulas in $T$} of depth $d$,  we solved them and computed the average execution time (CPU time in milliseconds). Once again, the performances (time and space) of our 16 rules are impressive comparing with those of the decision procedure for  decomposable theories.
\noindent
\begin{tabular}{|c|c|c|c|c|c|c|c}
 \hline  $ d$   & 4 &  8 & 
 12 & 22  & 26& 41  &    \\\hline CHR  (our 16 rules) & 1526 & 4212 & 16104 & $-$ & $-$& $-$ &
                        				\\\hline C++ \cite{moitplp}  & 108 & 375 & 1486 & 18973  & $-$ & $-$& 
						  \\\hline C++  (our 16 rules)  &  88 &  202 & 504 & 3552 & 11664 &  2142824 & 
                                                        \\\hline \end{tabular}

Note that for $d=42$, all the normalized formulas could not be solved and overflowed the memory.

 \section{Discussion and conclusion}
We gave in this paper a first-order axiomatization of an extended theory $T$ of
finite or infinite trees, built on a signature containing not only an infinite
set of function symbols but also a relation $\fini(t)$ which enables to
distinguish between finite or infinite trees. We showed that $T$ has at least
one model and proved its completeness by giving not only a decision procedure
but a full first-order constraint solver which transforms any first-order
constraint $\varphi$ into an equivalent disjunction $\phi$ of simple formulas
such that $\phi$ is either the formula $\vrai$, or the formula $\faux$, or a
formula having at least one free variable, being equivalent neither to $\vrai$
nor to $\faux$ and where the solutions of the free variables are expressed in a
clear and explicit way. This algorithm detects easily formulas that have free
variables but are always true or always false in $T$ and is able to solve any first-order 
constraint satisfaction problem in $T$.  Its correctness 
implies the completeness of $T$.

On the other hand S. Vorobyov~\cite{vo14} has shown that the problem of deciding
if a proposition is true or not in the theory of finite or infinite trees is
non-elementary, i.e. the complexity of all algorithms solving propositions is
not bounded by a tower of powers of $2's$ (top down evaluation) with a fixed
height. A. Colmerauer and T. Dao \cite{dao2} have also given a proof of
non-elementary complexity of solving constraints in this theory.  As a
consequence, our algorithm does not escape this huge complexity and the function
$\alpha(\varphi)$ used to show the termination of our rules illustrates this
result.

We implemented our algorithm in C++ and CHR and compared both performances with
those obtained using a recent decision procedure for decomposable theories
\cite{moitplp}.  This decision procedure is not able to present the solutions of
the free variables in a clear and explicit way and overflows the memory while
solving normalized formulas with depth $d>40$. Our C++ implementation is faster
than this decision procedure and can solve normalized formulas of depth
$d=80$. This is mainly due to the fact that our algorithm uses two steps: (1) a
top-down propagation of constraints and (2) a bottom-up elimination of
quantifiers and depth reduction of the working formulas. In particular, the
first step enables to minimize the number of application of costly distributions
 and avoids to lose time
with solving huge formulas which contradict their top-formulas.

Future implementation work will focus on our CHR implementation, since
from previous experience we are confident that we can get the performance
overhead down to a small constant factor while gaining the possibility to
prototype variations of our algorithm in a very high level language. 
%
Switching to a more recent optimizing CHR compiler from K.U. Leuven would most
likely improve performance.
We also think
that we can minimize the use of the debated negation-as-absence~\cite{chr06-neg}
by introducing reference counters for the two main constraints. This should also
give us the possibility to obtain a parallel implementation that is derived from
the existing one with little modification, similar to what has been done for
parallelizing the union-find algorithm in CHR~\cite{puf}.

\[\]

{\bfseries{Acknowledgments}}  We  thank  Alain Colmerauer for our
very long discussions about the theory of finite or infinite trees and its models. Many thanks also to the anonymous referees for their careful reading and suggestions which help us to improve this paper.  Khalil Djelloul thanks the DFG research project GLOB-CON for funding and supporting his research. Thanks also to Marc Meister and Hariolf Betz for their kind review of this article.

\end{document}